\documentclass[aps,prb,reprint,groupedaddress]{revtex4-2}
\usepackage{graphicx,multirow,gensymb,amsmath,mathtools}
\usepackage{makecell,tabularx}
\DeclareUnicodeCharacter{2212}{-}
\begin{document}

%\preprint{}

\title{Quantum oscillation study of the large magnetoresistance in Mo substituted WTe$_2$ single crystals}


\author{Sourabh Barua}
\email[corr. author ]{sourabh.barua@cup.edu.in, \\ baruasourabh@gmail.com}
%\thanks{Present Address: Birla Institute of Technology, Mesra, Ranchi, 835215, Jharkhand, India}
%\altaffiliation{ Birla Institute of Technology, Mesra, Ranchi, 835215, Jharkhand, India}
\altaffiliation[Present Address: ]{Central University Punjab, Bathinda, 151401, Punjab, India }
\author{M. R. Lees}
%\thanks{}
%\affiliation{}

\author{G. Balakrishnan}
%\email[]{}
%\thanks{}

\author{P. A. Goddard}
\email[corr. author ]{p.goddard@warwick.ac.uk}
%\thanks{}
\affiliation{Department of Physics, University of Warwick, Coventry, CV4 7AL, United Kingdom}


\date{\today}

\begin{abstract}
The list of interesting electrical properties exhibited by transition metal dichalcogenides has grown with the discovery of extremely large magnetoresistance (MR) and type-II Weyl semimetal behaviour in WTe$_2$ and MoTe$_2$. The extremely large MR in WTe$_2$ is still not adequately understood. Here, we systematically study the effect of Mo substitution on the quantum oscillations in the MR in WTe$_2$. The MR decreases with Mo substitution, however, the carrier concentrations extracted from the quantum oscillations show that the charge compensation improves. We believe that earlier interpretations based on the two-band theory, which attribute the decrease in MR to charge imbalance, could be incorrect due to over-parametrization. We attribute the decrease in MR in the presence of charge compensation to a fall in transport mobility, which is evident from the residual resistivity ratio data. The quantum scattering time and the effective masses do not change within experimental errors upon substitution. 
\end{abstract}

%\keywords{}

\maketitle

\section{\label{Intro}Introduction}
With the rise of graphene, other two-dimensional materials have emerged, among which the transition metal dichalcogenides (TMDCs) have proved to be particularly notable~\cite{Geim2009,Bhimanapati2015}. TMDCs such as MoS$_2$ offer enormous potential in photonics and optoelectronic devices, while others, including NbSe$_2$ and TaSe$_2$, display interesting many-particle electronic states like charge density waves and superconductivity~\cite{Manzeli2017}. TMDCs have received renewed interest after various unique properties were recently discovered in WTe$_2$ and MoTe$_2$. These include the discovery in WTe$_2$ of an extremely large non-saturating magnetoresistance (MR)~\cite{Ali2014} and superconductivity at high pressures \cite{Pan2015,Kang2015}, and in MoTe$_2$ of a large MR~\cite{Chen2016,Lee2018} and an enhancement of the superconducting transition temperature under pressure~\cite{Qi2016}. In addition, WTe$_2$ is predicted to be a type-II Weyl semimetal~\cite{Soluyanov2015} and this aspect has been probed successfully by angle resolved photoelectron spectroscopy (ARPES)~\cite{Wu2016,CWang2016,Bruno2016}. Similar studies have also confirmed the type-II Weyl character of MoTe$_2$ \cite{Deng2016,Tamai2016} and Mo substituted WTe$_2$ (Mo$_{x}$W$_{1-x}$Te$_2$)~\cite{Chang2016,BelopolskiPRB2016,BelopolskiNatComm2016}. Monolayer WTe$_2$ is also reported to be a quantum spin Hall insulator~\cite{Zheng2016,Wu2018} and charge-to-spin conversion~\cite{Zhao2020} as well as edge transport have been reported~\cite{Kononov2020}. Quantum oscillations and high-mobility electrons have recently been reported in topologically insulating monolayers of WTe$_2$~\cite{Wang2021}.

The extremely large MR in WTe$_2$ has been attributed to nearly perfect compensation of electrons and holes in the two-band model~\cite{Ali2014}. ARPES and quantum oscillations, which can give information about the carrier concentrations, mostly report four pockets; two electron and two hole pockets~\cite{Pletikosic2014,Jiang2015,Wu2015,Zhu2015,Rhodes2015,Xiang2015,Wang2015}. 
Proper mapping of the frequencies in the quantum oscillations with electron or hole pockets is vital in determining whether the electron and hole concentrations are equal. Most studies attribute the smallest and largest frequencies to hole pockets and the middle two frequencies to electron pockets. However, Cai~\textit{et al}.~\cite{Cai2015} have identified them differently in a pressure-dependent study. 

While the MR is typically discussed in terms of charge compensation, alternative mechanisms, like a field-induced metal-insulator transition~\cite{Zhao2015,Xiang2015} and Kohler's law~\cite{WangYL2015}, have also been proposed. Moreover, in the two-band model, there is no consensus on whether the mobility or the charge compensation is more important in accounting for the observed behaviour~\cite{Fatemi2017,Na2016,Wang2019,Yi2017,Wang2015}. Often, the MR remains large and non-saturating even when a significant charge imbalance is introduced via gating~\cite{Wang2016}. The MR decreases at higher temperatures, which is attributed to a deviation from charge compensation. However, an ARPES study has found no evidence of such a deviation in the electronic structure~\cite {thirupathaiah_2017} while another ARPES study found that the charge compensation is possible at best only in a very narrow range of temperature~\cite{wang_evidence_2017}. All this has raised questions about the appropriateness of the two-band model for the MR in WTe$_2$. Studies on the effect of non-stoichiometry caused by Te vacancies and Mo substitution (Mo$_{x}$W$_{1-x}$Te$_2$) can shed further light on the large MR in WTe$_2$. Although these studies report a decrease in MR with Mo substitution, the carrier concentrations extracted from the quantum oscillations have not previously been compared to those extracted from the fit to the two-band theory~\cite{Lv2016,Fu2018,Gong2018}. Here, we study the quantum oscillations and MR of four different compositions to shed light on the evolution of the Fermi surfaces and carrier concentrations with Mo substitution in WTe$_2$, and test the validity of the charge compensation theory.

\section{Experimental Details}

Single crystals of various compositions of Mo substituted WTe$_2$ (W$_{1-x}$Mo$_x$Te$_2$, $x = 0$, 0.05, 0.1, 0.2) were grown using chemical vapour transport technique with TeCl$_4$ as the transport agent. For pure WTe$_2$, the starting material was W and Te powder in stoichiometric ratio, while for the substituted compounds it was polycrystalline powder prepared earlier by heating stoichiometric ratio of the constituent elements W, Mo and Te at 700-760~$\degree$C for 5-7 days in evacuated quartz tubes. For the crystal growth, the starting material was taken in 20-25~cm long quartz tubes and evacuated and sealed. The tubes were then placed in horizontal three zone furnaces with the charge end and the growth end of the tubes kept at 900~$\degree$C and 700-800~$\degree$C, respectively. The growth was performed over a period of three weeks. X-ray diffraction measurements were performed on the single crystals using a high resolution Panalytical X'Pert Pro diffractometer with Cu K$\alpha_1$ radiation ($\lambda = 1.5406$~\AA) and Cu K$\alpha_1$ hybrid monochromator. The crystals were mounted on zero diffraction plates made of silicon cut at special orientation so that there were no background Bragg peaks from the sample holder. A Laue X-ray imaging system with a Photonic-Science Laue camera was used to acquire the Laue back-scattered diffraction images of the single crystals. The chemical composition of the single crystals was determined using an Oxford Instruments energy dispersive X-ray spectrometer fitted in a Zeiss SUPRA 55--VP scanning electron microscope. Electrical transport and MR measurements down to 2~K and in magnetic fields up to 9~T were performed in a Quantum Design Physical Property Measurement System (PPMS) using the AC transport measurement option. Four probe electrical contacts were made on the single crystals using silver wires and conductive silver paint. Currents of 1~mA were used in the AC resistivity measurements.

\section{Results}

\begin{figure}[t!]
\centering
\includegraphics[width=\linewidth]{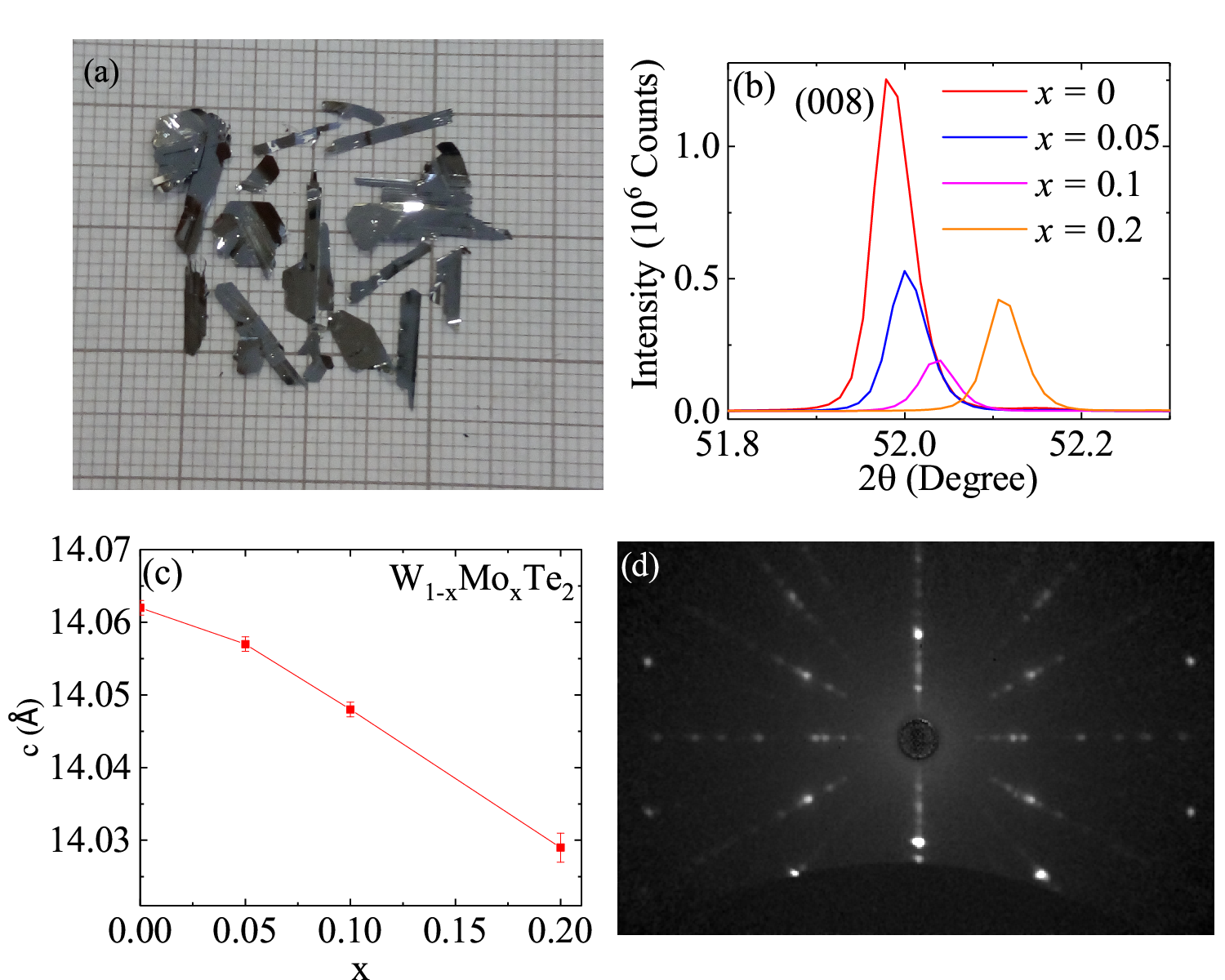}
\caption{(a) Single crystals of W$_{0.95}$Mo$_{0.05}$Te$_2$. (b) (008) Bragg peak seen in the XRD patterns of single crystals of W$_{1-x}$Mo$_x$Te$_2$ ($x = 0$, 0.05, 0.1, 0.2), showing a shift to higher 2$\theta$ values with increasing Mo substitution. (c) Lattice parameter along the $c$-axis as a function of Mo concentration ($x$) in the W$_{1-x}$Mo$_x$Te$_2$ single crystals. (d) Laue back-scattered diffraction image taken along the [001] direction for a single crystal of W$_{0.95}$Mo$_{0.05}$Te$_2$. }
\label{Figure_1}
\end{figure}

Fig.~\ref{Figure_1}(a) shows the representative image of the single crystals of W$_{1-x}$Mo$_x$Te$_2$ ($x = 0.05$), grown by the chemical vapour transport (CVT) method. WTe$_2$ grows in the distorted octahedral configuration, known as T$_{\text{d}}$-WTe$_2$~\cite{Lieth1977,Dawson1987}, with the space group $Pmn2_1$~\cite{Lee2015}. Fig.~\ref{Figure_1}(b) shows the (008) Bragg peak seen in the X-ray diffraction (XRD) pattern for the four compositions. The lattice parameter $c$, obtained from the XRD data, decreases with increasing Mo substitution, as seen in Fig.~\ref{Figure_1}(c). Since the radius of Mo$^{4+}$ ion (0.65~\AA) is smaller than that of the W$^{4+}$ ion (0.66~\AA)~\cite{Lv2016}, the lattice parameter along the $c$-axis decreases with Mo substitution and the Bragg peaks shift towards higher $2\theta$ values. The value for $c$ obtained for the non-substituted WTe$_2$ single crystal is $14.062\pm 0.002$~\AA~which is in agreement with that reported in literature~\cite{Lee2015,Brown1966}.

Figure~\ref{Figure_1}(d) shows the Laue back-scattered diffraction image along the [001] direction for a crystal with the composition W$_{0.95}$Mo$_{0.05}$Te$_2$ and demonstrates the very high quality of the crystals. The compositions of the substituted W$_{1-x}$Mo$_{x}$Te$_2$ crystals were determined using energy dispersive X-ray spectroscopy~(EDS) and were found to be close to their nominal values. 

\begin{figure}[tbh!]
\centering
\includegraphics[width=\linewidth]{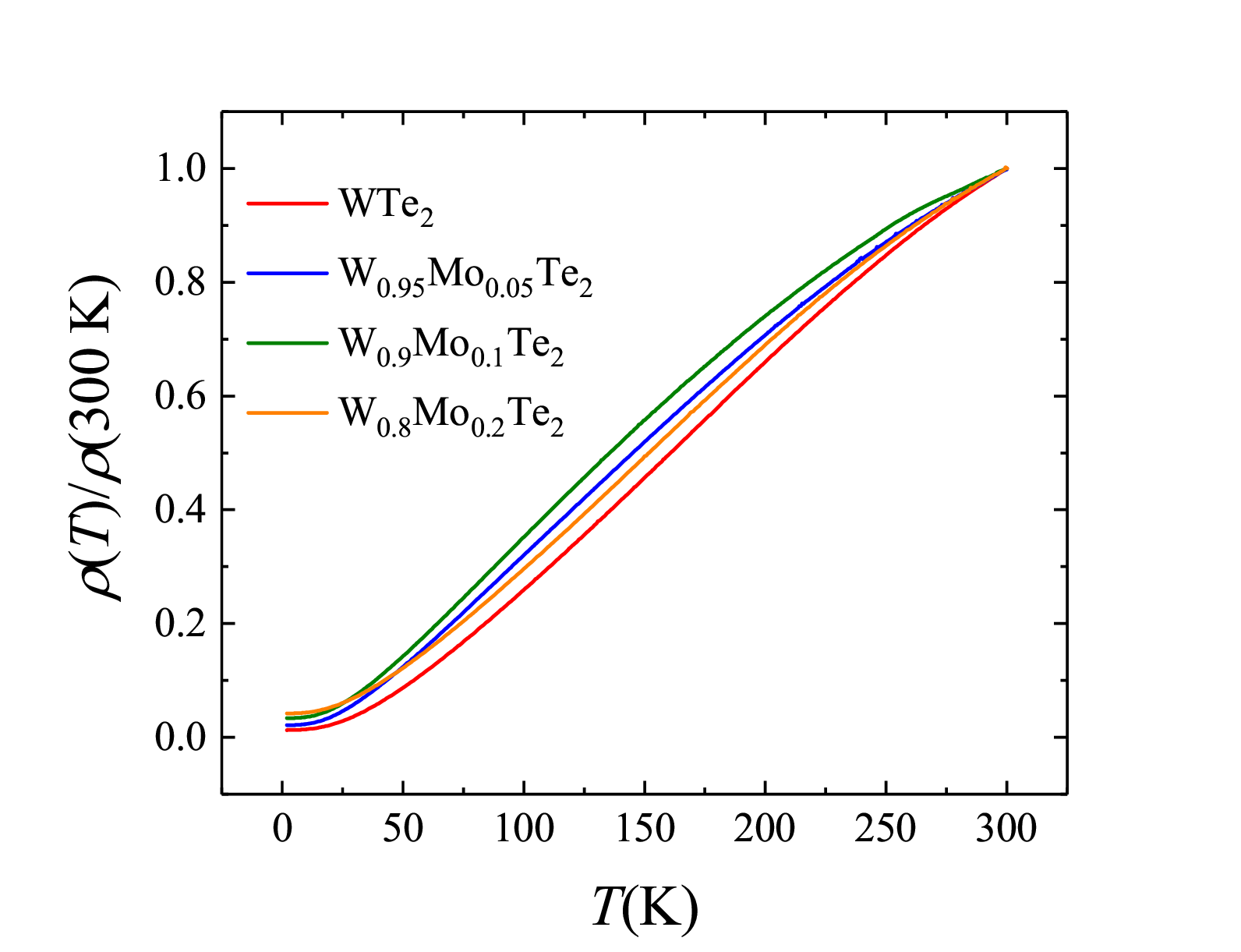}
\caption{Normalised resistivity $\rho(T)/\rho(300\mathrm{~K})$ versus temperature for the four different compositions of W$_{1-x}$Mo$_{x}$Te$_2$ studied.}
\label{Figure_2}
\end{figure}

\begin{table}[h!]
\caption{\label{Table_1} RRR and the relative change in MR at 9~T (as defined in the text) for single crystals of the different compositions of W$_{1-x}$Mo$_{x}$Te$_2$ ($x = 0$, 0.05, 0.1, 0.2). Both the RRR and MR decrease as the Mo substitution increases.}
\
\begin{ruledtabular}
\begin{tabular} {ccc}
Composition                  & RRR          & MR     \\
WTe$_2$                      & 79         & 8885   \\
W$_{0.95}$Mo$_{0.05}$Te$_2$  & 47           & 4871   \\
W$_{0.9}$Mo$_{0.1}$Te$_2$    & 30         & 1996   \\
W$_{0.8}$Mo$_{0.2}$Te$_2$    & 24         & 700    \\
\end{tabular}
\end{ruledtabular}
\end{table}

Electrical resistivity and MR measurements were performed on single crystal samples of the four compositions of W$_{1-x}$Mo$_{x}$Te$_2$. Fig.~\ref{Figure_2} shows the temperature variation of the normalised resistivity from room temperature down to 2~K. It is clear that all the compositions studied preserve the metallic nature of WTe$_2$, implying that the T$_{\text{d}}$~structure of WTe$_2$ is retained for these levels of Mo substitution. This is in agreement with the published phase diagram of the WTe$_2$-MoTe$_2$ system~\cite{Lv2017,Oliver2017}. The residual resistivity ratio (RRR) defined as $\rho(300 K)/\rho(2 K)$ for each of the four compositions is summarized in Table \ref{Table_1}. The RRR value for our pure WTe$_2$ sample is comparable to that previously reported for WTe$_2$ single crystals grown by CVT~\cite{Rhodes2015}. The RRR decreases with increasing Mo concentration in keeping with previous studies~\cite{Lv2016}. This is a sign of increased scattering at low temperatures due to an increase in the density of defects with substitution leading to a decrease in the transport scattering time and hence also the mobility.

\begin{figure}[t!]
\centering
\includegraphics[width=\linewidth]{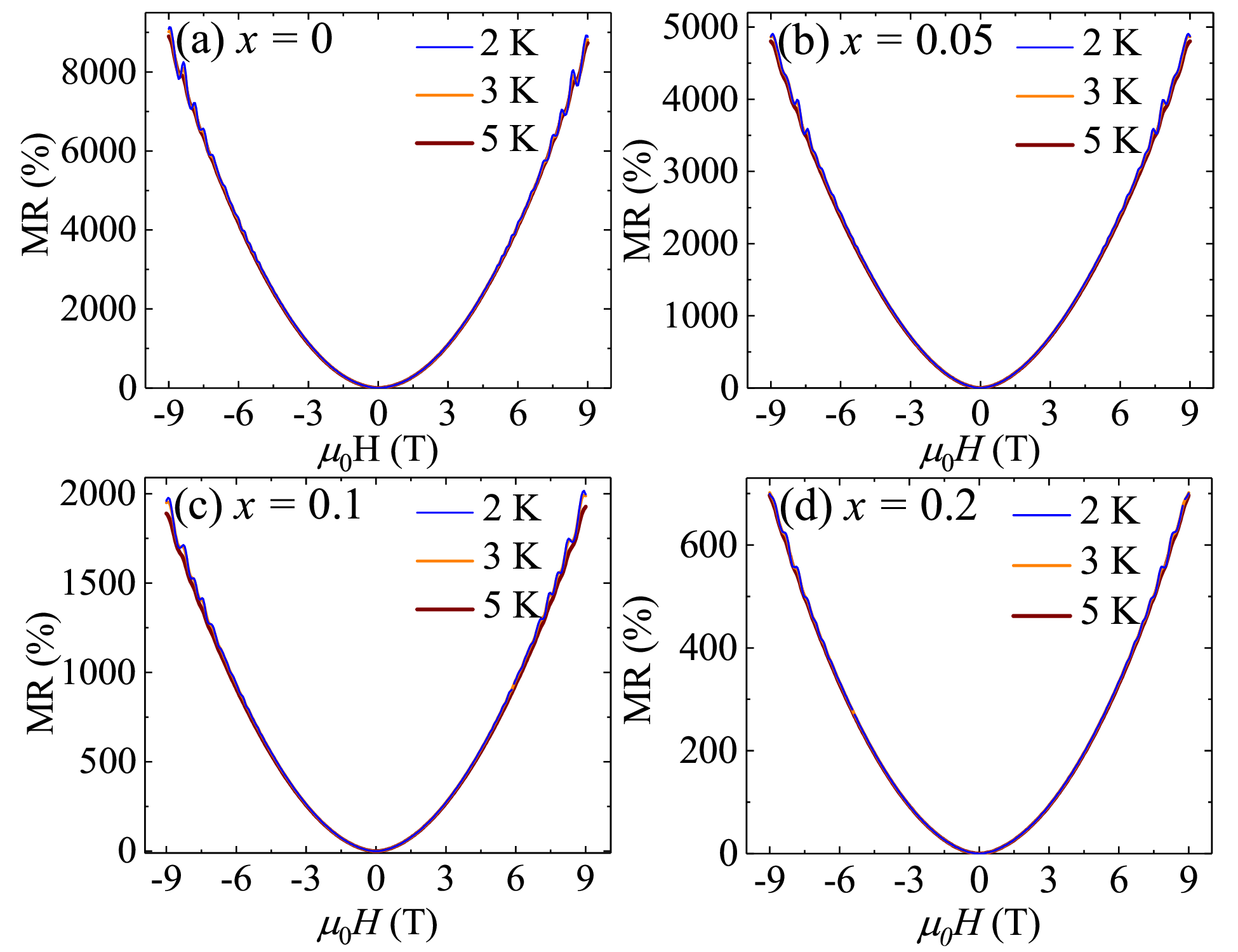}
\caption{Relative \% change in MR for the four different compositions of W$_{1-x}$Mo$_{x}$Te$_2$ at three different temperatures: (a) WTe$_2$ (b) W$_{0.95}$Mo$_{0.05}$Te$_2$ (c) W$_{0.9}$Mo$_{0.1}$Te$_2$ and (d) W$_{0.8}$Mo$_{0.2}$Te$_2$.}
\label{Figure_3}
\end{figure}

The MR of the single crystal samples of each of the four compositions was measured at low temperatures and in magnetic fields up to 9~T. All MR measurements were performed with the magnetic field applied parallel to the $c$-axis, easily identified as the direction perpendicular to the flat surfaces of the crystals. Fig.~\ref{Figure_3} shows the percentage relative change in MR at a few representative temperatures of 2,~3~and~5~K for the four compositions.

The relative change in MR is defined as,
\begin{equation}
\text{MR (\%)} = \frac{\rho_{xx}(\mu_0H)-\rho_{xx}(\mathrm{0})}{\rho_{xx}(\mathrm{0})}(\times 100)
\label{MRpercentage}
\end{equation}
where $\rho_{xx}$ is the longitudinal resistivity and $\mu_0H$ is the magnetic field. The values of \% relative change in MR at 9~T are given in Table~\ref{Table_1}. A large non-saturating MR is seen in all four different compositions, with a maximum of nearly 9000\% in case of the unsubstituted WTe$_2$ sample. The relative change in MR shows a trend of decreasing magnitude as one goes from the unsubstituted WTe$_2$ to the 20\% Mo substituted WTe$_2$ sample, which we propose is caused by the decreasing transport mobility. The MR also decreases as temperature increases for all the compositions.

\begin{figure}[tbh!]
\centering
\includegraphics[width=\linewidth]{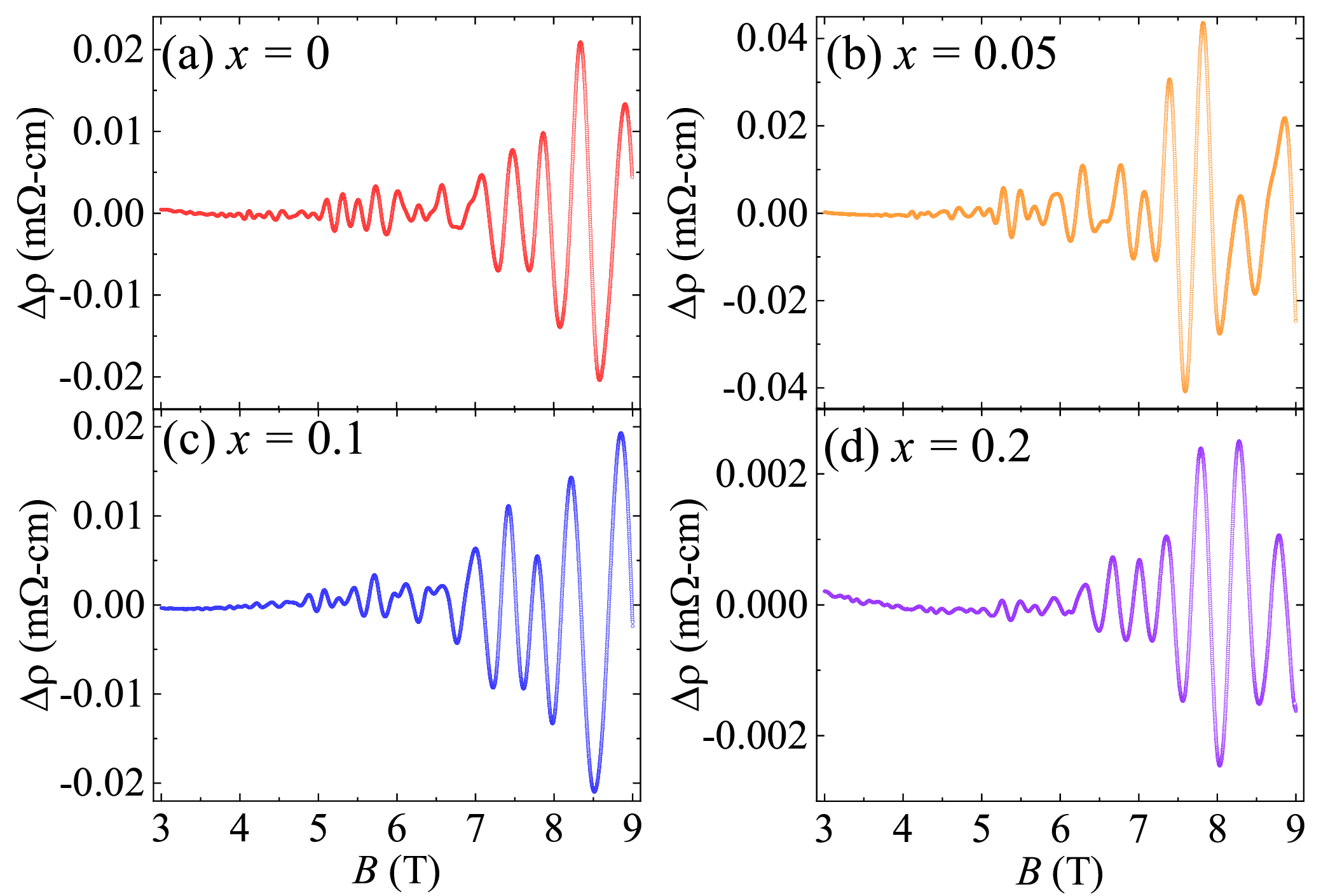}
\caption{Oscillatory component of MR at 2 K for the four different compositions: (a) WTe$_2$ (b) W$_{0.95}$Mo$_{0.05}$Te$_2$ (c) W$_{0.9}$Mo$_{0.1}$Te$_2$ and (d) W$_{0.8}$Mo$_{0.2}$Te$_2$.}
\label{Figure_4}
\end{figure}

All four compositions show prominent Shubnikov-de Haas oscillations in the MR. The oscillatory component of the MR is isolated by subtracting a second-order polynomial background from the raw data. Fig.~\ref{Figure_4} shows these extracted quantum oscillations for the four different compositions at 2~K. The individual oscillation frequencies correspond to extremal cross-sectional areas of different Fermi surfaces (electron or hole pockets) and are obtained by performing a fast Fourier transform (FFT) of an average of the oscillations in both rising and falling fields for $\mu_0H > 3$~T.

\begin{figure}[tbh!]
\centering
\includegraphics[width=\linewidth]{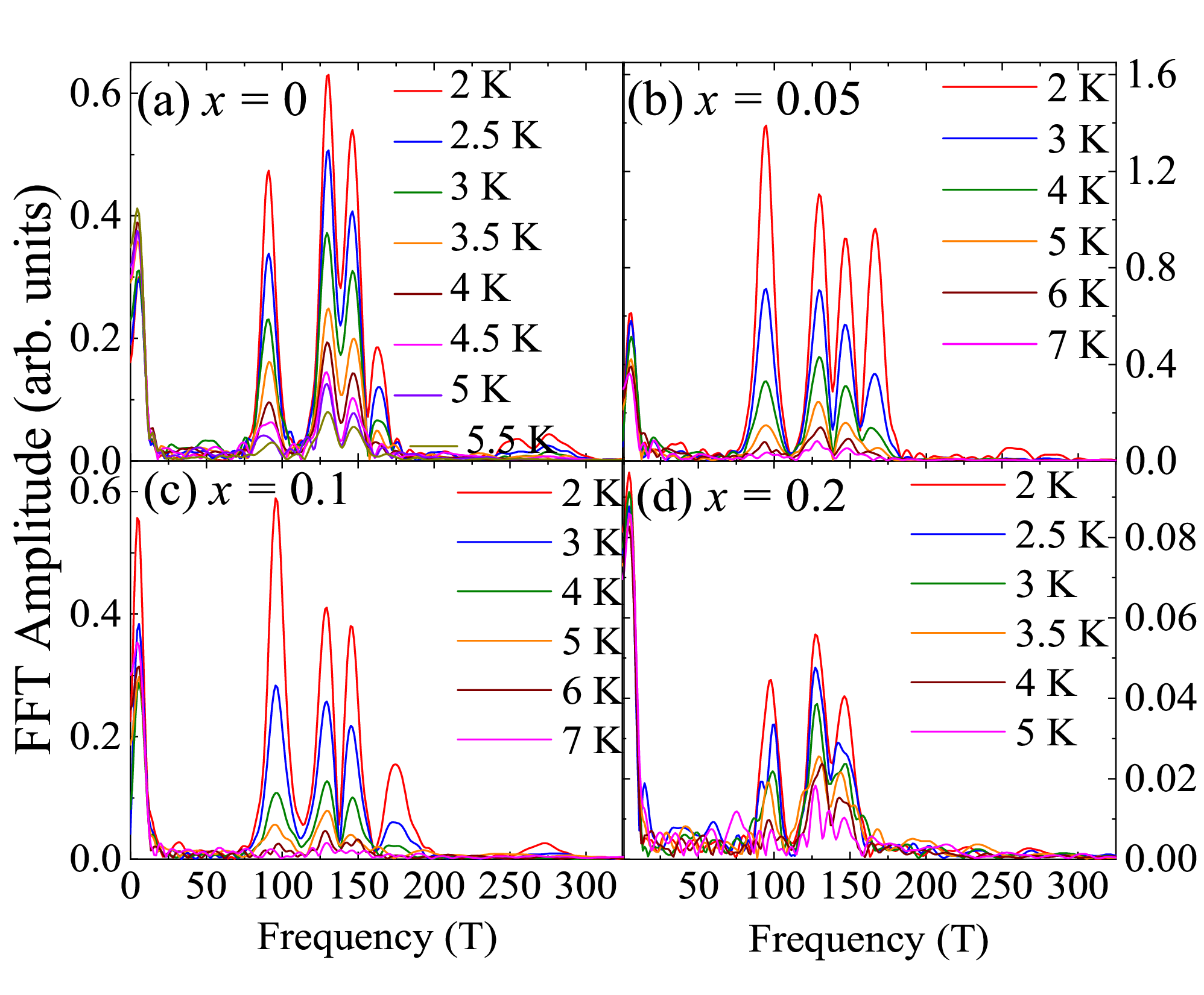}
\caption{FFT of the quantum oscillations in the MR of the single crystals of the four different compositions: (a) WTe$_2$ (b) W$_{0.95}$Mo$_{0.05}$Te$_2$ (c) W$_{0.9}$Mo$_{0.1}$Te$_2$ and (d) W$_{0.8}$Mo$_{0.2}$Te$_2$).}
\label{Figure_5}
\end{figure}

\begin{table*}
\caption{\label{Table_2} Carrier concentration $n_\mathrm{F}~(10^{18} \mathrm{cm}^{-3}$) corresponding to the different quantum oscillation frequencies (F$_{\alpha 1}$, F$_{\beta 1}$, F$_{\beta 2}$, F$_{\alpha 2}$) for W$_{1-x}$Mo$_{x}$Te$_2$ ($x = 0$, 0.05, 0.1, 0.2) assuming spherical Fermi pockets. Previously reported values of carrier concentration for unsubstituted WTe$_2$ is also given.} 
\begin{ruledtabular}
\begin{tabular}{cccccc}
Carrier concentration  & \multicolumn{2}{c}{WTe$_2$} & W$_{0.95}$Mo$_{0.05}$Te$_2$ & W$_{0.9}$Mo$_{0.1}$Te$_2$ & W$_{0.8}$Mo$_{0.2}$Te$_2$ \\
($10^{18} \mathrm{cm}^{-3}$) & This work & Previous work~\cite{Rhodes2015}& & & \\
$n_\mathrm{\mathrm{F}_{\alpha 1}}$   &  $4.87 \pm 0.03$ &  5.7  &  $5.14 \pm 0.03$ &  $5.31 \pm 0.05$ & $5.40 \pm 0.05$\\
$n_\mathrm{\mathrm{F}_{\beta 1}}$    &  $8.36 \pm 0.05$ &  8.4  &  $8.30 \pm 0.03$ &  $8.22 \pm 0.02$ & $8.18 \pm 0.03$\\
$n_\mathrm{\mathrm{F}_{\beta 2}}$    &  $9.97 \pm 0.06$ & 10    &  $10.0 \pm 0.1$ &  $9.96 \pm 0.05$ & $10.0 \pm 0.1$\\
$n_\mathrm{\mathrm{F}_{\alpha 2}}$   &  $11.9 \pm 0.1$ & 11.14 &   $12.1 \pm 0.1$ & $13.12 \pm 0.07$ &   -   \\
\end{tabular}
\end{ruledtabular}
\end{table*}

The results of the FFT for the pure sample at 2 K [see Fig.~\ref{Figure_5}(a)], show four peaks at the frequencies $\mathrm{F}_{\alpha 1} = \left(90.9 \pm 0.1\right)$~T, $\mathrm{F}_{\beta 1} = \left(130.0 \pm 0.1\right)$~T, $\mathrm{F}_{\beta 2} = \left(146.2 \pm 0.1\right)$~T and $\mathrm{F}_{\alpha 2} = \left(164.1 \pm 0.4\right)$~T. These agree very well with those previously reported in literature for WTe$_2$~\cite{Rhodes2015,Zhu2015,Cai2015}. We attribute the smallest and largest frequencies ($\mathrm{F}_{\alpha 1}$ and $\mathrm{F}_{\alpha 2}$) to a pair of nearly concentric hole pockets and the two middle frequencies ($\mathrm{F}_{\beta 1}$ and $\mathrm{F}_{\beta 2}$) to a pair of concentric electron pockets, following Refs.~\citenum{Rhodes2015,Zhu2015}, which rely on band structure calculations to match the frequencies with hole and electron pockets.

\begin{figure}[tbh!]
\centering
\includegraphics[width=\linewidth]{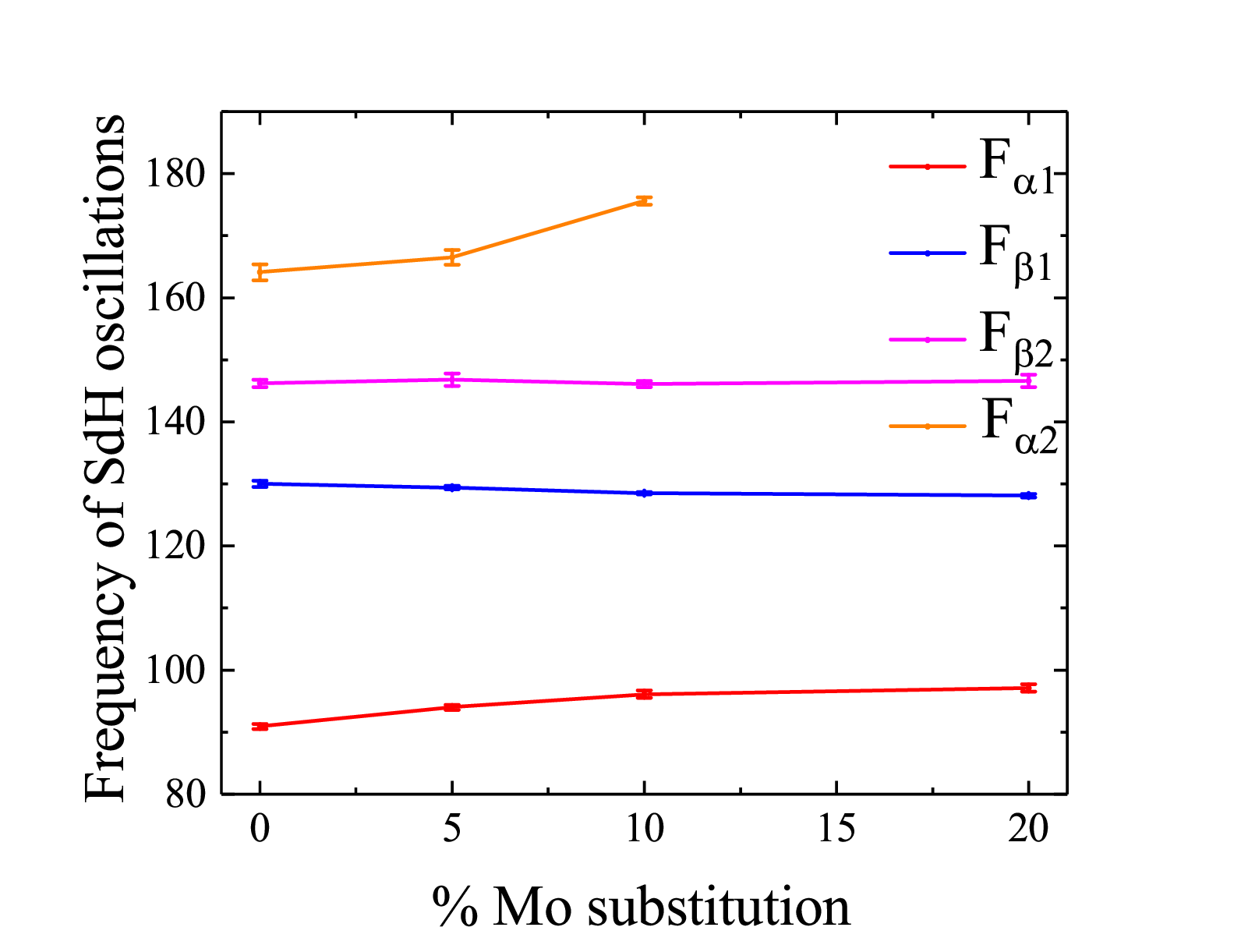}
\caption{Quantum oscillation frequencies as extracted from the FFT at 2~K for the four compositions W$_{1-x}$Mo$_{x}$Te$_2$ ($x = 0$, 0.05, 0.1, 0.2).}
\label{Figure_6}
\end{figure}

The frequencies of the quantum oscillations at 2~K for the four compositions as a function of Mo substitution are plotted in Fig.~\ref{Figure_6}. These values are also given in Supplemental Material Table~S1~\cite{SuppMat} along with previously reported values for unsubstituted WTe$_2$. While $\mathrm{F}_{\alpha 1}$ and $\mathrm{F}_{\alpha 2}$ increase somewhat with increased Mo content, $\mathrm{F}_{\beta 1}$ and $\mathrm{F}_{\beta 2}$ remain largely stable. The peak for the frequency $\mathrm{F}_{\alpha 2}$ is not resolved in the FFT of the oscillations in W$_{0.8}$Mo$_{0.2}$Te$_2$.  

The carrier concentration of the hole or electron pockets can be obtained via the Onsager relation~\cite{Zhu2015,Shoenberg1984}
\begin{equation}
\mathrm{F} = \frac{\hbar}{2 \pi e} A_{k},
\label{Equation_Onsager}
\end{equation}
which relates the frequency of oscillation F to the extremal cross-sectional area $A_{k}$ of the Fermi surface in a direction perpendicular to the applied field. Previous studies suggest that the Fermi pockets in WTe$_2$ are triaxial ellipsoids~\cite{Zhu2015,Rhodes2015}. However, angle-dependent quantum oscillation~\cite{Zhu2015} and anisotropic MR studies~\cite{Zhang_Xurui2021} have shown only mild anisotropy of the Fermi surface. The ratio of the Fermi wavevector ($k_\mathrm{F}$) along the $k_c$ and $k_a$ directions ($\gamma= k_c/k_a$) is expected to be less than 2.5 for all the four pockets~\cite{Zhu2015}. This is much less than seen in other 2D materials, e.g. graphite for which $\gamma = 12$~\cite{Zhang_Xurui2021}. The mild anisotropy has been used to justify the simplifying assumption of a spherical Fermi surface i.e. $A_{\mathrm{k}} = \pi {k_{\mathrm{F}}}^2$~\cite{Xiang2015,Lv2016} for WTe$_2$ which then allows an estimate to be made of the carrier density $n_{\mathrm{F}} = k_{\mathrm{F}}^3/3 \pi^2$.  

T$_\mathrm{d}$-MoTe$_2$ and T$_\mathrm{d}$-Mo$_{1-x}$W$_x$Te$_2$ are isostructural to WTe$_2$ and their the lattice parameters do not differ much~\cite{Rhodes_2017,Jin_2018}, suggesting that Mo substitution affects the lattice parameters isotropically, from which it can be assumed that the Fermi surface would also be affected isotropically. Indeed, ARPES studies have confirmed that the Fermi-surface topology in Mo$_{1-x}$W$_x$Te$_2$ is unchanged from that of WTe$_2$~\cite{Rhodes_2017,Jin_2018}, implying that if a spherical Fermi surface can be assumed for WTe$_2$ then it can also be assumed for Mo substituted WTe$_2$. We have calculated the carrier concentrations for each of the four frequencies for the compositions using the spherical surface assumption and these are presented in Table~\ref{Table_2}. 

\begin{figure}[tbh!]
\centering
\includegraphics[width=\linewidth]{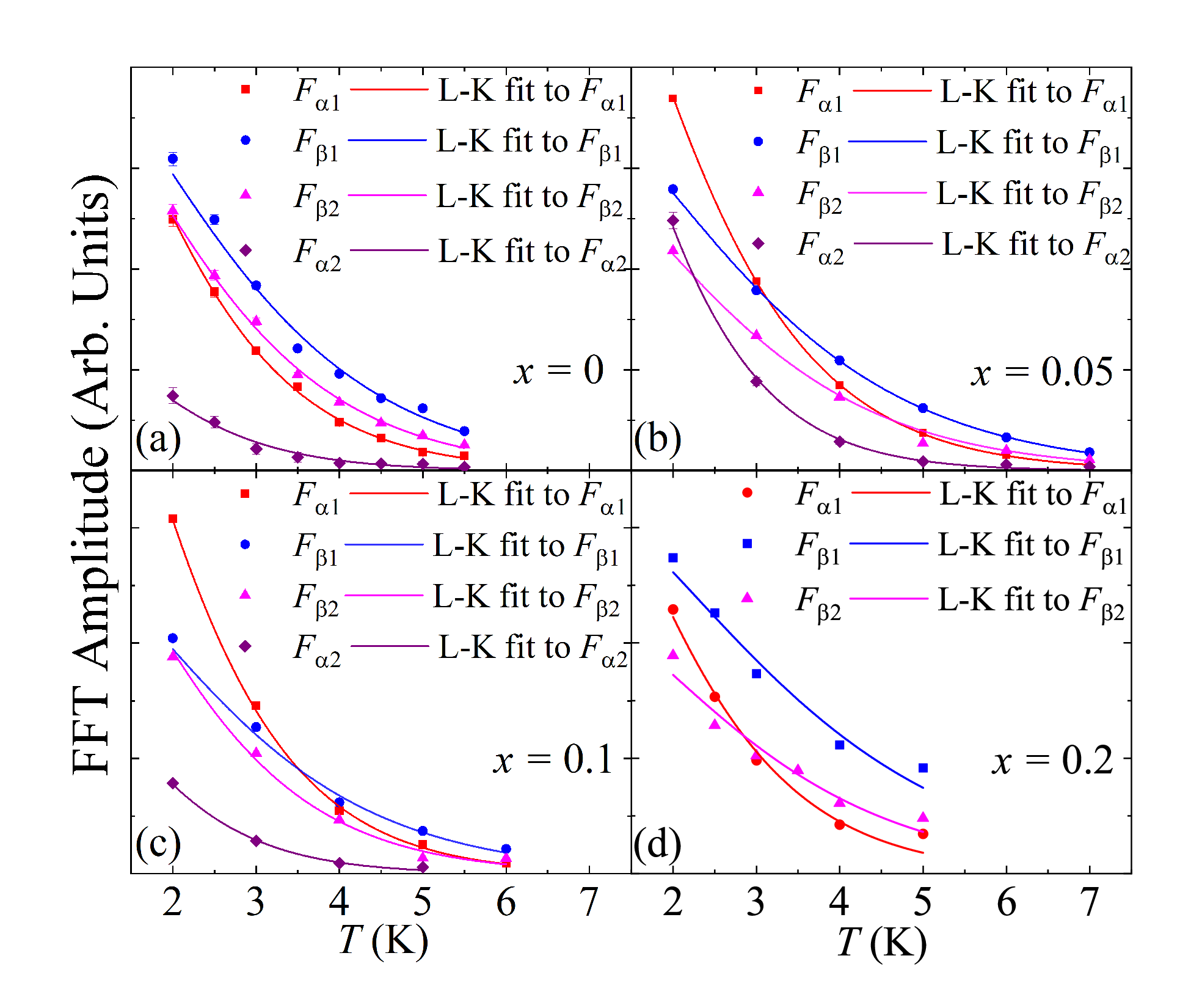}
\caption{Fit of the temperature dependence of the FFT amplitudes of the frequencies of oscillation to the temperature-dependent part of the Lifshitz-Kosevich theory for (a) WTe$_2$ (b) W$_{0.95}$Mo$_{0.05}$Te$_2$ (c) W$_{0.9}$Mo$_{0.1}$Te$_2$ and (d) W$_{0.8}$Mo$_{0.2}$Te$_2$.}
\label{Figure_7}
\end{figure} 

Two of each Fermi-surface pocket is known to be present within the Brillouin zone, one on each side of the $\Gamma$ point~\cite{Zhu2015,Rhodes2015,Xiang2015}. As a result, the total electron concentration $n_\mathrm{e} = 2(n_\mathrm{\mathrm{F}_{\beta 1}} + n_\mathrm{\mathrm{F}_{\beta 2}})$ and the total hole concentration $n_\mathrm{h} = 2(n_\mathrm{\mathrm{F}_{\alpha 1}} + n_\mathrm{\mathrm{F}_{\alpha 2}})$ and their values are given in Table~\ref{Table_3}. The electron concentration is slightly higher than the hole concentration in the unsubstituted sample. However, with increasing Mo substitution, the oscillation frequencies change as described above and hence the hole concentration increases slightly while the electron concentration remains the same. This would imply that the charge compensation actually becomes better with Mo substitution. The slightly diﬀerent eﬀect of the Mo substitution on the electron and hole pockets is likely due to diﬀerent curvatures of the bands.

\begin{figure}[tbh!]
\centering
\includegraphics[width=\linewidth]{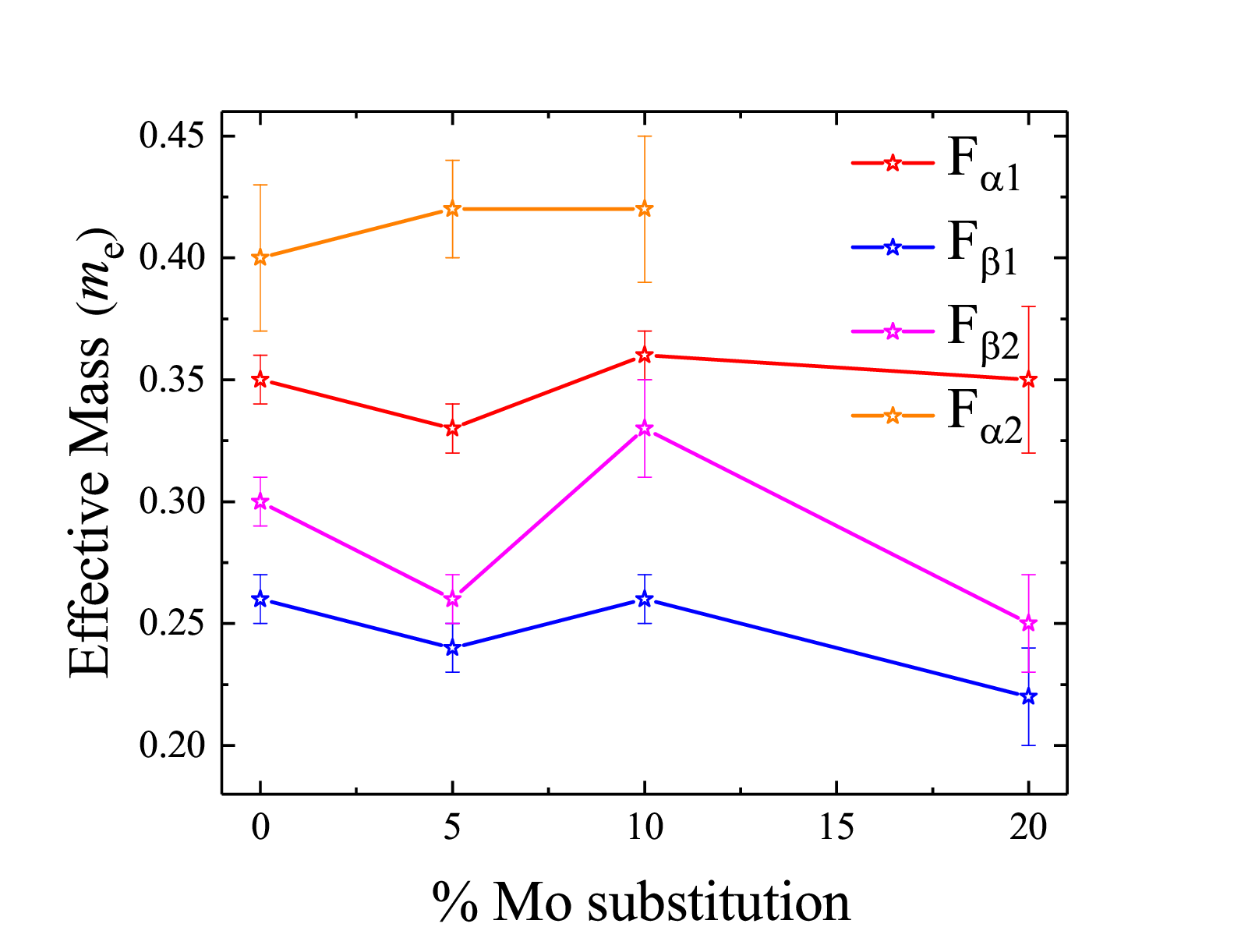}
\caption{Variation in the effective masses of the four pockets in units of $m_\mathrm{e}$ with Mo substitution.}
\label{Figure_8}
\end{figure}

Figure~\ref{Figure_7} shows the temperature dependence of the FFT amplitudes of each of the four peaks observed in the FFT spectra. The solid lines are fits to the Lifshitz-Kosevich formula 
%$A \frac{\chi}{\sinh \chi}$ 
$A\chi/(\sinh \chi)$, where $\chi = 2 \pi^{2} k_{\mathrm{B}}T m^{*}/\hbar e \mathrm{\overline{B}}$
with $\mathrm{\overline{B}} = 4.5$~T being the the average of the inverse magnetic-field window of the FFT~\cite{Lifshitz1956,Shoenberg1984}. From this we extract the effective masses $m^*$ given in Supplemental Material Table~S2~\cite{SuppMat} and plotted in Fig.~\ref{Figure_8}. There is not much change in the effective masses with Mo doping, further indicating that Mo doping does not affect the Fermi surface topology.

By performing a Dingle analysis, the quantum oscillations can also be used to find the quantum scattering time ($\tau_\mathrm{q}$), which is different from the transport scattering time ($\tau_\mathrm{t}$)~\cite{Culcer_2010}. It is the time an electron spends in a particular Bloch state and is affected equally by both small and large-angle scattering events. On the other hand, the transport scattering time represents the time between two collisions in electron transport. It is weighted by a factor $(1 - \cos{\theta})$ where $\theta$ is the angle of scattering. Thus, it is affected more by large-angle scattering than by small-angle scattering.

\begin{figure}[tbh!]
\centering
\includegraphics[width=\linewidth]{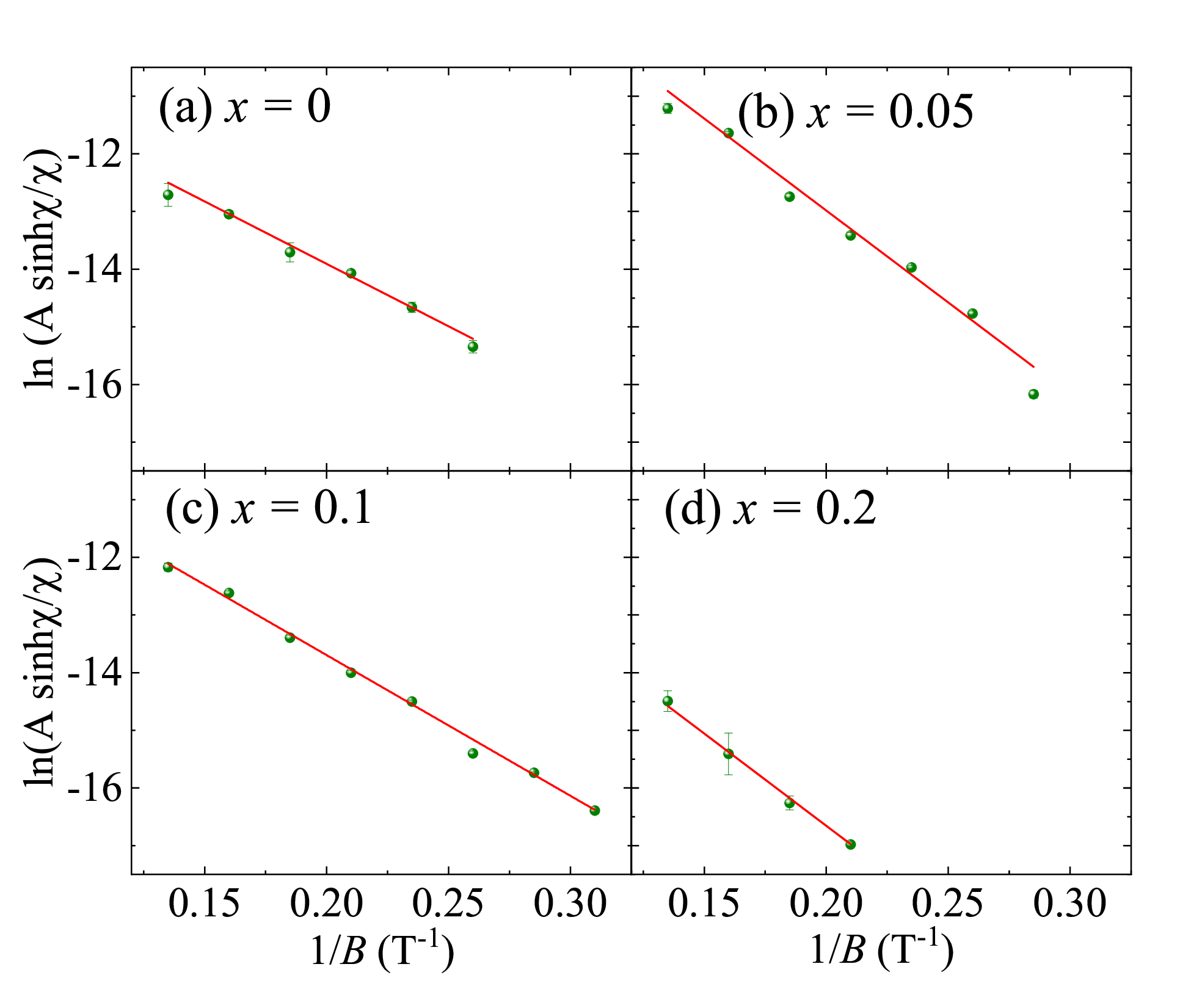}
\caption{Dingle plot for the carriers corresponding to the frequency $\mathrm{F}_{\alpha 1}$ for the four different compositions: (a) WTe$_2$ (b) W$_{0.95}$Mo$_{0.05}$Te$_2$ (c) W$_{0.9}$Mo$_{0.1}$Te$_2$ and (d) W$_{0.8}$Mo$_{0.2}$Te$_2$.}
\label{Figure_9}
\end{figure}

We perform the Dingle analysis~\cite{Shoenberg1984} for the pocket corresponding to the frequency $\mathrm{F}_{\alpha 1}$ for all four compositions and the relevant plots are shown in Fig.~\ref{Figure_9}. From the linear fit to the data, the Dingle temperature, $T_\mathrm{D}$ and the quantum scattering time $\tau_\mathrm{q}$, which are related according to $T_\mathrm{D} = \frac{\hbar}{2\pi k_\mathrm{B}\tau_\mathrm{q}}$ are obtained for the holes corresponding to the Fermi-surface pocket with frequency $\mathrm{F}_{\alpha 1}$ and the values are given in Table~\ref{Table_3}. Interestingly, the quantum scattering times obtained from the Dingle analysis do not show significant change across the four compositions, whereas a decrease in transport mobility implies a decrease in the transport scattering time. Significant differences between transport and quantum scattering times are expected and are seen, for example, in two-dimensional electron gases and graphene~\cite{Coleridge1991,Hong2009}.  

\begin{table*}[tbh!]
\caption{\label{Table_3} Dingle temperature $T_\mathrm{D}$ and the quantum scattering time $\tau_\mathrm{q}$ obtained from the quantum oscillations (QO) for the frequency $\mathrm{F}_{\alpha 1}$ for all four different compositions W$_{1-x}$Mo$_{x}$Te$_2$ (x = 0, 0.05, 0.1, 0.2). The total electron concentration $n_\mathrm{e}$(QO) and total hole concentration $n_\mathrm{h}$(QO) obtained by doubling the electron and hole concentrations obtained from the quantum oscillations are also given.}
\begin{ruledtabular}
\begin{tabular}{cccccc}
Composition & \multicolumn{2}{c}{WTe$_2$}  & W$_{0.95}$Mo$_{0.05}$Te$_2$ & W$_{0.9}$Mo$_{0.1}$Te$_2$  & W$_{0.8}$Mo$_{0.2}$Te$_2$ \\
& This work & Previous work & & & \\
 $T_\mathrm{D}$ (K)        & $4.2 \pm 0.2$  & 12.19 \cite{Fu2018}  &  $6.6 \pm 0.4$  &  $4.6 \pm 0.1$  & $6.2 \pm 0.3$ \\
 $\tau_\mathrm{q}$ (ps)               & $1.82 \pm 0.09$ & 0.1 \cite{Fu2018}  & $1.16 \pm 0.07$ &  $1.65 \pm 0.05$ & $1.23 \pm 0.06$ \\
 \makecell{$n_\mathrm{e}$(QO)(10$^{19}$ cm$^{-3}$)}                     & $3.666 \pm 0.02$ & 3.68 \cite{Rhodes2015}  & $3.666 \pm 0.02$ & $3.636 \pm 0.02$ & $3.638 \pm 0.02$ \\
 \makecell{$n_\mathrm{h}$(QO)(10$^{19}$ cm$^{-3}$)}                     & $3.346 \pm 0.04$ & 3.42 \cite{Rhodes2015}  & $3.452 \pm 0.02$ & $3.686 \pm 0.02$ & - \\
\end{tabular}
\end{ruledtabular}
\end{table*}

\section{Discussion}

Generally, the large MR in WTe$_2$ is discussed in the context of the two-band model~\cite{Ali2014,Wang2015,Lv2016,Fatemi2017,Luo2015,Fu2018}. This model, however, suffers from an oversimplification because it clubs together holes from two different pockets with possibly different mobilities into a single band and similarly for electrons from different pockets into another single band. The MR in this model is given by~\cite{Wang2015,Lv2016,Fu2018}

\begin{widetext}
\begin{equation}
\begin{aligned}
  \mathrm{MR} &= \dfrac{\rho_{xx}(B)-\rho_{xx}(\mathrm{0})}{\rho_{xx}(\mathrm{0})}\\
     &= \dfrac{(\mu_\mathrm{e} n_\mathrm{e} + \mu_\mathrm{h} n_\mathrm{h})^2+(\mu_\mathrm{e} n_\mathrm{e} + \mu_\mathrm{h} n_\mathrm{h})(\mu_\mathrm{e} n_\mathrm{h} + \mu_\mathrm{h} n_\mathrm{e}) \mu_\mathrm{e} \mu_\mathrm{h} B^2}{(\mu_\mathrm{e} n_\mathrm{e} + \mu_\mathrm{h} n_\mathrm{h})^2  + (n_\mathrm{h}-n_\mathrm{e})^2\mu_\mathrm{e}^2 \mu_\mathrm{h}^2 B^2}-1, \label{Equation_MR}
\end{aligned}
\end{equation}
\end{widetext}
where $B$ is the magnetic field and $n_\mathrm{e}$, $n_\mathrm{h}$, $\mu_\mathrm{e}$ and $\mu_\mathrm{h}$ are total electron concentration, total hole concentration, electron mobility and hole mobility, respectively (see Eqns. S1(a) and S1(b) in the Supplemental Material~\cite{SuppMat}).

The success of the two-band model lies in the fact that it predicts a parabolic and non-saturating MR ($MR \propto \mu_\mathrm{e} \mu_\mathrm{h} B^2$) in case of perfect compensation ($n_\mathrm{e} = n_\mathrm{h}$). In this model, when the charge compensation is imperfect, the MR saturates at high ﬁelds, the value of which depends on the mobility. In fact, a very large mobility is counterproductive to a large MR, as it will lead to saturation at low ﬁeld. For the optimum value of the mobility, the ﬁeld at which the MR saturates can be beyond that accessible experimentally. However, the two-band model suffers from over parametrization and can lead to wrong conclusions as has been seen in the case of ZrSiSe where a Lifshitz transition inferred from the two-band fit has been later proved to be incorrect~\cite{Xu_Jing_2023}.

Considering this mutual dependency in the two-band model and the fact that it discards much information about the Fermi surface, it is clear that the measured changes in quantum oscillation frequencies give a more accurate picture of the evolution of the Fermi surface. A simple analysis of quantum oscillations suggests that the charge compensation improves with Mo substitution, at least up to $x = 0.2$. This last statement relies on the reliability of two assumptions underlying the estimate of the carrier densities: (i) that the prior band-structure calculations are correct in their assignment of the holes and electrons to the different Fermi-surface pockets, and (ii) that the pockets evolve isotropically in all cases, thus allowing the use of the spherical approximation. In any case, the decrease in MR that accompanies the increasing Mo substitution should be attributed solely or in part due to a decrease in transport mobility, as evidenced by the RRR data. It is possible that doping with Mo, while nominally isoelectronic, might affect the stability of other charged defects in the material, and could lead to changes in the Fermi energy and hence the charge compensation in addition to those arising from changes in the lattice parameters. Any such changes will be captured by the evolution of the quantum oscillation frequencies.

While the transport scattering decreases, the quantum-oscillation analysis show that the quantum scattering time remains largely unchanged across the doping range investigated. A similar observation has been reported in Ni substituted WTe$_2$ where $\tau_\mathrm{q}$ increases while $\tau_\mathrm{t}$  decreases with Ni substitution \cite{Singh_2022}. These effects likely arise from an increase in large-angle scattering due to the introduction of the impurity ions. This will disproportionally affect the transport scattering time, while the quantum scattering time must instead be dominated by small-angle scattering.

In conclusion, we analyse the evolution of the quantum oscillations in WTe$_2$ with Mo substitution. The MR decreases with substitution, but the charge compensation improves based on the quantum oscillation data. We attribute the MR decrease to the decrease in transport mobility, which is evident from the RRR values. Thus, we show that it is possible that the MR in WTe$_2$ can decrease even with a very good level of charge compensation provided there is a fall in the transport mobility. Our quantum oscillation analysis further indicates that the quantum scattering time is unaffected by the Mo substitution. Ref~\onlinecite{Fu2018} reports changes in mobility in WTe$_2$ nanoflakes on Mo doping, similar to our results. However, they also use the two-band model to conclude that the charge balance remains undisturbed. Our data on bulk crystals shows that this approach is not valid for this material.

\section{Acknowledgments}
We acknowledge the EPSRC, UK for funding this work (Grant Nos. EP/L014963/1 and EP/M028771/1). We thank T. Orton and P. Ruddy for technical assistance.
For the purpose of open access, the author has applied a Creative Commons Attribution (CC-BY) licence to any Author Accepted Manuscript version arising from this submission. Data presented in this paper will be made available at XXXXXXX.

%\bibliography{References}

\begin{thebibliography}{61}%
\makeatletter
\providecommand \@ifxundefined [1]{%
 \@ifx{#1\undefined}
}%
\providecommand \@ifnum [1]{%
 \ifnum #1\expandafter \@firstoftwo
 \else \expandafter \@secondoftwo
 \fi
}%
\providecommand \@ifx [1]{%
 \ifx #1\expandafter \@firstoftwo
 \else \expandafter \@secondoftwo
 \fi
}%
\providecommand \natexlab [1]{#1}%
\providecommand \enquote  [1]{``#1''}%
\providecommand \bibnamefont  [1]{#1}%
\providecommand \bibfnamefont [1]{#1}%
\providecommand \citenamefont [1]{#1}%
\providecommand \href@noop [0]{\@secondoftwo}%
\providecommand \href [0]{\begingroup \@sanitize@url \@href}%
\providecommand \@href[1]{\@@startlink{#1}\@@href}%
\providecommand \@@href[1]{\endgroup#1\@@endlink}%
\providecommand \@sanitize@url [0]{\catcode `\\12\catcode `\$12\catcode
  `\&12\catcode `\#12\catcode `\^12\catcode `\_12\catcode `\%12\relax}%
\providecommand \@@startlink[1]{}%
\providecommand \@@endlink[0]{}%
\providecommand \url  [0]{\begingroup\@sanitize@url \@url }%
\providecommand \@url [1]{\endgroup\@href {#1}{\urlprefix }}%
\providecommand \urlprefix  [0]{URL }%
\providecommand \Eprint [0]{\href }%
\providecommand \doibase [0]{https://doi.org/}%
\providecommand \selectlanguage [0]{\@gobble}%
\providecommand \bibinfo  [0]{\@secondoftwo}%
\providecommand \bibfield  [0]{\@secondoftwo}%
\providecommand \translation [1]{[#1]}%
\providecommand \BibitemOpen [0]{}%
\providecommand \bibitemStop [0]{}%
\providecommand \bibitemNoStop [0]{.\EOS\space}%
\providecommand \EOS [0]{\spacefactor3000\relax}%
\providecommand \BibitemShut  [1]{\csname bibitem#1\endcsname}%
\let\auto@bib@innerbib\@empty
%</preamble>
\bibitem [{\citenamefont {Geim}(2009)}]{Geim2009}%
  \BibitemOpen
  \bibfield  {author} {\bibinfo {author} {\bibfnamefont {A.~K.}\ \bibnamefont
  {Geim}},\ }\bibfield  {title} {\bibinfo {title} {Graphene: Status and
  prospects},\ }\href {https://doi.org/10.1126/science.1158877} {\bibfield
  {journal} {\bibinfo  {journal} {Science}\ }\textbf {\bibinfo {volume}
  {324}},\ \bibinfo {pages} {1530} (\bibinfo {year} {2009})}\BibitemShut
  {NoStop}%
\bibitem [{\citenamefont {Bhimanapati}\ \emph {et~al.}(2015)\citenamefont
  {Bhimanapati}, \citenamefont {Lin}, \citenamefont {Meunier}, \citenamefont
  {Jung}, \citenamefont {Cha}, \citenamefont {Das}, \citenamefont {Xiao},
  \citenamefont {Son}, \citenamefont {Strano}, \citenamefont {Cooper},
  \citenamefont {Liang}, \citenamefont {Louie}, \citenamefont {Ringe},
  \citenamefont {Zhou}, \citenamefont {Kim}, \citenamefont {Naik},
  \citenamefont {Sumpter}, \citenamefont {Terrones}, \citenamefont {Xia},
  \citenamefont {Wang}, \citenamefont {Zhu}, \citenamefont {Akinwande},
  \citenamefont {Alem}, \citenamefont {Schuller}, \citenamefont {Schaak},
  \citenamefont {Terrones},\ and\ \citenamefont {Robinson}}]{Bhimanapati2015}%
  \BibitemOpen
  \bibfield  {author} {\bibinfo {author} {\bibfnamefont {G.~R.}\ \bibnamefont
  {Bhimanapati}}, \bibinfo {author} {\bibfnamefont {Z.}~\bibnamefont {Lin}},
  \bibinfo {author} {\bibfnamefont {V.}~\bibnamefont {Meunier}}, \bibinfo
  {author} {\bibfnamefont {Y.}~\bibnamefont {Jung}}, \bibinfo {author}
  {\bibfnamefont {J.}~\bibnamefont {Cha}}, \bibinfo {author} {\bibfnamefont
  {S.}~\bibnamefont {Das}}, \bibinfo {author} {\bibfnamefont {D.}~\bibnamefont
  {Xiao}}, \bibinfo {author} {\bibfnamefont {Y.}~\bibnamefont {Son}}, \bibinfo
  {author} {\bibfnamefont {M.~S.}\ \bibnamefont {Strano}}, \bibinfo {author}
  {\bibfnamefont {V.~R.}\ \bibnamefont {Cooper}}, \bibinfo {author}
  {\bibfnamefont {L.}~\bibnamefont {Liang}}, \bibinfo {author} {\bibfnamefont
  {S.~G.}\ \bibnamefont {Louie}}, \bibinfo {author} {\bibfnamefont
  {E.}~\bibnamefont {Ringe}}, \bibinfo {author} {\bibfnamefont
  {W.}~\bibnamefont {Zhou}}, \bibinfo {author} {\bibfnamefont {S.~S.}\
  \bibnamefont {Kim}}, \bibinfo {author} {\bibfnamefont {R.~R.}\ \bibnamefont
  {Naik}}, \bibinfo {author} {\bibfnamefont {B.~G.}\ \bibnamefont {Sumpter}},
  \bibinfo {author} {\bibfnamefont {H.}~\bibnamefont {Terrones}}, \bibinfo
  {author} {\bibfnamefont {F.}~\bibnamefont {Xia}}, \bibinfo {author}
  {\bibfnamefont {Y.}~\bibnamefont {Wang}}, \bibinfo {author} {\bibfnamefont
  {J.}~\bibnamefont {Zhu}}, \bibinfo {author} {\bibfnamefont {D.}~\bibnamefont
  {Akinwande}}, \bibinfo {author} {\bibfnamefont {N.}~\bibnamefont {Alem}},
  \bibinfo {author} {\bibfnamefont {J.~A.}\ \bibnamefont {Schuller}}, \bibinfo
  {author} {\bibfnamefont {R.~E.}\ \bibnamefont {Schaak}}, \bibinfo {author}
  {\bibfnamefont {M.}~\bibnamefont {Terrones}},\ and\ \bibinfo {author}
  {\bibfnamefont {J.~A.}\ \bibnamefont {Robinson}},\ }\bibfield  {title}
  {\bibinfo {title} {Recent advances in two-dimensional materials beyond
  graphene},\ }\href {https://doi.org/10.1021/acsnano.5b05556} {\bibfield
  {journal} {\bibinfo  {journal} {ACS Nano}\ }\textbf {\bibinfo {volume} {9}},\
  \bibinfo {pages} {11509} (\bibinfo {year} {2015})}\BibitemShut {NoStop}%
\bibitem [{\citenamefont {Manzeli}\ \emph {et~al.}(2017)\citenamefont
  {Manzeli}, \citenamefont {Ovchinnikov}, \citenamefont {Pasquier},
  \citenamefont {Yazyev},\ and\ \citenamefont {Kis}}]{Manzeli2017}%
  \BibitemOpen
  \bibfield  {author} {\bibinfo {author} {\bibfnamefont {S.}~\bibnamefont
  {Manzeli}}, \bibinfo {author} {\bibfnamefont {D.}~\bibnamefont
  {Ovchinnikov}}, \bibinfo {author} {\bibfnamefont {D.}~\bibnamefont
  {Pasquier}}, \bibinfo {author} {\bibfnamefont {O.~V.}\ \bibnamefont
  {Yazyev}},\ and\ \bibinfo {author} {\bibfnamefont {A.}~\bibnamefont {Kis}},\
  }\bibfield  {title} {\bibinfo {title} {2d transition metal dichalcogenides},\
  }\href {https://doi.org/10.1038/natrevmats.2017.33} {\bibfield  {journal}
  {\bibinfo  {journal} {Nature Reviews Materials}\ }\textbf {\bibinfo {volume}
  {2}},\ \bibinfo {pages} {17033} (\bibinfo {year} {2017})}\BibitemShut
  {NoStop}%
\bibitem [{\citenamefont {Ali}\ \emph {et~al.}(2014)\citenamefont {Ali},
  \citenamefont {Xiong}, \citenamefont {Flynn}, \citenamefont {Tao},
  \citenamefont {Gibson}, \citenamefont {Schoop}, \citenamefont {Liang},
  \citenamefont {Haldolaarachchige}, \citenamefont {Hirschberger},
  \citenamefont {Ong},\ and\ \citenamefont {Cava}}]{Ali2014}%
  \BibitemOpen
  \bibfield  {author} {\bibinfo {author} {\bibfnamefont {M.~N.}\ \bibnamefont
  {Ali}}, \bibinfo {author} {\bibfnamefont {J.}~\bibnamefont {Xiong}}, \bibinfo
  {author} {\bibfnamefont {S.}~\bibnamefont {Flynn}}, \bibinfo {author}
  {\bibfnamefont {J.}~\bibnamefont {Tao}}, \bibinfo {author} {\bibfnamefont
  {Q.~D.}\ \bibnamefont {Gibson}}, \bibinfo {author} {\bibfnamefont {L.~M.}\
  \bibnamefont {Schoop}}, \bibinfo {author} {\bibfnamefont {T.}~\bibnamefont
  {Liang}}, \bibinfo {author} {\bibfnamefont {N.}~\bibnamefont
  {Haldolaarachchige}}, \bibinfo {author} {\bibfnamefont {M.}~\bibnamefont
  {Hirschberger}}, \bibinfo {author} {\bibfnamefont {N.~P.}\ \bibnamefont
  {Ong}},\ and\ \bibinfo {author} {\bibfnamefont {R.~J.}\ \bibnamefont
  {Cava}},\ }\bibfield  {title} {\bibinfo {title} {Large, non-saturating
  magnetoresistance in {WT}e{$_2$}},\ }\href
  {https://doi.org/10.1038/nature13763} {\bibfield  {journal} {\bibinfo
  {journal} {Nature}\ }\textbf {\bibinfo {volume} {514}},\ \bibinfo {pages}
  {205} (\bibinfo {year} {2014})}\BibitemShut {NoStop}%
\bibitem [{\citenamefont {Pan}\ \emph {et~al.}(2015)\citenamefont {Pan},
  \citenamefont {Chen}, \citenamefont {Liu}, \citenamefont {Feng},
  \citenamefont {Wei}, \citenamefont {Zhou}, \citenamefont {Chi}, \citenamefont
  {Pi}, \citenamefont {Yen}, \citenamefont {Song}, \citenamefont {Wan},
  \citenamefont {Yang}, \citenamefont {Wang}, \citenamefont {Wang},\ and\
  \citenamefont {Zhang}}]{Pan2015}%
  \BibitemOpen
  \bibfield  {author} {\bibinfo {author} {\bibfnamefont {X.-C.}\ \bibnamefont
  {Pan}}, \bibinfo {author} {\bibfnamefont {X.}~\bibnamefont {Chen}}, \bibinfo
  {author} {\bibfnamefont {H.}~\bibnamefont {Liu}}, \bibinfo {author}
  {\bibfnamefont {Y.}~\bibnamefont {Feng}}, \bibinfo {author} {\bibfnamefont
  {Z.}~\bibnamefont {Wei}}, \bibinfo {author} {\bibfnamefont {Y.}~\bibnamefont
  {Zhou}}, \bibinfo {author} {\bibfnamefont {Z.}~\bibnamefont {Chi}}, \bibinfo
  {author} {\bibfnamefont {L.}~\bibnamefont {Pi}}, \bibinfo {author}
  {\bibfnamefont {F.}~\bibnamefont {Yen}}, \bibinfo {author} {\bibfnamefont
  {F.}~\bibnamefont {Song}}, \bibinfo {author} {\bibfnamefont {X.}~\bibnamefont
  {Wan}}, \bibinfo {author} {\bibfnamefont {Z.}~\bibnamefont {Yang}}, \bibinfo
  {author} {\bibfnamefont {B.}~\bibnamefont {Wang}}, \bibinfo {author}
  {\bibfnamefont {G.}~\bibnamefont {Wang}},\ and\ \bibinfo {author}
  {\bibfnamefont {Y.}~\bibnamefont {Zhang}},\ }\bibfield  {title} {\bibinfo
  {title} {Pressure-driven dome-shaped superconductivity and electronic
  structural evolution in tungsten ditelluride},\ }\href
  {https://doi.org/10.1038/ncomms8805} {\bibfield  {journal} {\bibinfo
  {journal} {Nature Communications}\ }\textbf {\bibinfo {volume} {6}},\
  \bibinfo {pages} {7805} (\bibinfo {year} {2015})}\BibitemShut {NoStop}%
\bibitem [{\citenamefont {Kang}\ \emph {et~al.}(2015)\citenamefont {Kang},
  \citenamefont {Zhou}, \citenamefont {Yi}, \citenamefont {Yang}, \citenamefont
  {Guo}, \citenamefont {Shi}, \citenamefont {Zhang}, \citenamefont {Wang},
  \citenamefont {Zhang}, \citenamefont {Jiang}, \citenamefont {Li},
  \citenamefont {Yang}, \citenamefont {Wu}, \citenamefont {Zhang},
  \citenamefont {Sun},\ and\ \citenamefont {Zhao}}]{Kang2015}%
  \BibitemOpen
  \bibfield  {author} {\bibinfo {author} {\bibfnamefont {D.}~\bibnamefont
  {Kang}}, \bibinfo {author} {\bibfnamefont {Y.}~\bibnamefont {Zhou}}, \bibinfo
  {author} {\bibfnamefont {W.}~\bibnamefont {Yi}}, \bibinfo {author}
  {\bibfnamefont {C.}~\bibnamefont {Yang}}, \bibinfo {author} {\bibfnamefont
  {J.}~\bibnamefont {Guo}}, \bibinfo {author} {\bibfnamefont {Y.}~\bibnamefont
  {Shi}}, \bibinfo {author} {\bibfnamefont {S.}~\bibnamefont {Zhang}}, \bibinfo
  {author} {\bibfnamefont {Z.}~\bibnamefont {Wang}}, \bibinfo {author}
  {\bibfnamefont {C.}~\bibnamefont {Zhang}}, \bibinfo {author} {\bibfnamefont
  {S.}~\bibnamefont {Jiang}}, \bibinfo {author} {\bibfnamefont
  {A.}~\bibnamefont {Li}}, \bibinfo {author} {\bibfnamefont {K.}~\bibnamefont
  {Yang}}, \bibinfo {author} {\bibfnamefont {Q.}~\bibnamefont {Wu}}, \bibinfo
  {author} {\bibfnamefont {G.}~\bibnamefont {Zhang}}, \bibinfo {author}
  {\bibfnamefont {L.}~\bibnamefont {Sun}},\ and\ \bibinfo {author}
  {\bibfnamefont {Z.}~\bibnamefont {Zhao}},\ }\bibfield  {title} {\bibinfo
  {title} {Superconductivity emerging from a suppressed large magnetoresistant
  state in tungsten ditelluride},\ }\href {https://doi.org/10.1038/ncomms8804}
  {\bibfield  {journal} {\bibinfo  {journal} {Nature Communications}\ }\textbf
  {\bibinfo {volume} {6}},\ \bibinfo {pages} {7804} (\bibinfo {year}
  {2015})}\BibitemShut {NoStop}%
\bibitem [{\citenamefont {Chen}\ \emph {et~al.}(2016)\citenamefont {Chen},
  \citenamefont {Lv}, \citenamefont {Luo}, \citenamefont {Lu}, \citenamefont
  {Pei}, \citenamefont {Lin}, \citenamefont {Han}, \citenamefont {Zhu},
  \citenamefont {Song},\ and\ \citenamefont {Sun}}]{Chen2016}%
  \BibitemOpen
  \bibfield  {author} {\bibinfo {author} {\bibfnamefont {F.~C.}\ \bibnamefont
  {Chen}}, \bibinfo {author} {\bibfnamefont {H.~Y.}\ \bibnamefont {Lv}},
  \bibinfo {author} {\bibfnamefont {X.}~\bibnamefont {Luo}}, \bibinfo {author}
  {\bibfnamefont {W.~J.}\ \bibnamefont {Lu}}, \bibinfo {author} {\bibfnamefont
  {Q.~L.}\ \bibnamefont {Pei}}, \bibinfo {author} {\bibfnamefont {G.~T.}\
  \bibnamefont {Lin}}, \bibinfo {author} {\bibfnamefont {Y.~Y.}\ \bibnamefont
  {Han}}, \bibinfo {author} {\bibfnamefont {X.~B.}\ \bibnamefont {Zhu}},
  \bibinfo {author} {\bibfnamefont {W.~H.}\ \bibnamefont {Song}},\ and\
  \bibinfo {author} {\bibfnamefont {Y.~P.}\ \bibnamefont {Sun}},\ }\bibfield
  {title} {\bibinfo {title} {Extremely large magnetoresistance in the type-{II}
  {W}eyl semimetal {M}o{T}e$_2$},\ }\href
  {https://doi.org/10.1103/PhysRevB.94.235154} {\bibfield  {journal} {\bibinfo
  {journal} {Physical Review B}\ }\textbf {\bibinfo {volume} {94}},\ \bibinfo
  {pages} {235154} (\bibinfo {year} {2016})}\BibitemShut {NoStop}%
\bibitem [{\citenamefont {Lee}\ \emph {et~al.}(2018)\citenamefont {Lee},
  \citenamefont {Jang}, \citenamefont {Kim}, \citenamefont {Jung},
  \citenamefont {Kim}, \citenamefont {Cho}, \citenamefont {Kim}, \citenamefont
  {Rhee}, \citenamefont {Park},\ and\ \citenamefont {Park}}]{Lee2018}%
  \BibitemOpen
  \bibfield  {author} {\bibinfo {author} {\bibfnamefont {S.}~\bibnamefont
  {Lee}}, \bibinfo {author} {\bibfnamefont {J.}~\bibnamefont {Jang}}, \bibinfo
  {author} {\bibfnamefont {S.-I.}\ \bibnamefont {Kim}}, \bibinfo {author}
  {\bibfnamefont {S.-G.}\ \bibnamefont {Jung}}, \bibinfo {author}
  {\bibfnamefont {J.}~\bibnamefont {Kim}}, \bibinfo {author} {\bibfnamefont
  {S.}~\bibnamefont {Cho}}, \bibinfo {author} {\bibfnamefont {S.~W.}\
  \bibnamefont {Kim}}, \bibinfo {author} {\bibfnamefont {J.~Y.}\ \bibnamefont
  {Rhee}}, \bibinfo {author} {\bibfnamefont {K.-S.}\ \bibnamefont {Park}},\
  and\ \bibinfo {author} {\bibfnamefont {T.}~\bibnamefont {Park}},\ }\bibfield
  {title} {\bibinfo {title} {Origin of extremely large magnetoresistance in the
  candidate type-{II} {W}eyl semimetal {M}o{T}e$_{2−x}$},\ }\href
  {https://doi.org/10.1038/s41598-018-32387-1} {\bibfield  {journal} {\bibinfo
  {journal} {Scientific Reports}\ }\textbf {\bibinfo {volume} {8}},\ \bibinfo
  {pages} {13937} (\bibinfo {year} {2018})}\BibitemShut {NoStop}%
\bibitem [{\citenamefont {Qi}\ \emph {et~al.}(2016)\citenamefont {Qi},
  \citenamefont {Naumov}, \citenamefont {Ali}, \citenamefont {Rajamathi},
  \citenamefont {Schnelle}, \citenamefont {Barkalov}, \citenamefont {Hanfland},
  \citenamefont {Wu}, \citenamefont {Shekhar}, \citenamefont {Sun},
  \citenamefont {Süß}, \citenamefont {Schmidt}, \citenamefont {Schwarz},
  \citenamefont {Pippel}, \citenamefont {Werner}, \citenamefont {Hillebrand},
  \citenamefont {Förster}, \citenamefont {Kampert}, \citenamefont {Parkin},
  \citenamefont {Cava}, \citenamefont {Felser}, \citenamefont {Yan},\ and\
  \citenamefont {Medvedev}}]{Qi2016}%
  \BibitemOpen
  \bibfield  {author} {\bibinfo {author} {\bibfnamefont {Y.}~\bibnamefont
  {Qi}}, \bibinfo {author} {\bibfnamefont {P.~G.}\ \bibnamefont {Naumov}},
  \bibinfo {author} {\bibfnamefont {M.~N.}\ \bibnamefont {Ali}}, \bibinfo
  {author} {\bibfnamefont {C.~R.}\ \bibnamefont {Rajamathi}}, \bibinfo {author}
  {\bibfnamefont {W.}~\bibnamefont {Schnelle}}, \bibinfo {author}
  {\bibfnamefont {O.}~\bibnamefont {Barkalov}}, \bibinfo {author}
  {\bibfnamefont {M.}~\bibnamefont {Hanfland}}, \bibinfo {author}
  {\bibfnamefont {S.-C.}\ \bibnamefont {Wu}}, \bibinfo {author} {\bibfnamefont
  {C.}~\bibnamefont {Shekhar}}, \bibinfo {author} {\bibfnamefont
  {Y.}~\bibnamefont {Sun}}, \bibinfo {author} {\bibfnamefont {V.}~\bibnamefont
  {Süß}}, \bibinfo {author} {\bibfnamefont {M.}~\bibnamefont {Schmidt}},
  \bibinfo {author} {\bibfnamefont {U.}~\bibnamefont {Schwarz}}, \bibinfo
  {author} {\bibfnamefont {E.}~\bibnamefont {Pippel}}, \bibinfo {author}
  {\bibfnamefont {P.}~\bibnamefont {Werner}}, \bibinfo {author} {\bibfnamefont
  {R.}~\bibnamefont {Hillebrand}}, \bibinfo {author} {\bibfnamefont
  {T.}~\bibnamefont {Förster}}, \bibinfo {author} {\bibfnamefont
  {E.}~\bibnamefont {Kampert}}, \bibinfo {author} {\bibfnamefont
  {S.}~\bibnamefont {Parkin}}, \bibinfo {author} {\bibfnamefont {R.~J.}\
  \bibnamefont {Cava}}, \bibinfo {author} {\bibfnamefont {C.}~\bibnamefont
  {Felser}}, \bibinfo {author} {\bibfnamefont {B.}~\bibnamefont {Yan}},\ and\
  \bibinfo {author} {\bibfnamefont {S.~A.}\ \bibnamefont {Medvedev}},\
  }\bibfield  {title} {\bibinfo {title} {Superconductivity in {W}eyl semimetal
  candidate {M}o{T}e$_2$},\ }\href {https://doi.org/10.1038/ncomms11038}
  {\bibfield  {journal} {\bibinfo  {journal} {Nature Communications}\ }\textbf
  {\bibinfo {volume} {7}},\ \bibinfo {pages} {11038} (\bibinfo {year}
  {2016})}\BibitemShut {NoStop}%
\bibitem [{\citenamefont {Soluyanov}\ \emph {et~al.}(2015)\citenamefont
  {Soluyanov}, \citenamefont {Gresch}, \citenamefont {Wang}, \citenamefont
  {Wu}, \citenamefont {Troyer}, \citenamefont {Dai},\ and\ \citenamefont
  {Bernevig}}]{Soluyanov2015}%
  \BibitemOpen
  \bibfield  {author} {\bibinfo {author} {\bibfnamefont {A.~A.}\ \bibnamefont
  {Soluyanov}}, \bibinfo {author} {\bibfnamefont {D.}~\bibnamefont {Gresch}},
  \bibinfo {author} {\bibfnamefont {Z.}~\bibnamefont {Wang}}, \bibinfo {author}
  {\bibfnamefont {Q.}~\bibnamefont {Wu}}, \bibinfo {author} {\bibfnamefont
  {M.}~\bibnamefont {Troyer}}, \bibinfo {author} {\bibfnamefont
  {X.}~\bibnamefont {Dai}},\ and\ \bibinfo {author} {\bibfnamefont {B.~A.}\
  \bibnamefont {Bernevig}},\ }\bibfield  {title} {\bibinfo {title} {Type-{II}
  {W}eyl semimetals},\ }\href {https://doi.org/10.1038/nature15768} {\bibfield
  {journal} {\bibinfo  {journal} {Nature}\ }\textbf {\bibinfo {volume} {527}},\
  \bibinfo {pages} {495} (\bibinfo {year} {2015})}\BibitemShut {NoStop}%
\bibitem [{\citenamefont {Wu}\ \emph {et~al.}(2016)\citenamefont {Wu},
  \citenamefont {Mou}, \citenamefont {Jo}, \citenamefont {Sun}, \citenamefont
  {Huang}, \citenamefont {Bud'ko}, \citenamefont {Canfield},\ and\
  \citenamefont {Kaminski}}]{Wu2016}%
  \BibitemOpen
  \bibfield  {author} {\bibinfo {author} {\bibfnamefont {Y.}~\bibnamefont
  {Wu}}, \bibinfo {author} {\bibfnamefont {D.}~\bibnamefont {Mou}}, \bibinfo
  {author} {\bibfnamefont {N.~H.}\ \bibnamefont {Jo}}, \bibinfo {author}
  {\bibfnamefont {K.}~\bibnamefont {Sun}}, \bibinfo {author} {\bibfnamefont
  {L.}~\bibnamefont {Huang}}, \bibinfo {author} {\bibfnamefont {S.~L.}\
  \bibnamefont {Bud'ko}}, \bibinfo {author} {\bibfnamefont {P.~C.}\
  \bibnamefont {Canfield}},\ and\ \bibinfo {author} {\bibfnamefont
  {A.}~\bibnamefont {Kaminski}},\ }\bibfield  {title} {\bibinfo {title}
  {Observation of {F}ermi arcs in the type-{II} {W}eyl semimetal candidate
  {W}{T}e$_2$},\ }\href {https://doi.org/10.1103/PhysRevB.94.121113} {\bibfield
   {journal} {\bibinfo  {journal} {Physical Review B}\ }\textbf {\bibinfo
  {volume} {94}},\ \bibinfo {pages} {121113(R)} (\bibinfo {year}
  {2016})}\BibitemShut {NoStop}%
\bibitem [{\citenamefont {Wang}\ \emph
  {et~al.}(2016{\natexlab{a}})\citenamefont {Wang}, \citenamefont {Zhang},
  \citenamefont {Huang}, \citenamefont {Nie}, \citenamefont {Liu},
  \citenamefont {Liang}, \citenamefont {Zhang}, \citenamefont {Shen},
  \citenamefont {Liu}, \citenamefont {Hu}, \citenamefont {Ding}, \citenamefont
  {Liu}, \citenamefont {Hu}, \citenamefont {He}, \citenamefont {Zhao},
  \citenamefont {Yu}, \citenamefont {Hu}, \citenamefont {Wei}, \citenamefont
  {Mao}, \citenamefont {Shi}, \citenamefont {Jia}, \citenamefont {Zhang},
  \citenamefont {Zhang}, \citenamefont {Yang}, \citenamefont {Wang},
  \citenamefont {Peng}, \citenamefont {Weng}, \citenamefont {Dai},
  \citenamefont {Fang}, \citenamefont {Xu}, \citenamefont {Chen},\ and\
  \citenamefont {Zhou}}]{CWang2016}%
  \BibitemOpen
  \bibfield  {author} {\bibinfo {author} {\bibfnamefont {C.}~\bibnamefont
  {Wang}}, \bibinfo {author} {\bibfnamefont {Y.}~\bibnamefont {Zhang}},
  \bibinfo {author} {\bibfnamefont {J.}~\bibnamefont {Huang}}, \bibinfo
  {author} {\bibfnamefont {S.}~\bibnamefont {Nie}}, \bibinfo {author}
  {\bibfnamefont {G.}~\bibnamefont {Liu}}, \bibinfo {author} {\bibfnamefont
  {A.}~\bibnamefont {Liang}}, \bibinfo {author} {\bibfnamefont
  {Y.}~\bibnamefont {Zhang}}, \bibinfo {author} {\bibfnamefont
  {B.}~\bibnamefont {Shen}}, \bibinfo {author} {\bibfnamefont {J.}~\bibnamefont
  {Liu}}, \bibinfo {author} {\bibfnamefont {C.}~\bibnamefont {Hu}}, \bibinfo
  {author} {\bibfnamefont {Y.}~\bibnamefont {Ding}}, \bibinfo {author}
  {\bibfnamefont {D.}~\bibnamefont {Liu}}, \bibinfo {author} {\bibfnamefont
  {Y.}~\bibnamefont {Hu}}, \bibinfo {author} {\bibfnamefont {S.}~\bibnamefont
  {He}}, \bibinfo {author} {\bibfnamefont {L.}~\bibnamefont {Zhao}}, \bibinfo
  {author} {\bibfnamefont {L.}~\bibnamefont {Yu}}, \bibinfo {author}
  {\bibfnamefont {J.}~\bibnamefont {Hu}}, \bibinfo {author} {\bibfnamefont
  {J.}~\bibnamefont {Wei}}, \bibinfo {author} {\bibfnamefont {Z.}~\bibnamefont
  {Mao}}, \bibinfo {author} {\bibfnamefont {Y.}~\bibnamefont {Shi}}, \bibinfo
  {author} {\bibfnamefont {X.}~\bibnamefont {Jia}}, \bibinfo {author}
  {\bibfnamefont {F.}~\bibnamefont {Zhang}}, \bibinfo {author} {\bibfnamefont
  {S.}~\bibnamefont {Zhang}}, \bibinfo {author} {\bibfnamefont
  {F.}~\bibnamefont {Yang}}, \bibinfo {author} {\bibfnamefont {Z.}~\bibnamefont
  {Wang}}, \bibinfo {author} {\bibfnamefont {Q.}~\bibnamefont {Peng}}, \bibinfo
  {author} {\bibfnamefont {H.}~\bibnamefont {Weng}}, \bibinfo {author}
  {\bibfnamefont {X.}~\bibnamefont {Dai}}, \bibinfo {author} {\bibfnamefont
  {Z.}~\bibnamefont {Fang}}, \bibinfo {author} {\bibfnamefont {Z.}~\bibnamefont
  {Xu}}, \bibinfo {author} {\bibfnamefont {C.}~\bibnamefont {Chen}},\ and\
  \bibinfo {author} {\bibfnamefont {X.~J.}\ \bibnamefont {Zhou}},\ }\bibfield
  {title} {\bibinfo {title} {Observation of {F}ermi arc and its connection with
  bulk states in the candidate type-{II} {W}eyl semimetal {W}{T}e$_2$},\ }\href
  {https://doi.org/10.1103/PhysRevB.94.241119} {\bibfield  {journal} {\bibinfo
  {journal} {Physical Review B}\ }\textbf {\bibinfo {volume} {94}},\ \bibinfo
  {pages} {241119(R)} (\bibinfo {year} {2016}{\natexlab{a}})}\BibitemShut
  {NoStop}%
\bibitem [{\citenamefont {Bruno}\ \emph {et~al.}(2016)\citenamefont {Bruno},
  \citenamefont {Tamai}, \citenamefont {Wu}, \citenamefont {Cucchi},
  \citenamefont {Barreteau}, \citenamefont {dela Torre}, \citenamefont
  {Walker}, \citenamefont {Riccò}, \citenamefont {Wang}, \citenamefont {Kim},
  \citenamefont {Hoesch}, \citenamefont {Shi}, \citenamefont {Plumb},
  \citenamefont {Giannini}, \citenamefont {Soluyanov},\ and\ \citenamefont
  {Baumberger}}]{Bruno2016}%
  \BibitemOpen
  \bibfield  {author} {\bibinfo {author} {\bibfnamefont {F.~Y.}\ \bibnamefont
  {Bruno}}, \bibinfo {author} {\bibfnamefont {A.}~\bibnamefont {Tamai}},
  \bibinfo {author} {\bibfnamefont {Q.~S.}\ \bibnamefont {Wu}}, \bibinfo
  {author} {\bibfnamefont {I.}~\bibnamefont {Cucchi}}, \bibinfo {author}
  {\bibfnamefont {C.}~\bibnamefont {Barreteau}}, \bibinfo {author}
  {\bibfnamefont {A.}~\bibnamefont {dela Torre}}, \bibinfo {author}
  {\bibfnamefont {S. }~\bibnamefont {McKeownWalker}}, \bibinfo {author}
  {\bibfnamefont {S.}~\bibnamefont {Ricco}}, \bibinfo {author} {\bibfnamefont
  {Z.}~\bibnamefont {Wang}}, \bibinfo {author} {\bibfnamefont {T.~K.}\
  \bibnamefont {Kim}}, \bibinfo {author} {\bibfnamefont {M.}~\bibnamefont
  {Hoesch}}, \bibinfo {author} {\bibfnamefont {M.}~\bibnamefont {Shi}},
  \bibinfo {author} {\bibfnamefont {N.~C.}\ \bibnamefont {Plumb}}, \bibinfo
  {author} {\bibfnamefont {E.}~\bibnamefont {Giannini}}, \bibinfo {author}
  {\bibfnamefont {A.~A.}\ \bibnamefont {Soluyanov}},\ and\ \bibinfo {author}
  {\bibfnamefont {F.}~\bibnamefont {Baumberger}},\ }\bibfield  {title}
  {\bibinfo {title} {Observation of large topologically trivial {F}ermi arcs in
  the candidate type-{II} {W}eyl semimetal {W}{T}e$_2$},\ }\href
  {https://doi.org/10.1103/PhysRevB.94.121112} {\bibfield  {journal} {\bibinfo
  {journal} {Physical Review B}\ }\textbf {\bibinfo {volume} {94}},\ \bibinfo
  {pages} {121112(R)} (\bibinfo {year} {2016})}\BibitemShut {NoStop}%
\bibitem [{\citenamefont {Deng}\ \emph {et~al.}(2016)\citenamefont {Deng},
  \citenamefont {Wan}, \citenamefont {Deng}, \citenamefont {Zhang},
  \citenamefont {Ding}, \citenamefont {Wang}, \citenamefont {Yan},
  \citenamefont {Huang}, \citenamefont {Zhang}, \citenamefont {Xu},
  \citenamefont {Denlinger}, \citenamefont {Fedorov}, \citenamefont {Yang},
  \citenamefont {Duan}, \citenamefont {Yao}, \citenamefont {Wu}, \citenamefont
  {Fan}, \citenamefont {Zhang}, \citenamefont {Chen},\ and\ \citenamefont
  {Zhou}}]{Deng2016}%
  \BibitemOpen
  \bibfield  {author} {\bibinfo {author} {\bibfnamefont {K.}~\bibnamefont
  {Deng}}, \bibinfo {author} {\bibfnamefont {G.}~\bibnamefont {Wan}}, \bibinfo
  {author} {\bibfnamefont {P.}~\bibnamefont {Deng}}, \bibinfo {author}
  {\bibfnamefont {K.}~\bibnamefont {Zhang}}, \bibinfo {author} {\bibfnamefont
  {S.}~\bibnamefont {Ding}}, \bibinfo {author} {\bibfnamefont {E.}~\bibnamefont
  {Wang}}, \bibinfo {author} {\bibfnamefont {M.}~\bibnamefont {Yan}}, \bibinfo
  {author} {\bibfnamefont {H.}~\bibnamefont {Huang}}, \bibinfo {author}
  {\bibfnamefont {H.}~\bibnamefont {Zhang}}, \bibinfo {author} {\bibfnamefont
  {Z.}~\bibnamefont {Xu}}, \bibinfo {author} {\bibfnamefont {J.}~\bibnamefont
  {Denlinger}}, \bibinfo {author} {\bibfnamefont {A.}~\bibnamefont {Fedorov}},
  \bibinfo {author} {\bibfnamefont {H.}~\bibnamefont {Yang}}, \bibinfo {author}
  {\bibfnamefont {W.}~\bibnamefont {Duan}}, \bibinfo {author} {\bibfnamefont
  {H.}~\bibnamefont {Yao}}, \bibinfo {author} {\bibfnamefont {Y.}~\bibnamefont
  {Wu}}, \bibinfo {author} {\bibfnamefont {S.}~\bibnamefont {Fan}}, \bibinfo
  {author} {\bibfnamefont {H.}~\bibnamefont {Zhang}}, \bibinfo {author}
  {\bibfnamefont {X.}~\bibnamefont {Chen}},\ and\ \bibinfo {author}
  {\bibfnamefont {S.}~\bibnamefont {Zhou}},\ }\bibfield  {title} {\bibinfo
  {title} {Experimental observation of topological {F}ermi arcs in type-{II}
  {W}eyl semimetal {M}o{T}e{$_2$}},\ }\href {https://doi.org/10.1038/nphys3871}
  {\bibfield  {journal} {\bibinfo  {journal} {Nature Physics}\ }\textbf
  {\bibinfo {volume} {12}},\ \bibinfo {pages} {1105} (\bibinfo {year}
  {2016})}\BibitemShut {NoStop}%
\bibitem [{\citenamefont {Tamai}\ \emph {et~al.}(2016)\citenamefont {Tamai},
  \citenamefont {Wu}, \citenamefont {Cucchi}, \citenamefont {Bruno},
  \citenamefont {Riccò}, \citenamefont {Kim}, \citenamefont {Hoesch},
  \citenamefont {Barreteau}, \citenamefont {Giannini}, \citenamefont {Besnard},
  \citenamefont {Soluyanov},\ and\ \citenamefont {Baumberger}}]{Tamai2016}%
  \BibitemOpen
  \bibfield  {author} {\bibinfo {author} {\bibfnamefont {A.}~\bibnamefont
  {Tamai}}, \bibinfo {author} {\bibfnamefont {Q.S.}~\bibnamefont {Wu}}, \bibinfo
  {author} {\bibfnamefont {I.}~\bibnamefont {Cucchi}}, \bibinfo {author}
  {\bibfnamefont {F.Y.}~\bibnamefont {Bruno}}, \bibinfo {author} {\bibfnamefont
  {S.}~\bibnamefont {Ricco}}, \bibinfo {author} {\bibfnamefont
  {T.K.}~\bibnamefont {Kim}}, \bibinfo {author} {\bibfnamefont {M.}~\bibnamefont
  {Hoesch}}, \bibinfo {author} {\bibfnamefont {C.}~\bibnamefont {Barreteau}},
  \bibinfo {author} {\bibfnamefont {E.}~\bibnamefont {Giannini}}, \bibinfo
  {author} {\bibfnamefont {C.}~\bibnamefont {Besnard}}, \bibinfo {author}
  {\bibfnamefont {A. A.}~\bibnamefont {Soluyanov}},\ and\ \bibinfo {author}
  {\bibfnamefont {F.}~\bibnamefont {Baumberger}},\ }\bibfield  {title}
  {\bibinfo {title} {Fermi arcs and their topological character in the
  candidate type-{II} {W}eyl semimetal {M}o{T}e$_2$},\ }\href
  {https://doi.org/10.1103/PhysRevX.6.031021} {\bibfield  {journal} {\bibinfo
  {journal} {Physical Review X}\ }\textbf {\bibinfo {volume} {6}},\ \bibinfo
  {pages} {031021} (\bibinfo {year} {2016})}\BibitemShut {NoStop}%
\bibitem [{\citenamefont {Chang}\ \emph {et~al.}(2016)\citenamefont {Chang},
  \citenamefont {Xu}, \citenamefont {Chang}, \citenamefont {Lee}, \citenamefont
  {Huang}, \citenamefont {Wang}, \citenamefont {Bian}, \citenamefont {Zheng},
  \citenamefont {Sanchez}, \citenamefont {Belopolski}, \citenamefont
  {Alidoust}, \citenamefont {Neupane}, \citenamefont {Bansil}, \citenamefont
  {Jeng}, \citenamefont {Lin},\ and\ \citenamefont {Hasan}}]{Chang2016}%
  \BibitemOpen
  \bibfield  {author} {\bibinfo {author} {\bibfnamefont {T.-R.}\ \bibnamefont
  {Chang}}, \bibinfo {author} {\bibfnamefont {S.-Y.}\ \bibnamefont {Xu}},
  \bibinfo {author} {\bibfnamefont {G.}~\bibnamefont {Chang}}, \bibinfo
  {author} {\bibfnamefont {C.-C.}\ \bibnamefont {Lee}}, \bibinfo {author}
  {\bibfnamefont {S.-M.}\ \bibnamefont {Huang}}, \bibinfo {author}
  {\bibfnamefont {B.}~\bibnamefont {Wang}}, \bibinfo {author} {\bibfnamefont
  {G.}~\bibnamefont {Bian}}, \bibinfo {author} {\bibfnamefont {H.}~\bibnamefont
  {Zheng}}, \bibinfo {author} {\bibfnamefont {D.~S.}\ \bibnamefont {Sanchez}},
  \bibinfo {author} {\bibfnamefont {I.}~\bibnamefont {Belopolski}}, \bibinfo
  {author} {\bibfnamefont {N.}~\bibnamefont {Alidoust}}, \bibinfo {author}
  {\bibfnamefont {M.}~\bibnamefont {Neupane}}, \bibinfo {author} {\bibfnamefont
  {A.}~\bibnamefont {Bansil}}, \bibinfo {author} {\bibfnamefont {H.-T.}\
  \bibnamefont {Jeng}}, \bibinfo {author} {\bibfnamefont {H.}~\bibnamefont
  {Lin}},\ and\ \bibinfo {author} {\bibfnamefont {M.~Z.}\ \bibnamefont
  {Hasan}},\ }\bibfield  {title} {\bibinfo {title} {Prediction of an
  arc-tunable weyl {F}ermion metallic state in {M}o$_x${W}$_{1−x}${T}e$_2$},\
  }\href {https://doi.org/10.1038/ncomms10639} {\bibfield  {journal} {\bibinfo
  {journal} {Nature Communications}\ }\textbf {\bibinfo {volume} {7}},\
  \bibinfo {pages} {10639} (\bibinfo {year} {2016})}\BibitemShut {NoStop}%
\bibitem [{\citenamefont {Belopolski}\ \emph
  {et~al.}(2016{\natexlab{a}})\citenamefont {Belopolski}, \citenamefont {Xu},
  \citenamefont {Ishida}, \citenamefont {Pan}, \citenamefont {Yu},
  \citenamefont {Sanchez}, \citenamefont {Zheng}, \citenamefont {Neupane},
  \citenamefont {Alidoust}, \citenamefont {Chang}, \citenamefont {Chang},
  \citenamefont {Wu}, \citenamefont {Bian}, \citenamefont {Huang},
  \citenamefont {Lee}, \citenamefont {Mou}, \citenamefont {Huang},
  \citenamefont {Song}, \citenamefont {Wang}, \citenamefont {Wang},
  \citenamefont {Yeh}, \citenamefont {Yao}, \citenamefont {Rault},
  \citenamefont {Fèvre}, \citenamefont {Bertran}, \citenamefont {Jeng},
  \citenamefont {Kondo}, \citenamefont {Kaminski}, \citenamefont {Lin},
  \citenamefont {Liu}, \citenamefont {Song}, \citenamefont {Shin},\ and\
  \citenamefont {Hasan}}]{BelopolskiPRB2016}%
  \BibitemOpen
  \bibfield  {author} {\bibinfo {author} {\bibfnamefont {I.}~\bibnamefont
  {Belopolski}}, \bibinfo {author} {\bibfnamefont {S.-Y.}\ \bibnamefont {Xu}},
  \bibinfo {author} {\bibfnamefont {Y.}~\bibnamefont {Ishida}}, \bibinfo
  {author} {\bibfnamefont {X.}~\bibnamefont {Pan}}, \bibinfo {author}
  {\bibfnamefont {P.}~\bibnamefont {Yu}}, \bibinfo {author} {\bibfnamefont
  {D.~S.}\ \bibnamefont {Sanchez}}, \bibinfo {author} {\bibfnamefont
  {H.}~\bibnamefont {Zheng}}, \bibinfo {author} {\bibfnamefont
  {M.}~\bibnamefont {Neupane}}, \bibinfo {author} {\bibfnamefont
  {N.}~\bibnamefont {Alidoust}}, \bibinfo {author} {\bibfnamefont
  {G.}~\bibnamefont {Chang}}, \bibinfo {author} {\bibfnamefont {T.-R.}\
  \bibnamefont {Chang}}, \bibinfo {author} {\bibfnamefont {Y.}~\bibnamefont
  {Wu}}, \bibinfo {author} {\bibfnamefont {G.}~\bibnamefont {Bian}}, \bibinfo
  {author} {\bibfnamefont {S.-M.}\ \bibnamefont {Huang}}, \bibinfo {author}
  {\bibfnamefont {C.-C.}\ \bibnamefont {Lee}}, \bibinfo {author} {\bibfnamefont
  {D.}~\bibnamefont {Mou}}, \bibinfo {author} {\bibfnamefont {L.}~\bibnamefont
  {Huang}}, \bibinfo {author} {\bibfnamefont {Y.}~\bibnamefont {Song}},
  \bibinfo {author} {\bibfnamefont {B.}~\bibnamefont {Wang}}, \bibinfo {author}
  {\bibfnamefont {G.}~\bibnamefont {Wang}}, \bibinfo {author} {\bibfnamefont
  {Y.-W.}\ \bibnamefont {Yeh}}, \bibinfo {author} {\bibfnamefont
  {N.}~\bibnamefont {Yao}}, \bibinfo {author} {\bibfnamefont {J.~E.}\
  \bibnamefont {Rault}}, \bibinfo {author} {\bibfnamefont {P.~L.}\ \bibnamefont
  {LeFevre}}, \bibinfo {author} {\bibfnamefont {F.}~\bibnamefont {Bertran}},
  \bibinfo {author} {\bibfnamefont {H.-T.}\ \bibnamefont {Jeng}}, \bibinfo
  {author} {\bibfnamefont {T.}~\bibnamefont {Kondo}}, \bibinfo {author}
  {\bibfnamefont {A.}~\bibnamefont {Kaminski}}, \bibinfo {author}
  {\bibfnamefont {H.}~\bibnamefont {Lin}}, \bibinfo {author} {\bibfnamefont
  {Z.}~\bibnamefont {Liu}}, \bibinfo {author} {\bibfnamefont {F.}~\bibnamefont
  {Song}}, \bibinfo {author} {\bibfnamefont {S.}~\bibnamefont {Shin}},\ and\
  \bibinfo {author} {\bibfnamefont {M.~Z.}\ \bibnamefont {Hasan}},\ }\bibfield
  {title} {\bibinfo {title} {{F}ermi arc electronic structure and {C}hern
  numbers in the type-{II} {W}eyl semimetal candidate
  {M}o{W}$_{1-x}${T}e$_2$},\ }\href
  {https://doi.org/10.1103/PhysRevB.94.085127} {\bibfield  {journal} {\bibinfo
  {journal} {Physical Review B}\ }\textbf {\bibinfo {volume} {94}},\ \bibinfo
  {pages} {085127} (\bibinfo {year} {2016}{\natexlab{a}})}\BibitemShut
  {NoStop}%
\bibitem [{\citenamefont {Belopolski}\ \emph
  {et~al.}(2016{\natexlab{b}})\citenamefont {Belopolski}, \citenamefont
  {Sanchez}, \citenamefont {Ishida}, \citenamefont {Pan}, \citenamefont {Yu},
  \citenamefont {Xu}, \citenamefont {Chang}, \citenamefont {Chang},
  \citenamefont {Zheng}, \citenamefont {Alidoust}, \citenamefont {Bian},
  \citenamefont {Neupane}, \citenamefont {Huang}, \citenamefont {Lee},
  \citenamefont {Song}, \citenamefont {Bu}, \citenamefont {Wang}, \citenamefont
  {Li}, \citenamefont {Eda}, \citenamefont {Jeng}, \citenamefont {Kondo},
  \citenamefont {Lin}, \citenamefont {Liu}, \citenamefont {Song}, \citenamefont
  {Shin},\ and\ \citenamefont {Hasan}}]{BelopolskiNatComm2016}%
  \BibitemOpen
  \bibfield  {author} {\bibinfo {author} {\bibfnamefont {I.}~\bibnamefont
  {Belopolski}}, \bibinfo {author} {\bibfnamefont {D.~S.}\ \bibnamefont
  {Sanchez}}, \bibinfo {author} {\bibfnamefont {Y.}~\bibnamefont {Ishida}},
  \bibinfo {author} {\bibfnamefont {X.}~\bibnamefont {Pan}}, \bibinfo {author}
  {\bibfnamefont {P.}~\bibnamefont {Yu}}, \bibinfo {author} {\bibfnamefont
  {S.-Y.}\ \bibnamefont {Xu}}, \bibinfo {author} {\bibfnamefont
  {G.}~\bibnamefont {Chang}}, \bibinfo {author} {\bibfnamefont {T.-R.}\
  \bibnamefont {Chang}}, \bibinfo {author} {\bibfnamefont {H.}~\bibnamefont
  {Zheng}}, \bibinfo {author} {\bibfnamefont {N.}~\bibnamefont {Alidoust}},
  \bibinfo {author} {\bibfnamefont {G.}~\bibnamefont {Bian}}, \bibinfo {author}
  {\bibfnamefont {M.}~\bibnamefont {Neupane}}, \bibinfo {author} {\bibfnamefont
  {S.-M.}\ \bibnamefont {Huang}}, \bibinfo {author} {\bibfnamefont {C.-C.}\
  \bibnamefont {Lee}}, \bibinfo {author} {\bibfnamefont {Y.}~\bibnamefont
  {Song}}, \bibinfo {author} {\bibfnamefont {H.}~\bibnamefont {Bu}}, \bibinfo
  {author} {\bibfnamefont {G.}~\bibnamefont {Wang}}, \bibinfo {author}
  {\bibfnamefont {S.}~\bibnamefont {Li}}, \bibinfo {author} {\bibfnamefont
  {G.}~\bibnamefont {Eda}}, \bibinfo {author} {\bibfnamefont {H.-T.}\
  \bibnamefont {Jeng}}, \bibinfo {author} {\bibfnamefont {T.}~\bibnamefont
  {Kondo}}, \bibinfo {author} {\bibfnamefont {H.}~\bibnamefont {Lin}}, \bibinfo
  {author} {\bibfnamefont {Z.}~\bibnamefont {Liu}}, \bibinfo {author}
  {\bibfnamefont {F.}~\bibnamefont {Song}}, \bibinfo {author} {\bibfnamefont
  {S.}~\bibnamefont {Shin}},\ and\ \bibinfo {author} {\bibfnamefont {M.~Z.}\
  \bibnamefont {Hasan}},\ }\bibfield  {title} {\bibinfo {title} {Discovery of a
  new type of topological {W}eyl fermion semimetal state in
  {M}o$_x${W}$_{1−x}${T}e$_2$},\ }\href {https://doi.org/10.1038/ncomms13643}
  {\bibfield  {journal} {\bibinfo  {journal} {Nature Communications}\ }\textbf
  {\bibinfo {volume} {7}},\ \bibinfo {pages} {13643} (\bibinfo {year}
  {2016}{\natexlab{b}})}\BibitemShut {NoStop}%
\bibitem [{\citenamefont {Zheng}\ \emph {et~al.}(2016)\citenamefont {Zheng},
  \citenamefont {Cai}, \citenamefont {Ge}, \citenamefont {Zhang}, \citenamefont
  {Liu}, \citenamefont {Lu}, \citenamefont {Zhang}, \citenamefont {Qiu},
  \citenamefont {Taniguchi}, \citenamefont {Watanabe}, \citenamefont {Jia},
  \citenamefont {Qi}, \citenamefont {Chen}, \citenamefont {Sun},\ and\
  \citenamefont {Feng}}]{Zheng2016}%
  \BibitemOpen
  \bibfield  {author} {\bibinfo {author} {\bibfnamefont {F.}~\bibnamefont
  {Zheng}}, \bibinfo {author} {\bibfnamefont {C.}~\bibnamefont {Cai}}, \bibinfo
  {author} {\bibfnamefont {S.}~\bibnamefont {Ge}}, \bibinfo {author}
  {\bibfnamefont {X.}~\bibnamefont {Zhang}}, \bibinfo {author} {\bibfnamefont
  {X.}~\bibnamefont {Liu}}, \bibinfo {author} {\bibfnamefont {H.}~\bibnamefont
  {Lu}}, \bibinfo {author} {\bibfnamefont {Y.}~\bibnamefont {Zhang}}, \bibinfo
  {author} {\bibfnamefont {J.}~\bibnamefont {Qiu}}, \bibinfo {author}
  {\bibfnamefont {T.}~\bibnamefont {Taniguchi}}, \bibinfo {author}
  {\bibfnamefont {K.}~\bibnamefont {Watanabe}}, \bibinfo {author}
  {\bibfnamefont {S.}~\bibnamefont {Jia}}, \bibinfo {author} {\bibfnamefont
  {J.}~\bibnamefont {Qi}}, \bibinfo {author} {\bibfnamefont {J.-H.}\
  \bibnamefont {Chen}}, \bibinfo {author} {\bibfnamefont {D.}~\bibnamefont
  {Sun}},\ and\ \bibinfo {author} {\bibfnamefont {J.}~\bibnamefont {Feng}},\
  }\bibfield  {title} {\bibinfo {title} {On the quantum spin {H}all gap of
  monolayer 1{T}'-{W}{T}e$_2$},\ }\href
  {https://doi.org/10.1002/adma.201600100} {\bibfield  {journal} {\bibinfo
  {journal} {Advanced Materials}\ }\textbf {\bibinfo {volume} {28}},\ \bibinfo
  {pages} {4845} (\bibinfo {year} {2016})}\BibitemShut {NoStop}%
\bibitem [{\citenamefont {Wu}\ \emph {et~al.}(2018)\citenamefont {Wu},
  \citenamefont {Fatemi}, \citenamefont {Gibson}, \citenamefont {Watanabe},
  \citenamefont {Taniguchi}, \citenamefont {Cava},\ and\ \citenamefont
  {Jarillo-Herrero}}]{Wu2018}%
  \BibitemOpen
  \bibfield  {author} {\bibinfo {author} {\bibfnamefont {S.}~\bibnamefont
  {Wu}}, \bibinfo {author} {\bibfnamefont {V.}~\bibnamefont {Fatemi}}, \bibinfo
  {author} {\bibfnamefont {Q.~D.}\ \bibnamefont {Gibson}}, \bibinfo {author}
  {\bibfnamefont {K.}~\bibnamefont {Watanabe}}, \bibinfo {author}
  {\bibfnamefont {T.}~\bibnamefont {Taniguchi}}, \bibinfo {author}
  {\bibfnamefont {R.~J.}\ \bibnamefont {Cava}},\ and\ \bibinfo {author}
  {\bibfnamefont {P.}~\bibnamefont {Jarillo-Herrero}},\ }\bibfield  {title}
  {\bibinfo {title} {Observation of the quantum spin {H}all effect up to 100
  kelvin in a monolayer crystal},\ }\href
  {https://doi.org/10.1126/science.aan6003} {\bibfield  {journal} {\bibinfo
  {journal} {Science}\ }\textbf {\bibinfo {volume} {359}},\ \bibinfo {pages}
  {76} (\bibinfo {year} {2018})}\BibitemShut {NoStop}%
\bibitem [{\citenamefont {Zhao}\ \emph {et~al.}(2020)\citenamefont {Zhao},
  \citenamefont {Khokhriakov}, \citenamefont {Zhang}, \citenamefont {Fu},
  \citenamefont {Karpiak}, \citenamefont {Hoque}, \citenamefont {Xu},
  \citenamefont {Jiang}, \citenamefont {Yan},\ and\ \citenamefont
  {Dash}}]{Zhao2020}%
  \BibitemOpen
  \bibfield  {author} {\bibinfo {author} {\bibfnamefont {B.}~\bibnamefont
  {Zhao}}, \bibinfo {author} {\bibfnamefont {D.}~\bibnamefont {Khokhriakov}},
  \bibinfo {author} {\bibfnamefont {Y.}~\bibnamefont {Zhang}}, \bibinfo
  {author} {\bibfnamefont {H.}~\bibnamefont {Fu}}, \bibinfo {author}
  {\bibfnamefont {B.}~\bibnamefont {Karpiak}}, \bibinfo {author} {\bibfnamefont
  {A.~M.}\ \bibnamefont {Hoque}}, \bibinfo {author} {\bibfnamefont
  {X.}~\bibnamefont {Xu}}, \bibinfo {author} {\bibfnamefont {Y.}~\bibnamefont
  {Jiang}}, \bibinfo {author} {\bibfnamefont {B.}~\bibnamefont {Yan}},\ and\
  \bibinfo {author} {\bibfnamefont {S.~P.}\ \bibnamefont {Dash}},\ }\bibfield
  {title} {\bibinfo {title} {Observation of charge to spin conversion in {W}eyl
  semimetal {W}{T}e$_2$ at room temperature},\ }\href
  {https://doi.org/10.1103/PhysRevResearch.2.013286} {\bibfield  {journal}
  {\bibinfo  {journal} {Physical Review Research}\ }\textbf {\bibinfo {volume}
  {2}},\ \bibinfo {pages} {013286} (\bibinfo {year} {2020})}\BibitemShut
  {NoStop}%
\bibitem [{\citenamefont {Kononov}\ \emph {et~al.}(2020)\citenamefont
  {Kononov}, \citenamefont {Abulizi}, \citenamefont {Qu}, \citenamefont {Yan},
  \citenamefont {Mandrus}, \citenamefont {Watanabe}, \citenamefont
  {Taniguchi},\ and\ \citenamefont {Schönenberger}}]{Kononov2020}%
  \BibitemOpen
  \bibfield  {author} {\bibinfo {author} {\bibfnamefont {A.}~\bibnamefont
  {Kononov}}, \bibinfo {author} {\bibfnamefont {G.}~\bibnamefont {Abulizi}},
  \bibinfo {author} {\bibfnamefont {K.}~\bibnamefont {Qu}}, \bibinfo {author}
  {\bibfnamefont {J.}~\bibnamefont {Yan}}, \bibinfo {author} {\bibfnamefont
  {D.}~\bibnamefont {Mandrus}}, \bibinfo {author} {\bibfnamefont
  {K.}~\bibnamefont {Watanabe}}, \bibinfo {author} {\bibfnamefont
  {T.}~\bibnamefont {Taniguchi}},\ and\ \bibinfo {author} {\bibfnamefont
  {C.}~\bibnamefont {Schönenberger}},\ }\bibfield  {title} {\bibinfo {title}
  {One-dimensional edge transport in few-layer {W}{T}e$_2$},\ }\href
  {https://doi.org/10.1021/acs.nanolett.0c00658} {\bibfield  {journal}
  {\bibinfo  {journal} {Nano Letters}\ }\textbf {\bibinfo {volume} {20}},\
  \bibinfo {pages} {4228} (\bibinfo {year} {2020})}\BibitemShut {NoStop}%
\bibitem [{\citenamefont {Wang}\ \emph {et~al.}(2021)\citenamefont {Wang},
  \citenamefont {Yu}, \citenamefont {Jia}, \citenamefont {Onyszczak},
  \citenamefont {Cevallos}, \citenamefont {Lei}, \citenamefont {Klemenz},
  \citenamefont {Watanabe}, \citenamefont {Taniguchi}, \citenamefont {Cava},
  \citenamefont {Schoop},\ and\ \citenamefont {Wu}}]{Wang2021}%
  \BibitemOpen
  \bibfield  {author} {\bibinfo {author} {\bibfnamefont {P.}~\bibnamefont
  {Wang}}, \bibinfo {author} {\bibfnamefont {G.}~\bibnamefont {Yu}}, \bibinfo
  {author} {\bibfnamefont {Y.}~\bibnamefont {Jia}}, \bibinfo {author}
  {\bibfnamefont {M.}~\bibnamefont {Onyszczak}}, \bibinfo {author}
  {\bibfnamefont {F.~A.}\ \bibnamefont {Cevallos}}, \bibinfo {author}
  {\bibfnamefont {S.}~\bibnamefont {Lei}}, \bibinfo {author} {\bibfnamefont
  {S.}~\bibnamefont {Klemenz}}, \bibinfo {author} {\bibfnamefont
  {K.}~\bibnamefont {Watanabe}}, \bibinfo {author} {\bibfnamefont
  {T.}~\bibnamefont {Taniguchi}}, \bibinfo {author} {\bibfnamefont {R.~J.}\
  \bibnamefont {Cava}}, \bibinfo {author} {\bibfnamefont {L.~M.}\ \bibnamefont
  {Schoop}},\ and\ \bibinfo {author} {\bibfnamefont {S.}~\bibnamefont {Wu}},\
  }\bibfield  {title} {\bibinfo {title} {Landau quantization and highly mobile
  fermions in an insulator},\ }\href
  {https://doi.org/10.1038/s41586-020-03084-9} {\bibfield  {journal} {\bibinfo
  {journal} {Nature}\ }\textbf {\bibinfo {volume} {589}},\ \bibinfo {pages}
  {225} (\bibinfo {year} {2021})}\BibitemShut {NoStop}%
\bibitem [{\citenamefont {Pletikosić}\ \emph {et~al.}(2014)\citenamefont
  {Pletikosić}, \citenamefont {Ali}, \citenamefont {Fedorov}, \citenamefont
  {Cava},\ and\ \citenamefont {Valla}}]{Pletikosic2014}%
  \BibitemOpen
  \bibfield  {author} {\bibinfo {author} {\bibfnamefont {I.}~\bibnamefont
  {Pletikosic}}, \bibinfo {author} {\bibfnamefont {M.~N.}\ \bibnamefont
  {Ali}}, \bibinfo {author} {\bibfnamefont {A. V.}~\bibnamefont {Fedorov}},
  \bibinfo {author} {\bibfnamefont {R.J.}~\bibnamefont {Cava}},\ and\ \bibinfo
  {author} {\bibfnamefont {T.}~\bibnamefont {Valla}},\ }\bibfield  {title}
  {\bibinfo {title} {Electronic structure basis for the extraordinary
  magnetoresistance in {W}{T}e$_2$},\ }\href
  {https://doi.org/10.1103/PhysRevLett.113.216601} {\bibfield  {journal}
  {\bibinfo  {journal} {Physical Review Letters}\ }\textbf {\bibinfo {volume}
  {113}},\ \bibinfo {pages} {216601} (\bibinfo {year} {2014})}\BibitemShut
  {NoStop}%
\bibitem [{\citenamefont {Jiang}\ \emph {et~al.}(2015)\citenamefont {Jiang},
  \citenamefont {Tang}, \citenamefont {Pan}, \citenamefont {Liu}, \citenamefont
  {Niu}, \citenamefont {Wang}, \citenamefont {Xu}, \citenamefont {Yang},
  \citenamefont {Xie}, \citenamefont {Song}, \citenamefont {Dudin},
  \citenamefont {Kim}, \citenamefont {Hoesch}, \citenamefont {Das},
  \citenamefont {Vobornik}, \citenamefont {Wan},\ and\ \citenamefont
  {Feng}}]{Jiang2015}%
  \BibitemOpen
  \bibfield  {author} {\bibinfo {author} {\bibfnamefont {J.}~\bibnamefont
  {Jiang}}, \bibinfo {author} {\bibfnamefont {F.}~\bibnamefont {Tang}},
  \bibinfo {author} {\bibfnamefont {X. C.}~\bibnamefont {Pan}}, \bibinfo {author}
  {\bibfnamefont {H. M.}~\bibnamefont {Liu}}, \bibinfo {author} {\bibfnamefont
  {X. H.}~\bibnamefont {Niu}}, \bibinfo {author} {\bibfnamefont {Y. X.}~\bibnamefont
  {Wang}}, \bibinfo {author} {\bibfnamefont {D. F.}~\bibnamefont {Xu}}, \bibinfo
  {author} {\bibfnamefont {H. F.}~\bibnamefont {Yang}}, \bibinfo {author}
  {\bibfnamefont {B. P.}~\bibnamefont {Xie}}, \bibinfo {author} {\bibfnamefont
  {F. Q.}~\bibnamefont {Song}}, \bibinfo {author} {\bibfnamefont {P.}~\bibnamefont
  {Dudin}}, \bibinfo {author} {\bibfnamefont {T. K.}~\bibnamefont {Kim}}, \bibinfo
  {author} {\bibfnamefont {M.}~\bibnamefont {Hoesch}}, \bibinfo {author}
  {\bibfnamefont {P.~K.}\ \bibnamefont {Das}}, \bibinfo {author} {\bibfnamefont
  {I.}~\bibnamefont {Vobornik}}, \bibinfo {author} {\bibfnamefont
  {X. G.}~\bibnamefont {Wan}},\ and\ \bibinfo {author} {\bibfnamefont
  {D. L.}~\bibnamefont {Feng}},\ }\bibfield  {title} {\bibinfo {title} {Signature
  of strong spin-orbital coupling in the large nonsaturating magnetoresistance
  material {W}{T}e$_2$},\ }\href
  {https://doi.org/10.1103/PhysRevLett.115.166601} {\bibfield  {journal}
  {\bibinfo  {journal} {Physical Review Letters}\ }\textbf {\bibinfo {volume}
  {115}},\ \bibinfo {pages} {166601} (\bibinfo {year} {2015})}\BibitemShut
  {NoStop}%
\bibitem [{\citenamefont {Wu}\ \emph {et~al.}(2015)\citenamefont {Wu},
  \citenamefont {Jo}, \citenamefont {Ochi}, \citenamefont {Huang},
  \citenamefont {Mou}, \citenamefont {Bud’ko}, \citenamefont {Canfield},
  \citenamefont {Trivedi}, \citenamefont {Arita},\ and\ \citenamefont
  {Kaminski}}]{Wu2015}%
  \BibitemOpen
  \bibfield  {author} {\bibinfo {author} {\bibfnamefont {Y.}~\bibnamefont
  {Wu}}, \bibinfo {author} {\bibfnamefont {N.~H.}\ \bibnamefont {Jo}}, \bibinfo
  {author} {\bibfnamefont {M.}~\bibnamefont {Ochi}}, \bibinfo {author}
  {\bibfnamefont {L.}~\bibnamefont {Huang}}, \bibinfo {author} {\bibfnamefont
  {D.}~\bibnamefont {Mou}}, \bibinfo {author} {\bibfnamefont {S.~L.}\
  \bibnamefont {Bud’ko}}, \bibinfo {author} {\bibfnamefont {P. C.}~\bibnamefont
  {Canfield}}, \bibinfo {author} {\bibfnamefont {N.}~\bibnamefont {Trivedi}},
  \bibinfo {author} {\bibfnamefont {R.}~\bibnamefont {Arita}},\ and\ \bibinfo
  {author} {\bibfnamefont {A.}~\bibnamefont {Kaminski}},\ }\bibfield  {title}
  {\bibinfo {title} {Temperature-induced {L}ifshitz transition in
  {W}{T}e$_2$},\ }\href {https://doi.org/10.1103/PhysRevLett.115.166602}
  {\bibfield  {journal} {\bibinfo  {journal} {Physical Review Letters}\
  }\textbf {\bibinfo {volume} {115}},\ \bibinfo {pages} {166602} (\bibinfo
  {year} {2015})}\BibitemShut {NoStop}%
\bibitem [{\citenamefont {Zhu}\ \emph {et~al.}(2015)\citenamefont {Zhu},
  \citenamefont {Lin}, \citenamefont {Liu}, \citenamefont {Fauque},
  \citenamefont {Tao}, \citenamefont {Yang}, \citenamefont {Shi},\ and\
  \citenamefont {Behnia}}]{Zhu2015}%
  \BibitemOpen
  \bibfield  {author} {\bibinfo {author} {\bibfnamefont {Z.}~\bibnamefont
  {Zhu}}, \bibinfo {author} {\bibfnamefont {X.}~\bibnamefont {Lin}}, \bibinfo
  {author} {\bibfnamefont {J.}~\bibnamefont {Liu}}, \bibinfo {author}
  {\bibfnamefont {B.}~\bibnamefont {Fauque}}, \bibinfo {author} {\bibfnamefont
  {Q.}~\bibnamefont {Tao}}, \bibinfo {author} {\bibfnamefont {C.}~\bibnamefont
  {Yang}}, \bibinfo {author} {\bibfnamefont {Y.}~\bibnamefont {Shi}},\ and\
  \bibinfo {author} {\bibfnamefont {K.}~\bibnamefont {Behnia}},\ }\bibfield
  {title} {\bibinfo {title} {Quantum oscillations, thermoelectric coefficients,
  and the {F}ermi surface of semimetallic {W}{T}e$_2$},\ }\href
  {https://doi.org/10.1103/PhysRevLett.114.176601} {\bibfield  {journal}
  {\bibinfo  {journal} {Physical Review Letters}\ }\textbf {\bibinfo {volume}
  {114}},\ \bibinfo {pages} {176601} (\bibinfo {year} {2015})}\BibitemShut
  {NoStop}%
\bibitem [{\citenamefont {Rhodes}\ \emph {et~al.}(2015)\citenamefont {Rhodes},
  \citenamefont {Das}, \citenamefont {Zhang}, \citenamefont {Zeng},
  \citenamefont {Pradhan}, \citenamefont {Kikugawa}, \citenamefont
  {Manousakis},\ and\ \citenamefont {Balicas}}]{Rhodes2015}%
  \BibitemOpen
  \bibfield  {author} {\bibinfo {author} {\bibfnamefont {D.}~\bibnamefont
  {Rhodes}}, \bibinfo {author} {\bibfnamefont {S.}~\bibnamefont {Das}},
  \bibinfo {author} {\bibfnamefont {Q.~R.}\ \bibnamefont {Zhang}}, \bibinfo
  {author} {\bibfnamefont {B.}~\bibnamefont {Zeng}}, \bibinfo {author}
  {\bibfnamefont {N.~R.}\ \bibnamefont {Pradhan}}, \bibinfo {author}
  {\bibfnamefont {N.}~\bibnamefont {Kikugawa}}, \bibinfo {author}
  {\bibfnamefont {E.}~\bibnamefont {Manousakis}},\ and\ \bibinfo {author}
  {\bibfnamefont {L.}~\bibnamefont {Balicas}},\ }\bibfield  {title} {\bibinfo
  {title} {Role of spin-orbit coupling and evolution of the electronic
  structure of {W}{T}e$_2$ under an external magnetic field},\ }\href
  {https://doi.org/10.1103/PhysRevB.92.125152} {\bibfield  {journal} {\bibinfo
  {journal} {Physical Review B}\ }\textbf {\bibinfo {volume} {92}},\ \bibinfo
  {pages} {125152} (\bibinfo {year} {2015})}\BibitemShut {NoStop}%
\bibitem [{\citenamefont {Xiang}\ \emph {et~al.}(2015)\citenamefont {Xiang},
  \citenamefont {Veldhorst}, \citenamefont {Dou},\ and\ \citenamefont
  {Wang}}]{Xiang2015}%
  \BibitemOpen
  \bibfield  {author} {\bibinfo {author} {\bibfnamefont {F.-X.}\ \bibnamefont
  {Xiang}}, \bibinfo {author} {\bibfnamefont {M.}~\bibnamefont {Veldhorst}},
  \bibinfo {author} {\bibfnamefont {S.-X.}\ \bibnamefont {Dou}},\ and\ \bibinfo
  {author} {\bibfnamefont {X.-L.}\ \bibnamefont {Wang}},\ }\bibfield  {title}
  {\bibinfo {title} {Multiple fermi pockets revealed by {S}hubnikov-de {H}aas
  oscillations in {W}{T}e$_2$},\ }\href
  {https://doi.org/10.1209/0295-5075/112/37009} {\bibfield  {journal} {\bibinfo
   {journal} {EPL (Europhysics Letters)}\ }\textbf {\bibinfo {volume} {112}},\
  \bibinfo {pages} {37009} (\bibinfo {year} {2015})}\BibitemShut {NoStop}%
\bibitem [{\citenamefont {Wang}\ \emph
  {et~al.}(2015{\natexlab{a}})\citenamefont {Wang}, \citenamefont
  {Gutiérrez-Lezama}, \citenamefont {Barreteau}, \citenamefont {Ubrig},
  \citenamefont {Giannini},\ and\ \citenamefont {Morpurgo}}]{Wang2015}%
  \BibitemOpen
  \bibfield  {author} {\bibinfo {author} {\bibfnamefont {L.}~\bibnamefont
  {Wang}}, \bibinfo {author} {\bibfnamefont {I.}~\bibnamefont
  {Gutiérrez-Lezama}}, \bibinfo {author} {\bibfnamefont {C.}~\bibnamefont
  {Barreteau}}, \bibinfo {author} {\bibfnamefont {N.}~\bibnamefont {Ubrig}},
  \bibinfo {author} {\bibfnamefont {E.}~\bibnamefont {Giannini}},\ and\
  \bibinfo {author} {\bibfnamefont {A.~F.}\ \bibnamefont {Morpurgo}},\
  }\bibfield  {title} {\bibinfo {title} {Tuning magnetotransport in a
  compensated semimetal at the atomic scale},\ }\href
  {https://doi.org/10.1038/ncomms9892} {\bibfield  {journal} {\bibinfo
  {journal} {Nature Communications}\ }\textbf {\bibinfo {volume} {6}},\
  \bibinfo {pages} {8892} (\bibinfo {year} {2015}{\natexlab{a}})}\BibitemShut
  {NoStop}%
\bibitem [{\citenamefont {Cai}\ \emph {et~al.}(2015)\citenamefont {Cai},
  \citenamefont {Hu}, \citenamefont {He}, \citenamefont {Pan}, \citenamefont
  {Hong}, \citenamefont {Zhang}, \citenamefont {Zhang}, \citenamefont {Wei},
  \citenamefont {Mao},\ and\ \citenamefont {Li}}]{Cai2015}%
  \BibitemOpen
  \bibfield  {author} {\bibinfo {author} {\bibfnamefont {P. L.}~\bibnamefont
  {Cai}}, \bibinfo {author} {\bibfnamefont {J.}~\bibnamefont {Hu}}, \bibinfo
  {author} {\bibfnamefont {L. P.}~\bibnamefont {He}}, \bibinfo {author}
  {\bibfnamefont {J.}~\bibnamefont {Pan}}, \bibinfo {author} {\bibfnamefont
  {X. C.}~\bibnamefont {Hong}}, \bibinfo {author} {\bibfnamefont {Z.}~\bibnamefont
  {Zhang}}, \bibinfo {author} {\bibfnamefont {J.}~\bibnamefont {Zhang}},
  \bibinfo {author} {\bibfnamefont {J.}~\bibnamefont {Wei}}, \bibinfo {author}
  {\bibfnamefont {Z. Q.}~\bibnamefont {Mao}},\ and\ \bibinfo {author}
  {\bibfnamefont {S. Y.}~\bibnamefont {Li}},\ }\bibfield  {title} {\bibinfo
  {title} {Drastic pressure effect on the extremely large magnetoresistance in
  {W}{T}e$_2$ : Quantum oscillation study},\ }\href
  {https://doi.org/10.1103/PhysRevLett.115.057202} {\bibfield  {journal}
  {\bibinfo  {journal} {Physical Review Letters}\ }\textbf {\bibinfo {volume}
  {115}},\ \bibinfo {pages} {057202} (\bibinfo {year} {2015})}\BibitemShut
  {NoStop}%
\bibitem [{\citenamefont {Zhao}\ \emph {et~al.}(2015)\citenamefont {Zhao},
  \citenamefont {Liu}, \citenamefont {Yan}, \citenamefont {An}, \citenamefont
  {Liu}, \citenamefont {Zhang}, \citenamefont {Wang}, \citenamefont {Liu},
  \citenamefont {Jiang}, \citenamefont {Li}, \citenamefont {Wang},
  \citenamefont {Li}, \citenamefont {Mandrus}, \citenamefont {Xie},
  \citenamefont {Pan},\ and\ \citenamefont {Wang}}]{Zhao2015}%
  \BibitemOpen
  \bibfield  {author} {\bibinfo {author} {\bibfnamefont {Y.}~\bibnamefont
  {Zhao}}, \bibinfo {author} {\bibfnamefont {H.}~\bibnamefont {Liu}}, \bibinfo
  {author} {\bibfnamefont {J.}~\bibnamefont {Yan}}, \bibinfo {author}
  {\bibfnamefont {W.}~\bibnamefont {An}}, \bibinfo {author} {\bibfnamefont
  {J.}~\bibnamefont {Liu}}, \bibinfo {author} {\bibfnamefont {X.}~\bibnamefont
  {Zhang}}, \bibinfo {author} {\bibfnamefont {H.}~\bibnamefont {Wang}},
  \bibinfo {author} {\bibfnamefont {Y.}~\bibnamefont {Liu}}, \bibinfo {author}
  {\bibfnamefont {H.}~\bibnamefont {Jiang}}, \bibinfo {author} {\bibfnamefont
  {Q.}~\bibnamefont {Li}}, \bibinfo {author} {\bibfnamefont {Y.}~\bibnamefont
  {Wang}}, \bibinfo {author} {\bibfnamefont {X.-Z.}\ \bibnamefont {Li}},
  \bibinfo {author} {\bibfnamefont {D.}~\bibnamefont {Mandrus}}, \bibinfo
  {author} {\bibfnamefont {X.~C.}\ \bibnamefont {Xie}}, \bibinfo {author}
  {\bibfnamefont {M.}~\bibnamefont {Pan}},\ and\ \bibinfo {author}
  {\bibfnamefont {J.}~\bibnamefont {Wang}},\ }\bibfield  {title} {\bibinfo
  {title} {Anisotropic magnetotransport and exotic longitudinal linear
  magnetoresistance in {W}{T}e$_2$ crystals},\ }\href
  {https://doi.org/10.1103/PhysRevB.92.041104} {\bibfield  {journal} {\bibinfo
  {journal} {Physical Review B}\ }\textbf {\bibinfo {volume} {92}},\ \bibinfo
  {pages} {041104(R)} (\bibinfo {year} {2015})}\BibitemShut {NoStop}%
\bibitem [{\citenamefont {Wang}\ \emph
  {et~al.}(2015{\natexlab{b}})\citenamefont {Wang}, \citenamefont {Thoutam},
  \citenamefont {Xiao}, \citenamefont {Hu}, \citenamefont {Das}, \citenamefont
  {Mao}, \citenamefont {Wei}, \citenamefont {Divan}, \citenamefont
  {Luican-Mayer}, \citenamefont {Crabtree},\ and\ \citenamefont
  {Kwok}}]{WangYL2015}%
  \BibitemOpen
  \bibfield  {author} {\bibinfo {author} {\bibfnamefont {Y.~L.}\ \bibnamefont
  {Wang}}, \bibinfo {author} {\bibfnamefont {L.~R.}\ \bibnamefont {Thoutam}},
  \bibinfo {author} {\bibfnamefont {Z.~L.}\ \bibnamefont {Xiao}}, \bibinfo
  {author} {\bibfnamefont {J.}~\bibnamefont {Hu}}, \bibinfo {author}
  {\bibfnamefont {S.}~\bibnamefont {Das}}, \bibinfo {author} {\bibfnamefont
  {Z.~Q.}\ \bibnamefont {Mao}}, \bibinfo {author} {\bibfnamefont
  {J.}~\bibnamefont {Wei}}, \bibinfo {author} {\bibfnamefont {R.}~\bibnamefont
  {Divan}}, \bibinfo {author} {\bibfnamefont {A.}~\bibnamefont {Luican-Mayer}},
  \bibinfo {author} {\bibfnamefont {G.~W.}\ \bibnamefont {Crabtree}},\ and\
  \bibinfo {author} {\bibfnamefont {W.~K.}\ \bibnamefont {Kwok}},\ }\bibfield
  {title} {\bibinfo {title} {Origin of the turn-on temperature behavior in
  {W}{T}e$_2$},\ }\href {https://doi.org/10.1103/PhysRevB.92.180402} {\bibfield
   {journal} {\bibinfo  {journal} {Physical Review B}\ }\textbf {\bibinfo
  {volume} {92}},\ \bibinfo {pages} {180402(R)} (\bibinfo {year}
  {2015}{\natexlab{b}})}\BibitemShut {NoStop}%
\bibitem [{\citenamefont {Fatemi}\ \emph {et~al.}(2017)\citenamefont {Fatemi},
  \citenamefont {Gibson}, \citenamefont {Watanabe}, \citenamefont {Taniguchi},
  \citenamefont {Cava},\ and\ \citenamefont {Jarillo-Herrero}}]{Fatemi2017}%
  \BibitemOpen
  \bibfield  {author} {\bibinfo {author} {\bibfnamefont {V.}~\bibnamefont
  {Fatemi}}, \bibinfo {author} {\bibfnamefont {Q.~D.}\ \bibnamefont {Gibson}},
  \bibinfo {author} {\bibfnamefont {K.}~\bibnamefont {Watanabe}}, \bibinfo
  {author} {\bibfnamefont {T.}~\bibnamefont {Taniguchi}}, \bibinfo {author}
  {\bibfnamefont {R.~J.}\ \bibnamefont {Cava}},\ and\ \bibinfo {author}
  {\bibfnamefont {P.}~\bibnamefont {Jarillo-Herrero}},\ }\bibfield  {title}
  {\bibinfo {title} {Magnetoresistance and quantum oscillations of an
  electrostatically tuned semimetal-to-metal transition in ultrathin
  {W}{T}e$_2$},\ }\href {https://doi.org/10.1103/PhysRevB.95.041410} {\bibfield
   {journal} {\bibinfo  {journal} {Physical Review B}\ }\textbf {\bibinfo
  {volume} {95}},\ \bibinfo {pages} { 041410(R)} (\bibinfo {year}
  {2017})}\BibitemShut {NoStop}%
\bibitem [{\citenamefont {Na}\ \emph {et~al.}(2016)\citenamefont {Na},
  \citenamefont {Hoyer}, \citenamefont {Schoop}, \citenamefont {Weber},
  \citenamefont {Lotsch}, \citenamefont {Burghard},\ and\ \citenamefont
  {Kern}}]{Na2016}%
  \BibitemOpen
  \bibfield  {author} {\bibinfo {author} {\bibfnamefont {J.}~\bibnamefont
  {Na}}, \bibinfo {author} {\bibfnamefont {A.}~\bibnamefont {Hoyer}}, \bibinfo
  {author} {\bibfnamefont {L.}~\bibnamefont {Schoop}}, \bibinfo {author}
  {\bibfnamefont {D.}~\bibnamefont {Weber}}, \bibinfo {author} {\bibfnamefont
  {B.~V.}\ \bibnamefont {Lotsch}}, \bibinfo {author} {\bibfnamefont
  {M.}~\bibnamefont {Burghard}},\ and\ \bibinfo {author} {\bibfnamefont
  {K.}~\bibnamefont {Kern}},\ }\bibfield  {title} {\bibinfo {title} {Tuning the
  magnetoresistance of ultrathin {W}{T}e$_2$ sheets by electrostatic gating},\
  }\href {https://doi.org/10.1039/C6NR06327F} {\bibfield  {journal} {\bibinfo
  {journal} {Nanoscale}\ }\textbf {\bibinfo {volume} {8}},\ \bibinfo {pages}
  {18703} (\bibinfo {year} {2016})}\BibitemShut {NoStop}%
\bibitem [{\citenamefont {Wang}\ \emph {et~al.}(2019)\citenamefont {Wang},
  \citenamefont {Wang}, \citenamefont {Liu}, \citenamefont {Wu}, \citenamefont
  {Wang}, \citenamefont {Yan}, \citenamefont {Cheng}, \citenamefont {Shi},
  \citenamefont {Watanabe}, \citenamefont {Taniguchi}, \citenamefont {Liang},\
  and\ \citenamefont {Miao}}]{Wang2019}%
  \BibitemOpen
  \bibfield  {author} {\bibinfo {author} {\bibfnamefont {Y.}~\bibnamefont
  {Wang}}, \bibinfo {author} {\bibfnamefont {L.}~\bibnamefont {Wang}}, \bibinfo
  {author} {\bibfnamefont {X.}~\bibnamefont {Liu}}, \bibinfo {author}
  {\bibfnamefont {H.}~\bibnamefont {Wu}}, \bibinfo {author} {\bibfnamefont
  {P.}~\bibnamefont {Wang}}, \bibinfo {author} {\bibfnamefont {D.}~\bibnamefont
  {Yan}}, \bibinfo {author} {\bibfnamefont {B.}~\bibnamefont {Cheng}}, \bibinfo
  {author} {\bibfnamefont {Y.}~\bibnamefont {Shi}}, \bibinfo {author}
  {\bibfnamefont {K.}~\bibnamefont {Watanabe}}, \bibinfo {author}
  {\bibfnamefont {T.}~\bibnamefont {Taniguchi}}, \bibinfo {author}
  {\bibfnamefont {S.-J.}\ \bibnamefont {Liang}},\ and\ \bibinfo {author}
  {\bibfnamefont {F.}~\bibnamefont {Miao}},\ }\bibfield  {title} {\bibinfo
  {title} {Direct evidence for charge compensation-induced large
  magnetoresistance in thin {W}{T}e$_2$},\ }\href
  {https://doi.org/10.1021/acs.nanolett.9b01275} {\bibfield  {journal}
  {\bibinfo  {journal} {Nano Letters}\ }\textbf {\bibinfo {volume} {19}},\
  \bibinfo {pages} {3969} (\bibinfo {year} {2019})}\BibitemShut {NoStop}%
\bibitem [{\citenamefont {Yi}\ \emph {et~al.}(2017)\citenamefont {Yi},
  \citenamefont {Wu}, \citenamefont {Wang}, \citenamefont {Liu}, \citenamefont
  {Li}, \citenamefont {Zhang}, \citenamefont {He},\ and\ \citenamefont
  {Wang}}]{Yi2017}%
  \BibitemOpen
  \bibfield  {author} {\bibinfo {author} {\bibfnamefont {Y.}~\bibnamefont
  {Yi}}, \bibinfo {author} {\bibfnamefont {C.}~\bibnamefont {Wu}}, \bibinfo
  {author} {\bibfnamefont {H.}~\bibnamefont {Wang}}, \bibinfo {author}
  {\bibfnamefont {H.}~\bibnamefont {Liu}}, \bibinfo {author} {\bibfnamefont
  {H.}~\bibnamefont {Li}}, \bibinfo {author} {\bibfnamefont {H.}~\bibnamefont
  {Zhang}}, \bibinfo {author} {\bibfnamefont {H.}~\bibnamefont {He}},\ and\
  \bibinfo {author} {\bibfnamefont {J.}~\bibnamefont {Wang}},\ }\bibfield
  {title} {\bibinfo {title} {Thickness dependent magneto transport properties
  of {W}{T}e$_2$ thin films},\ }\href
  {https://doi.org/10.1016/j.ssc.2017.05.017} {\bibfield  {journal} {\bibinfo
  {journal} {Solid State Communications}\ }\textbf {\bibinfo {volume} {260}},\
  \bibinfo {pages} {45} (\bibinfo {year} {2017})}\BibitemShut {NoStop}%
\bibitem [{\citenamefont {Wang}\ \emph
  {et~al.}(2016{\natexlab{b}})\citenamefont {Wang}, \citenamefont {Wang},
  \citenamefont {Reutt-Robey}, \citenamefont {Paglione},\ and\ \citenamefont
  {Fuhrer}}]{Wang2016}%
  \BibitemOpen
  \bibfield  {author} {\bibinfo {author} {\bibfnamefont {Y.}~\bibnamefont
  {Wang}}, \bibinfo {author} {\bibfnamefont {K.}~\bibnamefont {Wang}}, \bibinfo
  {author} {\bibfnamefont {J.}~\bibnamefont {Reutt-Robey}}, \bibinfo {author}
  {\bibfnamefont {J.}~\bibnamefont {Paglione}},\ and\ \bibinfo {author}
  {\bibfnamefont {M.~S.}\ \bibnamefont {Fuhrer}},\ }\bibfield  {title}
  {\bibinfo {title} {Breakdown of compensation and persistence of nonsaturating
  magnetoresistance in gated {W}{T}e$_2$ thin flakes},\ }\href
  {https://doi.org/10.1103/PhysRevB.93.121108} {\bibfield  {journal} {\bibinfo
  {journal} {Physical Review B}\ }\textbf {\bibinfo {volume} {93}},\ \bibinfo
  {pages} {121108(R)} (\bibinfo {year} {2016}{\natexlab{b}})}\BibitemShut
  {NoStop}%
\bibitem [{\citenamefont {{S. Thirupathaiah}}\ \emph
  {et~al.}(2017)\citenamefont {{S. Thirupathaiah}}, \citenamefont {{S.
  Thirupathaiah}}, \citenamefont {Thirupathaiah}, \citenamefont {{Rajveer
  Jha}}, \citenamefont {Jha}, \citenamefont {{Banabir Pal}}, \citenamefont
  {Pal}, \citenamefont {{J. S. Matias}}, \citenamefont {Matias}, \citenamefont
  {{Pranab K. Das}}, \citenamefont {Das}, \citenamefont {{I. Vobornik}},
  \citenamefont {Vobornik}, \citenamefont {{R. A. Ribeiro}}, \citenamefont
  {Ribeiro}, \citenamefont {{D. D. Sarma}},\ and\ \citenamefont
  {Sarma}}]{thirupathaiah_2017}%
  \BibitemOpen
  \bibfield  {author} {\bibinfo {author} {\bibnamefont {{S. Thirupathaiah}}},
    \bibinfo {author} {\bibfnamefont
  {R.}~\bibnamefont {Jha}}, 
  \bibinfo {author} {\bibfnamefont {B.}~\bibnamefont {Pal}}, \bibinfo {author} {\bibfnamefont {J.~S.}\
  \bibnamefont {Matias}}, 
  \bibinfo {author} {\bibfnamefont {P.~K.}\ \bibnamefont {Das}}, \bibinfo {author} {\bibfnamefont
  {I.}~\bibnamefont {Vobornik}},  \bibinfo {author} {\bibfnamefont {R.~A.}\ \bibnamefont
  {Ribeiro}}, \ and\ \bibinfo
  {author} {\bibfnamefont {D.~D.}\ \bibnamefont {Sarma}},\ }\bibfield  {title}
  {\bibinfo {title} {Temperature-independent band structure of {W}{T}e{$_2$} as
  observed from angle-resolved photoemission spectroscopy},\ }\href
  {https://doi.org/10.1103/physrevb.96.165149} {\bibfield  {journal} {\bibinfo
  {journal} {Physical Review B}\ }\textbf {\bibinfo {volume} {96}},\ \bibinfo
  {pages} {165149} (\bibinfo {year} {2017})}\BibitemShut {NoStop}%
\bibitem [{\citenamefont {Wang}\ \emph {et~al.}(2017)\citenamefont {Wang},
  \citenamefont {{Yan Zhang}}, \citenamefont {Zhang}, \citenamefont {Huang},
  \citenamefont {Huang}, \citenamefont {Liu}, \citenamefont {Liang},
  \citenamefont {{Yuxiao Zhang}}, \citenamefont {Zhang}, \citenamefont {{Yu
  Xiao Zhang}}, \citenamefont {Zhang}, \citenamefont {Shen}, \citenamefont
  {{Bing Shen}}, \citenamefont {Shen}, \citenamefont {{Jing Liu}},
  \citenamefont {Liu}, \citenamefont {Hu}, \citenamefont {Ding}, \citenamefont
  {Liu}, \citenamefont {Hu}, \citenamefont {{Yong Hu}}, \citenamefont {He},
  \citenamefont {Zhao}, \citenamefont {Zhao}, \citenamefont {Yu}, \citenamefont
  {Yu}, \citenamefont {Yu}, \citenamefont {Hu}, \citenamefont {Wei},
  \citenamefont {Mao}, \citenamefont {Shi}, \citenamefont {Jia}, \citenamefont
  {Jia}, \citenamefont {Zhang}, \citenamefont {Zhang}, \citenamefont {Zhang},
  \citenamefont {Zhang}, \citenamefont {Yang}, \citenamefont {Wang},
  \citenamefont {Peng}, \citenamefont {Xu}, \citenamefont {Xu}, \citenamefont
  {Chen},\ and\ \citenamefont {Zhou}}]{wang_evidence_2017}%
  \BibitemOpen
  \bibfield  {author} {\bibinfo {author} {\bibfnamefont {C.}~\bibnamefont
  {Wang}}, \bibinfo {author} {\bibnamefont {{Yan Zhang}}}, \bibinfo {author}
  {\bibfnamefont {Y.}~\bibnamefont {Zhang}}, \bibinfo {author} {\bibfnamefont
  {J.}~\bibnamefont {Huang}}, \bibinfo {author} {\bibfnamefont
  {J.}~\bibnamefont {Huang}}, \bibinfo {author} {\bibfnamefont
  {G.}~\bibnamefont {Liu}}, \bibinfo {author} {\bibfnamefont {A.}~\bibnamefont
  {Liang}}, \bibinfo {author} {\bibnamefont {{Yuxiao Zhang}}}, \bibinfo
  {author} {\bibfnamefont {Y.}~\bibnamefont {Zhang}}, \bibinfo {author}
  {\bibnamefont {{Yu Xiao Zhang}}}, \bibinfo {author} {\bibfnamefont
  {Y.}~\bibnamefont {Zhang}}, \bibinfo {author} {\bibfnamefont
  {B.}~\bibnamefont {Shen}}, \bibinfo {author} {\bibnamefont {{Bing Shen}}},
  \bibinfo {author} {\bibfnamefont {B.}~\bibnamefont {Shen}}, \bibinfo {author}
  {\bibnamefont {{Jing Liu}}}, \bibinfo {author} {\bibfnamefont
  {J.}~\bibnamefont {Liu}}, \bibinfo {author} {\bibfnamefont {C.}~\bibnamefont
  {Hu}}, \bibinfo {author} {\bibfnamefont {Y.}~\bibnamefont {Ding}}, \bibinfo
  {author} {\bibfnamefont {D.}~\bibnamefont {Liu}}, \bibinfo {author}
  {\bibfnamefont {Y.}~\bibnamefont {Hu}}, \bibinfo {author} {\bibnamefont
  {{Yong Hu}}}, \bibinfo {author} {\bibfnamefont {S.}~\bibnamefont {He}},
  \bibinfo {author} {\bibfnamefont {L.}~\bibnamefont {Zhao}}, \bibinfo {author}
  {\bibfnamefont {L.}~\bibnamefont {Zhao}}, \bibinfo {author} {\bibfnamefont
  {L.}~\bibnamefont {Yu}}, \bibinfo {author} {\bibfnamefont {L.}~\bibnamefont
  {Yu}}, \bibinfo {author} {\bibfnamefont {L.}~\bibnamefont {Yu}}, \bibinfo
  {author} {\bibfnamefont {J.}~\bibnamefont {Hu}}, \bibinfo {author}
  {\bibfnamefont {J.}~\bibnamefont {Wei}}, \bibinfo {author} {\bibfnamefont
  {Z.}~\bibnamefont {Mao}}, \bibinfo {author} {\bibfnamefont {Y.}~\bibnamefont
  {Shi}}, \bibinfo {author} {\bibfnamefont {X.~W.}\ \bibnamefont {Jia}},
  \bibinfo {author} {\bibfnamefont {X.~W.}\ \bibnamefont {Jia}}, \bibinfo
  {author} {\bibfnamefont {F.}~\bibnamefont {Zhang}}, \bibinfo {author}
  {\bibfnamefont {F.~F.}\ \bibnamefont {Zhang}}, \bibinfo {author}
  {\bibfnamefont {S.~J.}\ \bibnamefont {Zhang}}, \bibinfo {author}
  {\bibfnamefont {S.}~\bibnamefont {Zhang}}, \bibinfo {author} {\bibfnamefont
  {F.}~\bibnamefont {Yang}}, \bibinfo {author} {\bibfnamefont {Z.}~\bibnamefont
  {Wang}}, \bibinfo {author} {\bibfnamefont {Q.}~\bibnamefont {Peng}}, \bibinfo
  {author} {\bibfnamefont {Z.}~\bibnamefont {Xu}}, \bibinfo {author}
  {\bibfnamefont {Z.}~\bibnamefont {Xu}}, \bibinfo {author} {\bibfnamefont
  {C.}~\bibnamefont {Chen}},\ and\ \bibinfo {author} {\bibfnamefont
  {X.}~\bibnamefont {Zhou}},\ }\bibfield  {title} {\bibinfo {title} {Evidence
  of electron-hole imbalance in {WT}e{$_2$} from high-resolution angle-resolved
  photoemission spectroscopy},\ }\href
  {https://doi.org/10.1088/0256-307x/34/9/097305} {\bibfield  {journal}
  {\bibinfo  {journal} {Chinese Physics Letters}\ }\textbf {\bibinfo {volume}
  {34}},\ \bibinfo {pages} {097305} (\bibinfo {year} {2017})}\BibitemShut
  {NoStop}%
\bibitem [{\citenamefont {Lv}\ \emph {et~al.}(2016)\citenamefont {Lv},
  \citenamefont {Zhang}, \citenamefont {Li}, \citenamefont {Pang},
  \citenamefont {Zhang}, \citenamefont {Lin}, \citenamefont {Zhou},
  \citenamefont {Yao}, \citenamefont {Chen}, \citenamefont {Zhang},
  \citenamefont {Lu}, \citenamefont {Liu}, \citenamefont {Chen},\ and\
  \citenamefont {Chen}}]{Lv2016}%
  \BibitemOpen
  \bibfield  {author} {\bibinfo {author} {\bibfnamefont {Y.-Y.}\ \bibnamefont
  {Lv}}, \bibinfo {author} {\bibfnamefont {B.-B.}\ \bibnamefont {Zhang}},
  \bibinfo {author} {\bibfnamefont {X.}~\bibnamefont {Li}}, \bibinfo {author}
  {\bibfnamefont {B.}~\bibnamefont {Pang}}, \bibinfo {author} {\bibfnamefont
  {F.}~\bibnamefont {Zhang}}, \bibinfo {author} {\bibfnamefont {D.-J.}\
  \bibnamefont {Lin}}, \bibinfo {author} {\bibfnamefont {J.}~\bibnamefont
  {Zhou}}, \bibinfo {author} {\bibfnamefont {S.-H.}\ \bibnamefont {Yao}},
  \bibinfo {author} {\bibfnamefont {Y.~B.}\ \bibnamefont {Chen}}, \bibinfo
  {author} {\bibfnamefont {S.-T.}\ \bibnamefont {Zhang}}, \bibinfo {author}
  {\bibfnamefont {M.}~\bibnamefont {Lu}}, \bibinfo {author} {\bibfnamefont
  {Z.}~\bibnamefont {Liu}}, \bibinfo {author} {\bibfnamefont {Y.}~\bibnamefont
  {Chen}},\ and\ \bibinfo {author} {\bibfnamefont {Y.-F.}\ \bibnamefont
  {Chen}},\ }\bibfield  {title} {\bibinfo {title} {Dramatically decreased
  magnetoresistance in non-stoichiometric {W}{T}e$_2$ crystals},\ }\href
  {https://doi.org/10.1038/srep26903} {\bibfield  {journal} {\bibinfo
  {journal} {Scientific Reports}\ }\textbf {\bibinfo {volume} {6}},\ \bibinfo
  {pages} {26903} (\bibinfo {year} {2016})}\BibitemShut {NoStop}%
\bibitem [{\citenamefont {Fu}\ \emph {et~al.}(2018)\citenamefont {Fu},
  \citenamefont {Pan}, \citenamefont {Bai}, \citenamefont {Fei}, \citenamefont
  {Umana-Membreno}, \citenamefont {Song}, \citenamefont {Wang}, \citenamefont
  {Wang},\ and\ \citenamefont {Song}}]{Fu2018}%
  \BibitemOpen
  \bibfield  {author} {\bibinfo {author} {\bibfnamefont {D.}~\bibnamefont
  {Fu}}, \bibinfo {author} {\bibfnamefont {X.}~\bibnamefont {Pan}}, \bibinfo
  {author} {\bibfnamefont {Z.}~\bibnamefont {Bai}}, \bibinfo {author}
  {\bibfnamefont {F.}~\bibnamefont {Fei}}, \bibinfo {author} {\bibfnamefont
  {G.~A.}\ \bibnamefont {Umana-Membreno}}, \bibinfo {author} {\bibfnamefont
  {H.}~\bibnamefont {Song}}, \bibinfo {author} {\bibfnamefont {X.}~\bibnamefont
  {Wang}}, \bibinfo {author} {\bibfnamefont {B.}~\bibnamefont {Wang}},\ and\
  \bibinfo {author} {\bibfnamefont {F.}~\bibnamefont {Song}},\ }\bibfield
  {title} {\bibinfo {title} {Tuning the electrical transport of type {II}
  {W}eyl semimetal {W}{T}e$_2$ nanodevices by {M}o doping},\ }\href
  {https://doi.org/10.1088/1361-6528/aaa811} {\bibfield  {journal} {\bibinfo
  {journal} {Nanotechnology}\ }\textbf {\bibinfo {volume} {29}},\ \bibinfo
  {pages} {135705} (\bibinfo {year} {2018})}\BibitemShut {NoStop}%
\bibitem [{\citenamefont {Gong}\ \emph {et~al.}(2018)\citenamefont {Gong},
  \citenamefont {Yang}, \citenamefont {Ge}, \citenamefont {Wang}, \citenamefont
  {Liang}, \citenamefont {Luo}, \citenamefont {Yan}, \citenamefont {Zhen},
  \citenamefont {Weng}, \citenamefont {Pi}, \citenamefont {Zhang},\ and\
  \citenamefont {Zhu}}]{Gong2018}%
  \BibitemOpen
  \bibfield  {author} {\bibinfo {author} {\bibfnamefont {J.-X.}\ \bibnamefont
  {Gong}}, \bibinfo {author} {\bibfnamefont {J.}~\bibnamefont {Yang}}, \bibinfo
  {author} {\bibfnamefont {M.}~\bibnamefont {Ge}}, \bibinfo {author}
  {\bibfnamefont {Y.-J.}\ \bibnamefont {Wang}}, \bibinfo {author}
  {\bibfnamefont {D.-D.}\ \bibnamefont {Liang}}, \bibinfo {author}
  {\bibfnamefont {L.}~\bibnamefont {Luo}}, \bibinfo {author} {\bibfnamefont
  {X.}~\bibnamefont {Yan}}, \bibinfo {author} {\bibfnamefont {W.-L.}\
  \bibnamefont {Zhen}}, \bibinfo {author} {\bibfnamefont {S.-R.}\ \bibnamefont
  {Weng}}, \bibinfo {author} {\bibfnamefont {L.}~\bibnamefont {Pi}}, \bibinfo
  {author} {\bibfnamefont {C.-J.}\ \bibnamefont {Zhang}},\ and\ \bibinfo
  {author} {\bibfnamefont {W.-K.}\ \bibnamefont {Zhu}},\ }\bibfield  {title}
  {\bibinfo {title} {Non-stoichiometry effects on the extreme magnetoresistance
  in {W}eyl semimetal {W}{T}e$_2$},\ }\href
  {https://doi.org/10.1088/0256-307X/35/9/097101} {\bibfield  {journal}
  {\bibinfo  {journal} {Chinese Physics Letters}\ }\textbf {\bibinfo {volume}
  {35}},\ \bibinfo {pages} {097101} (\bibinfo {year} {2018})}\BibitemShut
  {NoStop}%
\bibitem [{\citenamefont {Lieth}\ and\ \citenamefont {JCJM}(1977)}]{Lieth1977}%
  \BibitemOpen
  \bibfield  {author} {\bibinfo {author} {\bibfnamefont {R.~M.~A.}\
  \bibnamefont {Lieth}}\ and\ \bibinfo {author} {\bibfnamefont
  {T.}~\bibnamefont {JCJM}},\ }\href@noop {} {\emph {\bibinfo {title}
  {Preparation and Crystal Growth of Materials with Layered Structures}}},\
  edited by\ \bibinfo {editor} {\bibfnamefont {R.}~\bibnamefont {Lieth}}\
  (\bibinfo  {publisher} {Springer},\ \bibinfo {year} {1977})\BibitemShut
  {NoStop}%
\bibitem [{\citenamefont {Dawson}\ and\ \citenamefont
  {Bullett}(1987)}]{Dawson1987}%
  \BibitemOpen
  \bibfield  {author} {\bibinfo {author} {\bibfnamefont {W.~G.}\ \bibnamefont
  {Dawson}}\ and\ \bibinfo {author} {\bibfnamefont {D.~W.}\ \bibnamefont
  {Bullett}},\ }\bibfield  {title} {\bibinfo {title} {Electronic structure and
  crystallography of {M}o{T}e$_2$ and {W}{T}e$_2$},\ }\href
  {https://doi.org/10.1088/0022-3719/20/36/017} {\bibfield  {journal} {\bibinfo
   {journal} {Journal of Physics C: Solid State Physics}\ }\textbf {\bibinfo
  {volume} {20}},\ \bibinfo {pages} {6159} (\bibinfo {year}
  {1987})}\BibitemShut {NoStop}%
\bibitem [{\citenamefont {Lee}\ \emph {et~al.}(2015)\citenamefont {Lee},
  \citenamefont {Silva}, \citenamefont {Calderin}, \citenamefont {Nguyen},
  \citenamefont {Hollander}, \citenamefont {Bersch}, \citenamefont {Mallouk},\
  and\ \citenamefont {Robinson}}]{Lee2015}%
  \BibitemOpen
  \bibfield  {author} {\bibinfo {author} {\bibfnamefont {C.-H.}\ \bibnamefont
  {Lee}}, \bibinfo {author} {\bibfnamefont {E.~C.}\ \bibnamefont {Silva}},
  \bibinfo {author} {\bibfnamefont {L.}~\bibnamefont {Calderin}}, \bibinfo
  {author} {\bibfnamefont {M.~A.~T.}\ \bibnamefont {Nguyen}}, \bibinfo {author}
  {\bibfnamefont {M.~J.}\ \bibnamefont {Hollander}}, \bibinfo {author}
  {\bibfnamefont {B.}~\bibnamefont {Bersch}}, \bibinfo {author} {\bibfnamefont
  {T.~E.}\ \bibnamefont {Mallouk}},\ and\ \bibinfo {author} {\bibfnamefont
  {J.~A.}\ \bibnamefont {Robinson}},\ }\bibfield  {title} {\bibinfo {title}
  {Tungsten ditelluride: a layered semimetal},\ }\href
  {https://doi.org/10.1038/srep10013} {\bibfield  {journal} {\bibinfo
  {journal} {Scientific Reports}\ }\textbf {\bibinfo {volume} {5}},\ \bibinfo
  {pages} {10013} (\bibinfo {year} {2015})}\BibitemShut {NoStop}%
\bibitem [{\citenamefont {Brown}(1966)}]{Brown1966}%
  \BibitemOpen
  \bibfield  {author} {\bibinfo {author} {\bibfnamefont {B.~E.}\ \bibnamefont
  {Brown}},\ }\bibfield  {title} {\bibinfo {title} {The crystal structures of
  {W}{T}e$_2$ and high-temperature {M}o{T}e$_2$},\ }\href
  {https://doi.org/10.1107/S0365110X66000513} {\bibfield  {journal} {\bibinfo
  {journal} {Acta Crystallographica}\ }\textbf {\bibinfo {volume} {20}},\
  \bibinfo {pages} {268} (\bibinfo {year} {1966})}\BibitemShut {NoStop}%
\bibitem [{\citenamefont {Lv}\ \emph {et~al.}(2017)\citenamefont {Lv},
  \citenamefont {Cao}, \citenamefont {Li}, \citenamefont {Zhang}, \citenamefont
  {Wang}, \citenamefont {Pang}, \citenamefont {Ma}, \citenamefont {Lin},
  \citenamefont {Yao}, \citenamefont {Zhou}, \citenamefont {Chen},
  \citenamefont {Dong}, \citenamefont {Liu}, \citenamefont {Lu}, \citenamefont
  {Chen},\ and\ \citenamefont {Chen}}]{Lv2017}%
  \BibitemOpen
  \bibfield  {author} {\bibinfo {author} {\bibfnamefont {Y.-Y.}\ \bibnamefont
  {Lv}}, \bibinfo {author} {\bibfnamefont {L.}~\bibnamefont {Cao}}, \bibinfo
  {author} {\bibfnamefont {X.}~\bibnamefont {Li}}, \bibinfo {author}
  {\bibfnamefont {B.-B.}\ \bibnamefont {Zhang}}, \bibinfo {author}
  {\bibfnamefont {K.}~\bibnamefont {Wang}}, \bibinfo {author} {\bibfnamefont
  {B.}~\bibnamefont {Pang}}, \bibinfo {author} {\bibfnamefont {L.}~\bibnamefont
  {Ma}}, \bibinfo {author} {\bibfnamefont {D.}~\bibnamefont {Lin}}, \bibinfo
  {author} {\bibfnamefont {S.-H.}\ \bibnamefont {Yao}}, \bibinfo {author}
  {\bibfnamefont {J.}~\bibnamefont {Zhou}}, \bibinfo {author} {\bibfnamefont
  {Y.~B.}\ \bibnamefont {Chen}}, \bibinfo {author} {\bibfnamefont {S.-T.}\
  \bibnamefont {Dong}}, \bibinfo {author} {\bibfnamefont {W.}~\bibnamefont
  {Liu}}, \bibinfo {author} {\bibfnamefont {M.-H.}\ \bibnamefont {Lu}},
  \bibinfo {author} {\bibfnamefont {Y.}~\bibnamefont {Chen}},\ and\ \bibinfo
  {author} {\bibfnamefont {Y.-F.}\ \bibnamefont {Chen}},\ }\bibfield  {title}
  {\bibinfo {title} {Composition and temperature-dependent phase transition in
  miscible {M}o$_{1−x}${W}$_x${T}e$_2$ single crystals},\ }\href
  {https://doi.org/10.1038/srep44587} {\bibfield  {journal} {\bibinfo
  {journal} {Scientific Reports}\ }\textbf {\bibinfo {volume} {7}},\ \bibinfo
  {pages} {44587} (\bibinfo {year} {2017})}\BibitemShut {NoStop}%
\bibitem [{\citenamefont {Oliver}\ \emph {et~al.}(2017)\citenamefont {Oliver},
  \citenamefont {Beams}, \citenamefont {Krylyuk}, \citenamefont {Kalish},
  \citenamefont {Singh}, \citenamefont {Bruma}, \citenamefont {Tavazza},
  \citenamefont {Joshi}, \citenamefont {Stone}, \citenamefont {Stranick},
  \citenamefont {Davydov},\ and\ \citenamefont {Vora}}]{Oliver2017}%
  \BibitemOpen
  \bibfield  {author} {\bibinfo {author} {\bibfnamefont {S.~M.}\ \bibnamefont
  {Oliver}}, \bibinfo {author} {\bibfnamefont {R.}~\bibnamefont {Beams}},
  \bibinfo {author} {\bibfnamefont {S.}~\bibnamefont {Krylyuk}}, \bibinfo
  {author} {\bibfnamefont {I.}~\bibnamefont {Kalish}}, \bibinfo {author}
  {\bibfnamefont {A.~K.}\ \bibnamefont {Singh}}, \bibinfo {author}
  {\bibfnamefont {A.}~\bibnamefont {Bruma}}, \bibinfo {author} {\bibfnamefont
  {F.}~\bibnamefont {Tavazza}}, \bibinfo {author} {\bibfnamefont
  {J.}~\bibnamefont {Joshi}}, \bibinfo {author} {\bibfnamefont {I.~R.}\
  \bibnamefont {Stone}}, \bibinfo {author} {\bibfnamefont {S.~J.}\ \bibnamefont
  {Stranick}}, \bibinfo {author} {\bibfnamefont {A.~V.}\ \bibnamefont
  {Davydov}},\ and\ \bibinfo {author} {\bibfnamefont {P.~M.}\ \bibnamefont
  {Vora}},\ }\bibfield  {title} {\bibinfo {title} {The structural phases and
  vibrational properties of {M}o$_{1−x}${W}$_x${T}e$_2$ alloys},\ }\href
  {https://doi.org/10.1088/2053-1583/aa7a32} {\bibfield  {journal} {\bibinfo
  {journal} {2D Materials}\ }\textbf {\bibinfo {volume} {4}},\ \bibinfo {pages}
  {045008} (\bibinfo {year} {2017})}\BibitemShut {NoStop}%
\bibitem [{Sup()}]{SuppMat}%
  \BibitemOpen
  \href@noop {} {}\bibinfo {note} {See Supplemental Material at [URL will be
  inserted by publisher] for the frequencies of the quantum oscillations and
  the effective masses of the carriers in {W$_{1-x}$Mo$_x$Te$_2$}}\BibitemShut
  {NoStop}%
\bibitem [{\citenamefont {Shoenberg}(1984)}]{Shoenberg1984}%
  \BibitemOpen
  \bibfield  {author} {\bibinfo {author} {\bibfnamefont {D.}~\bibnamefont
  {Shoenberg}},\ }\href {https://doi.org/10.1017/CBO9780511897870} {\emph
  {\bibinfo {title} {Magnetic Oscillations in Metals}}}\ (\bibinfo  {publisher}
  {Cambridge University Press},\ \bibinfo {year} {1984})\BibitemShut {NoStop}%
\bibitem [{\citenamefont {Zhang}\ \emph {et~al.}(2021)\citenamefont {Zhang},
  \citenamefont {Kakani}, \citenamefont {Woods}, \citenamefont {Cha},\ and\
  \citenamefont {Shi}}]{Zhang_Xurui2021}%
  \BibitemOpen
  \bibfield  {author} {\bibinfo {author} {\bibfnamefont {X.}~\bibnamefont
  {Zhang}}, \bibinfo {author} {\bibfnamefont {V.}~\bibnamefont {Kakani}},
  \bibinfo {author} {\bibfnamefont {J.~M.}\ \bibnamefont {Woods}}, \bibinfo
  {author} {\bibfnamefont {J.~J.}\ \bibnamefont {Cha}},\ and\ \bibinfo {author}
  {\bibfnamefont {X.}~\bibnamefont {Shi}},\ }\bibfield  {title} {\bibinfo
  {title} {Thickness dependence of magnetotransport properties of tungsten
  ditelluride},\ }\href {https://doi.org/10.1103/PhysRevB.104.165126}
  {\bibfield  {journal} {\bibinfo  {journal} {Physical Review B}\ }\textbf
  {\bibinfo {volume} {104}},\ \bibinfo {pages} {165126} (\bibinfo {year}
  {2021})}\BibitemShut {NoStop}%
\bibitem [{\citenamefont {Rhodes}\ \emph {et~al.}(2017)\citenamefont {Rhodes},
  \citenamefont {Chenet}, \citenamefont {Janicek}, \citenamefont {Nyby},
  \citenamefont {Lin}, \citenamefont {Jin}, \citenamefont {Edelberg},
  \citenamefont {Mannebach}, \citenamefont {Finney}, \citenamefont {Antony},
  \citenamefont {Schiros}, \citenamefont {Klarr}, \citenamefont {Mazzoni},
  \citenamefont {Chin}, \citenamefont {Chiu}, \citenamefont {Zheng},
  \citenamefont {Zhang}, \citenamefont {Ernst}, \citenamefont {Dadap},
  \citenamefont {Tong}, \citenamefont {Ma}, \citenamefont {Lou}, \citenamefont
  {Wang}, \citenamefont {Qian}, \citenamefont {Ding}, \citenamefont {Osgood},
  \citenamefont {Paley}, \citenamefont {Lindenberg}, \citenamefont {Huang},
  \citenamefont {Pasupathy}, \citenamefont {Dubey}, \citenamefont {Hone},\ and\
  \citenamefont {Balicas}}]{Rhodes_2017}%
  \BibitemOpen
  \bibfield  {author} {\bibinfo {author} {\bibfnamefont {D.}~\bibnamefont
  {Rhodes}}, \bibinfo {author} {\bibfnamefont {D.~A.}\ \bibnamefont {Chenet}},
  \bibinfo {author} {\bibfnamefont {B.~E.}\ \bibnamefont {Janicek}}, \bibinfo
  {author} {\bibfnamefont {C.}~\bibnamefont {Nyby}}, \bibinfo {author}
  {\bibfnamefont {Y.}~\bibnamefont {Lin}}, \bibinfo {author} {\bibfnamefont
  {W.}~\bibnamefont {Jin}}, \bibinfo {author} {\bibfnamefont {D.}~\bibnamefont
  {Edelberg}}, \bibinfo {author} {\bibfnamefont {E.}~\bibnamefont {Mannebach}},
  \bibinfo {author} {\bibfnamefont {N.}~\bibnamefont {Finney}}, \bibinfo
  {author} {\bibfnamefont {A.}~\bibnamefont {Antony}}, \bibinfo {author}
  {\bibfnamefont {T.}~\bibnamefont {Schiros}}, \bibinfo {author} {\bibfnamefont
  {T.}~\bibnamefont {Klarr}}, \bibinfo {author} {\bibfnamefont
  {A.}~\bibnamefont {Mazzoni}}, \bibinfo {author} {\bibfnamefont
  {M.}~\bibnamefont {Chin}}, \bibinfo {author} {\bibfnamefont {Y.-c.}\
  \bibnamefont {Chiu}}, \bibinfo {author} {\bibfnamefont {W.}~\bibnamefont
  {Zheng}}, \bibinfo {author} {\bibfnamefont {Q.~R.}\ \bibnamefont {Zhang}},
  \bibinfo {author} {\bibfnamefont {F.}~\bibnamefont {Ernst}}, \bibinfo
  {author} {\bibfnamefont {J.~I.}\ \bibnamefont {Dadap}}, \bibinfo {author}
  {\bibfnamefont {X.}~\bibnamefont {Tong}}, \bibinfo {author} {\bibfnamefont
  {J.}~\bibnamefont {Ma}}, \bibinfo {author} {\bibfnamefont {R.}~\bibnamefont
  {Lou}}, \bibinfo {author} {\bibfnamefont {S.}~\bibnamefont {Wang}}, \bibinfo
  {author} {\bibfnamefont {T.}~\bibnamefont {Qian}}, \bibinfo {author}
  {\bibfnamefont {H.}~\bibnamefont {Ding}}, \bibinfo {author} {\bibfnamefont
  {R.~M.~J.}\ \bibnamefont {Osgood}}, \bibinfo {author} {\bibfnamefont {D.~W.}\
  \bibnamefont {Paley}}, \bibinfo {author} {\bibfnamefont {A.~M.}\ \bibnamefont
  {Lindenberg}}, \bibinfo {author} {\bibfnamefont {P.~Y.}\ \bibnamefont
  {Huang}}, \bibinfo {author} {\bibfnamefont {A.~N.}\ \bibnamefont
  {Pasupathy}}, \bibinfo {author} {\bibfnamefont {M.}~\bibnamefont {Dubey}},
  \bibinfo {author} {\bibfnamefont {J.}~\bibnamefont {Hone}},\ and\ \bibinfo
  {author} {\bibfnamefont {L.}~\bibnamefont {Balicas}},\ }\bibfield  {title}
  {\bibinfo {title} {Engineering the structural and electronic phases of
  {M}o{T}e{$_2$} through {W} substitution},\ }\href
  {https://doi.org/10.1021/acs.nanolett.6b04814} {\bibfield  {journal}
  {\bibinfo  {journal} {Nano Letters}\ }\textbf {\bibinfo {volume} {17}},\
  \bibinfo {pages} {1616} (\bibinfo {year} {2017})}\BibitemShut {NoStop}%
\bibitem [{\citenamefont {Jin}\ \emph {et~al.}(2018)\citenamefont {Jin},
  \citenamefont {Schiros}, \citenamefont {Lin}, \citenamefont {Ma},
  \citenamefont {Lou}, \citenamefont {Dai}, \citenamefont {Yu}, \citenamefont
  {Rhodes}, \citenamefont {Sadowski}, \citenamefont {Tong}, \citenamefont
  {Qian}, \citenamefont {Hashimoto}, \citenamefont {Lu}, \citenamefont {Dadap},
  \citenamefont {Wang}, \citenamefont {Santos}, \citenamefont {Zang},
  \citenamefont {Pohl}, \citenamefont {Ding}, \citenamefont {Hone},
  \citenamefont {Balicas}, \citenamefont {Pasupathy},\ and\ \citenamefont
  {Osgood}}]{Jin_2018}%
  \BibitemOpen
  \bibfield  {author} {\bibinfo {author} {\bibfnamefont {W.}~\bibnamefont
  {Jin}}, \bibinfo {author} {\bibfnamefont {T.}~\bibnamefont {Schiros}},
  \bibinfo {author} {\bibfnamefont {Y.}~\bibnamefont {Lin}}, \bibinfo {author}
  {\bibfnamefont {J.}~\bibnamefont {Ma}}, \bibinfo {author} {\bibfnamefont
  {R.}~\bibnamefont {Lou}}, \bibinfo {author} {\bibfnamefont {Z.}~\bibnamefont
  {Dai}}, \bibinfo {author} {\bibfnamefont {J.-X.}\ \bibnamefont {Yu}},
  \bibinfo {author} {\bibfnamefont {D.}~\bibnamefont {Rhodes}}, \bibinfo
  {author} {\bibfnamefont {J.~T.}\ \bibnamefont {Sadowski}}, \bibinfo {author}
  {\bibfnamefont {X.}~\bibnamefont {Tong}}, \bibinfo {author} {\bibfnamefont
  {T.}~\bibnamefont {Qian}}, \bibinfo {author} {\bibfnamefont {M.}~\bibnamefont
  {Hashimoto}}, \bibinfo {author} {\bibfnamefont {D.}~\bibnamefont {Lu}},
  \bibinfo {author} {\bibfnamefont {J.~I.}\ \bibnamefont {Dadap}}, \bibinfo
  {author} {\bibfnamefont {S.}~\bibnamefont {Wang}}, \bibinfo {author}
  {\bibfnamefont {E.~J.~G.}\ \bibnamefont {Santos}}, \bibinfo {author}
  {\bibfnamefont {J.}~\bibnamefont {Zang}}, \bibinfo {author} {\bibfnamefont
  {K.}~\bibnamefont {Pohl}}, \bibinfo {author} {\bibfnamefont {H.}~\bibnamefont
  {Ding}}, \bibinfo {author} {\bibfnamefont {J.}~\bibnamefont {Hone}}, \bibinfo
  {author} {\bibfnamefont {L.}~\bibnamefont {Balicas}}, \bibinfo {author}
  {\bibfnamefont {A.~N.}\ \bibnamefont {Pasupathy}},\ and\ \bibinfo {author}
  {\bibfnamefont {R.~M.}\ \bibnamefont {Osgood}},\ }\bibfield  {title}
  {\bibinfo {title} {Phase transition and electronic structure evolution of
  {M}o{T}e{$_2$} induced by {W} substitution},\ }\href
  {https://doi.org/10.1103/PhysRevB.98.144114} {\bibfield  {journal} {\bibinfo
  {journal} {Physical Review B}\ }\textbf {\bibinfo {volume} {98}},\ \bibinfo
  {pages} {144114} (\bibinfo {year} {2018})}\BibitemShut {NoStop}%
\bibitem [{\citenamefont {Lifshitz}\ and\ \citenamefont
  {Kosevich}(1956)}]{Lifshitz1956}%
  \BibitemOpen
  \bibfield  {author} {\bibinfo {author} {\bibfnamefont {I.~M.}\ \bibnamefont
  {Lifshitz}}\ and\ \bibinfo {author} {\bibfnamefont {A.~M.}\ \bibnamefont
  {Kosevich}},\ }\href@noop {} {\bibfield  {journal} {\bibinfo  {journal}
  {Zhurnal Eksperimental'noi i Teoreticheskoi Fiziki}\ }\textbf {\bibinfo
  {volume} {29}},\ \bibinfo {pages} {730} (\bibinfo {year} {1956})}\BibitemShut
  {NoStop}%
\bibitem [{\citenamefont {Culcer}\ \emph {et~al.}(2010)\citenamefont {Culcer},
  \citenamefont {Hwang}, \citenamefont {Stanescu},\ and\ \citenamefont
  {Das~Sarma}}]{Culcer_2010}%
  \BibitemOpen
  \bibfield  {author} {\bibinfo {author} {\bibfnamefont {D.}~\bibnamefont
  {Culcer}}, \bibinfo {author} {\bibfnamefont {E.~H.}\ \bibnamefont {Hwang}},
  \bibinfo {author} {\bibfnamefont {T.~D.}\ \bibnamefont {Stanescu}},\ and\
  \bibinfo {author} {\bibfnamefont {S.}~\bibnamefont {Das~Sarma}},\ }\bibfield
  {title} {\bibinfo {title} {Two-dimensional surface charge transport in
  topological insulators},\ }\href {https://doi.org/10.1103/PhysRevB.82.155457}
  {\bibfield  {journal} {\bibinfo  {journal} {Physical Review B}\ }\textbf
  {\bibinfo {volume} {82}},\ \bibinfo {pages} {155457} (\bibinfo {year}
  {2010})}\BibitemShut {NoStop}%
\bibitem [{\citenamefont {Coleridge}(1991)}]{Coleridge1991}%
  \BibitemOpen
  \bibfield  {author} {\bibinfo {author} {\bibfnamefont {P.~T.}\ \bibnamefont
  {Coleridge}},\ }\bibfield  {title} {\bibinfo {title} {Small-angle scattering
  in two-dimensional electron gases},\ }\href
  {https://doi.org/10.1103/PhysRevB.44.3793} {\bibfield  {journal} {\bibinfo
  {journal} {Physical Review B}\ }\textbf {\bibinfo {volume} {44}},\ \bibinfo
  {pages} {3793} (\bibinfo {year} {1991})}\BibitemShut {NoStop}%
\bibitem [{\citenamefont {Hong}\ \emph {et~al.}(2009)\citenamefont {Hong},
  \citenamefont {Zou},\ and\ \citenamefont {Zhu}}]{Hong2009}%
  \BibitemOpen
  \bibfield  {author} {\bibinfo {author} {\bibfnamefont {X.}~\bibnamefont
  {Hong}}, \bibinfo {author} {\bibfnamefont {K.}~\bibnamefont {Zou}},\ and\
  \bibinfo {author} {\bibfnamefont {J.}~\bibnamefont {Zhu}},\ }\bibfield
  {title} {\bibinfo {title} {Quantum scattering time and its implications on
  scattering sources in graphene},\ }\href
  {https://doi.org/10.1103/PhysRevB.80.241415} {\bibfield  {journal} {\bibinfo
  {journal} {Physical Review B}\ }\textbf {\bibinfo {volume} {80}},\ \bibinfo
  {pages} {241415(R)} (\bibinfo {year} {2009})}\BibitemShut {NoStop}%
\bibitem [{\citenamefont {Luo}\ \emph {et~al.}(2015)\citenamefont {Luo},
  \citenamefont {Li}, \citenamefont {Dai}, \citenamefont {Miao}, \citenamefont
  {Shi}, \citenamefont {Ding}, \citenamefont {Taylor}, \citenamefont
  {Yarotski}, \citenamefont {Prasankumar},\ and\ \citenamefont
  {Thompson}}]{Luo2015}%
  \BibitemOpen
  \bibfield  {author} {\bibinfo {author} {\bibfnamefont {Y.}~\bibnamefont
  {Luo}}, \bibinfo {author} {\bibfnamefont {H.}~\bibnamefont {Li}}, \bibinfo
  {author} {\bibfnamefont {Y.~M.}\ \bibnamefont {Dai}}, \bibinfo {author}
  {\bibfnamefont {H.}~\bibnamefont {Miao}}, \bibinfo {author} {\bibfnamefont
  {Y.~G.}\ \bibnamefont {Shi}}, \bibinfo {author} {\bibfnamefont
  {H.}~\bibnamefont {Ding}}, \bibinfo {author} {\bibfnamefont {A.~J.}\
  \bibnamefont {Taylor}}, \bibinfo {author} {\bibfnamefont {D.~A.}\
  \bibnamefont {Yarotski}}, \bibinfo {author} {\bibfnamefont {R.~P.}\
  \bibnamefont {Prasankumar}},\ and\ \bibinfo {author} {\bibfnamefont {J.~D.}\
  \bibnamefont {Thompson}},\ }\bibfield  {title} {\bibinfo {title} {Hall effect
  in the extremely large magnetoresistance semimetal {W}{T}e$_2$},\ }\href
  {https://doi.org/10.1063/1.4935240} {\bibfield  {journal} {\bibinfo
  {journal} {Applied Physics Letters}\ }\textbf {\bibinfo {volume} {107}},\
  \bibinfo {pages} {182411} (\bibinfo {year} {2015})}\BibitemShut {NoStop}%
\bibitem [{\citenamefont {Xu}\ \emph {et~al.}(2023)\citenamefont {Xu},
  \citenamefont {Wang}, \citenamefont {Pate}, \citenamefont {Zhu},
  \citenamefont {Mao}, \citenamefont {Zhang}, \citenamefont {Zhou},
  \citenamefont {Welp}, \citenamefont {Kwok}, \citenamefont {Chung},
  \citenamefont {Kanatzidis},\ and\ \citenamefont {Xiao}}]{Xu_Jing_2023}%
  \BibitemOpen
  \bibfield  {author} {\bibinfo {author} {\bibfnamefont {J.}~\bibnamefont
  {Xu}}, \bibinfo {author} {\bibfnamefont {Y.}~\bibnamefont {Wang}}, \bibinfo
  {author} {\bibfnamefont {S.~E.}\ \bibnamefont {Pate}}, \bibinfo {author}
  {\bibfnamefont {Y.}~\bibnamefont {Zhu}}, \bibinfo {author} {\bibfnamefont
  {Z.}~\bibnamefont {Mao}}, \bibinfo {author} {\bibfnamefont {X.}~\bibnamefont
  {Zhang}}, \bibinfo {author} {\bibfnamefont {X.}~\bibnamefont {Zhou}},
  \bibinfo {author} {\bibfnamefont {U.}~\bibnamefont {Welp}}, \bibinfo {author}
  {\bibfnamefont {W.-K.}\ \bibnamefont {Kwok}}, \bibinfo {author}
  {\bibfnamefont {D.~Y.}\ \bibnamefont {Chung}}, \bibinfo {author}
  {\bibfnamefont {M.~G.}\ \bibnamefont {Kanatzidis}},\ and\ \bibinfo {author}
  {\bibfnamefont {Z.-L.}\ \bibnamefont {Xiao}},\ }\bibfield  {title} {\bibinfo
  {title} {Unreliability of two-band model analysis of magnetoresistivities in
  unveiling temperature-driven {L}ifshitz transition},\ }\href
  {https://doi.org/10.1103/PhysRevB.107.035104} {\bibfield  {journal} {\bibinfo
   {journal} {Physical Review B}\ }\textbf {\bibinfo {volume} {107}},\ \bibinfo
  {pages} {035104} (\bibinfo {year} {2023})}\BibitemShut {NoStop}%
\bibitem [{\citenamefont {Singh}\ \emph {et~al.}(2022)\citenamefont {Singh},
  \citenamefont {Sasmal}, \citenamefont {Iyer}, \citenamefont {Thamizhavel},\
  and\ \citenamefont {Maiti}}]{Singh_2022}%
  \BibitemOpen
  \bibfield  {author} {\bibinfo {author} {\bibfnamefont {A.}~\bibnamefont
  {Singh}}, \bibinfo {author} {\bibfnamefont {S.}~\bibnamefont {Sasmal}},
  \bibinfo {author} {\bibfnamefont {K.~K.}\ \bibnamefont {Iyer}}, \bibinfo
  {author} {\bibfnamefont {A.}~\bibnamefont {Thamizhavel}},\ and\ \bibinfo
  {author} {\bibfnamefont {K.}~\bibnamefont {Maiti}},\ }\bibfield  {title}
  {\bibinfo {title} {Evolution of extremely large magnetoresistance in a {W}eyl
  semimetal, {WT}e{$_2$} with {N}i-doping},\ }\href
  {https://doi.org/10.1103/PhysRevMaterials.6.124202} {\bibfield  {journal}
  {\bibinfo  {journal} {Physical Review Materials}\ }\textbf {\bibinfo {volume}
  {6}},\ \bibinfo {pages} {124202} (\bibinfo {year} {2022})}\BibitemShut
  {NoStop}%
\end{thebibliography}


\end{document}

\renewcommand{\thefigure}{S\arabic{figure}}
\renewcommand{\thetable}{S\arabic{table}}
\renewcommand{\theequation}{S\arabic{equation}}

\makeatletter
\renewcommand\@bibitem[1]{\item\if@filesw \immediate\write\@auxout
    {\string\bibcite{#1}{S\the\value{\@listctr}}}\fi\ignorespaces}% <------------
\def\@biblabel#1{[S#1]}% <-------------------
\makeatother

%\preprint{}

\title{Supplemental Material: Quantum oscillation study of the large magnetoresistance in Mo substituted WTe$_2$ single crystals}


\author{Sourabh Barua}
\email[Corresponding author~]{sourabh.barua@cup.edu.in, baruasourabh@gmail.com}
%\thanks{}
\altaffiliation[Present Address: ]{ Central University Punjab, Bathinda, 151401, Punjab, India}
%\affiliation{}

\author{M. R. Lees}
%\thanks{}
%\affiliation{}

\author{G. Balakrishnan}
%\thanks{}
%\affiliation{}

\author{P. A. Goddard}
\email[Corresponding author~]{p.goddard@warwick.ac.uk}
%\thanks{}
\affiliation{Department of Physics, University of Warwick, Coventry, CV4 7AL, United Kingdom}


\date{\today}

\pacs{}
%\keywords{}

\maketitle

\newpage

\section{\label{sec:R_vs_T} Absolute resistivity versus temperature}

\begin{figure}[tbh!]
\centering
\includegraphics[width=\linewidth]{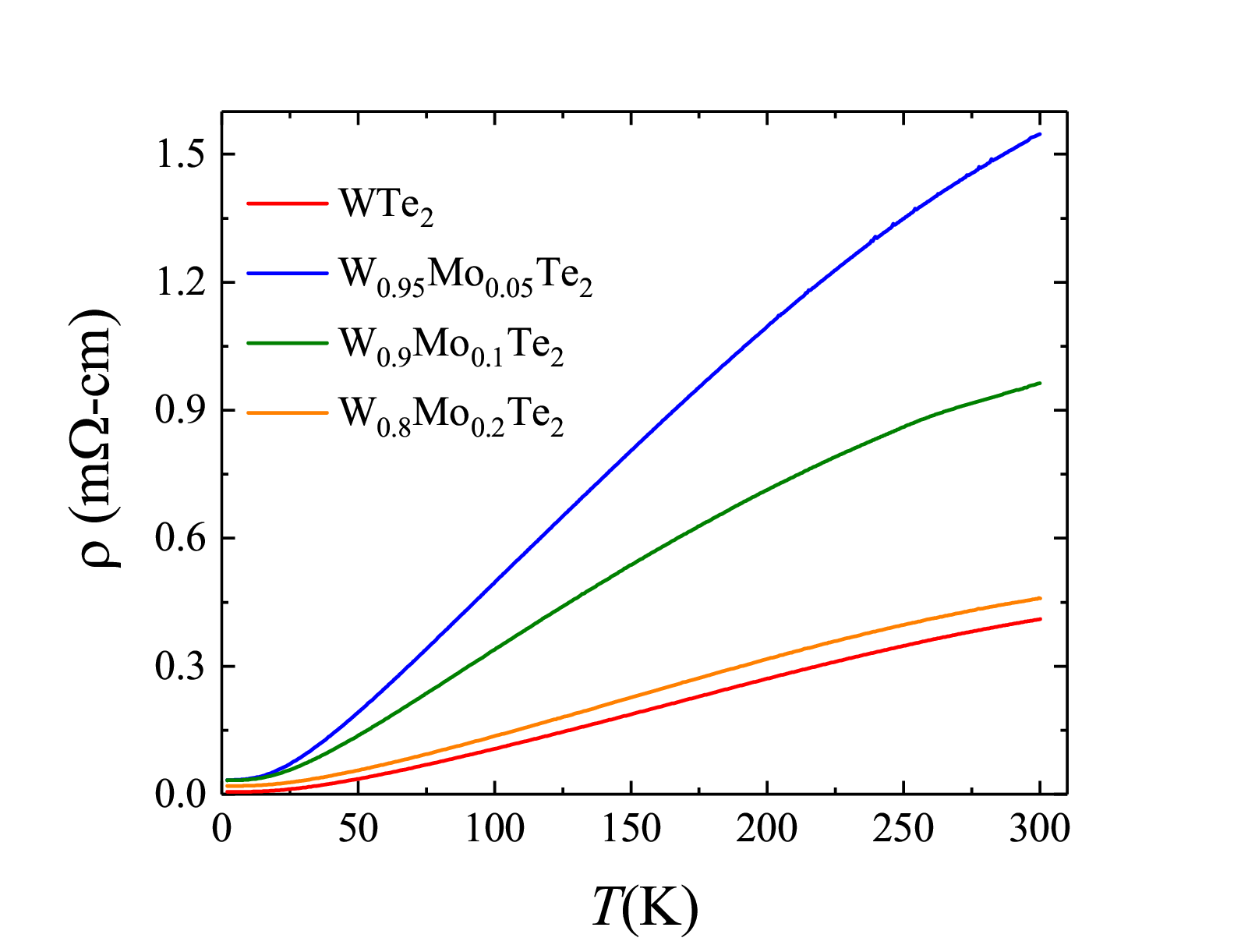}
\caption{Absolute resistivity versus temperature for the four different compositions of W$_{1-x}$Mo$_{x}$Te$_2$.}
\label{Figure_S1}
\end{figure}

\clearpage

\section{\label{sec:Mag}Quantum oscillations and effective masses}
\begin{table*}[tbh!]
\caption{\label{Table_S1} Frequencies of the quantum oscillations for each of the four different compositions of W$_{1-x}$Mo$_{x}$Te$_2$ ($x=0$, 0.05, 0.1, 0.2) obtained from the FFT at 2~K. These oscillations were measured for $H \parallel c$. Previously reported values for unsubstituted WTe$_2$ are also given.}
\begin{ruledtabular}
\begin{tabular}{cccccc}
Frequency & \multicolumn{2}{c}{WTe$_2$} & W$_{0.95}$Mo$_{0.05}$Te$_2$ & W$_{0.9}$Mo$_{0.1}$Te$_2$ & W$_{0.8}$Mo$_{0.2}$Te$_2$ \\
& This work & Previous work~\cite{Rhodes2015,Zhu2015,Cai2015} &&& \\
$\mathrm{F}_{\alpha 1}$ (T) &  $90.9 \pm 0.4$ &  90--94.7   &  $94.0 \pm 0.4$ &  $96.1 \pm 0.6$ &  $97.1  \pm  0.6$ \\
$\mathrm{F}_{\beta 1}$ (T)  & $130.0  \pm  0.5$ & 125--132    & $129.4 \pm  0.3$ & $128.5 \pm  0.2$ & $128.1 \pm  0.3$ \\
$\mathrm{F}_{\beta 2}$ (T)  & $146.2 \pm  0.6$ & 142--148    & $146.8 \pm  1.0$ & $146.1 \pm  0.5$ & $146.6 \pm  1.0$ \\
$\mathrm{F}_{\alpha 2}$ (T) & $164.1 \pm  1.3$ & 160--166    & $166.5 \pm  1.2$ & $175.6 \pm  0.6$ &   -             \\
\end{tabular} 
\end{ruledtabular}
\end{table*}

\begin{table*}[tbh!]
\caption{\label{Table_S2} Effective masses of the carriers in units of $m_\mathrm{e}$ corresponding to the different frequencies in the oscillations, obtained by fitting the temperature dependence of the FFT amplitudes for each of the four compositions W$_{1-x}$Mo$_{x}$Te$_2$ ($x = 0$, 0.05, 0.1, 0.2). The frequency $\mathrm{F}_{\alpha 2}$ was not resolved for W$_{0.8}$Mo$_{0.2}$Te$_2$. Previously reported values for the effective masses for unsubstituted WTe$_2$ are also given.}
\begin{ruledtabular}
\begin{tabular}{cccccc}
Effective masses & \multicolumn{2}{c}{WTe$_2$}  & W$_{0.95}$Mo$_{0.05}$Te$_2$ & W$_{0.9}$Mo$_{0.1}$Te$_2$ & W$_{0.8}$Mo$_{0.2}$Te$_2$ \\
& This work & Previous work~\cite{Cai2015,Rhodes2015,Xiang2015}& & & \\
$m^*_{\mathrm{F}_{\alpha 1}}$ & $0.35 \pm 0.01$ & 0.42 -- 0.462 & $0.33 \pm 0.01$ & $0.36 \pm 0.01$ &  $0.35 \pm 0.03$  \\
$m^*_{\mathrm{F}_{\beta 1}}$  & $0.26 \pm 0.01$ & 0.322 -- 0.387 & $0.24 \pm 0.01$ & $0.26 \pm 0.01$ &  $0.22 \pm 0.02$  \\
$m^*_{\mathrm{F}_{\beta 2}}$  & $0.30 \pm 0.01$ & 0.304 -- 0.414 & $0.26 \pm 0.01$ & $0.33 \pm 0.02$ &  $0.25 \pm 0.02$  \\
$m^*_{\mathrm{F}_{\alpha 2}}$ & $0.40 \pm 0.03$ & 0.313 -- 0.47 & $0.42 \pm 0.02$ & $0.42 \pm 0.03$ &    -              \\
\end{tabular}
\end{ruledtabular}
\end{table*}

\section{Two-band model}

The longitudinal ($\rho_{xx}$) and transverse ($\rho_{xy}$) magnetoresistivities  in the two-band model are given by 

\begin{subequations}
\begin{equation}
\rho_{xx} = \frac{(n_\mathrm{e} \mu_\mathrm{e} + n_\mathrm{h} \mu_\mathrm{h}) + (n_\mathrm{e} \mu_\mathrm{e} \mu_\mathrm{h}^2 + n_\mathrm{h} \mu_\mathrm{h} \mu_\mathrm{e}^2) B^2}{e[(n_\mathrm{e} \mu_\mathrm{e} + n_\mathrm{h} \mu_\mathrm{h})^2 + (n_\mathrm{h} - n_\mathrm{e})^2 \mu_\mathrm{e}^2 \mu_\mathrm{h}^2 B^2]},
\label{Equation_1a}
\end{equation}
\begin{equation}
\rho_{xy} = \frac{(n_\mathrm{h} \mu_\mathrm{h}^2 - n_\mathrm{e} \mu_\mathrm{e}^2) B + \mu_\mathrm{e}^2 \mu_\mathrm{h}^2 (n_\mathrm{h} - n_\mathrm{e}) B^3}{e[(n_\mathrm{e} \mu_\mathrm{e} + n_\mathrm{h} \mu_\mathrm{h})^2 + (n_\mathrm{h} - n_\mathrm{e})^2 \mu_\mathrm{e}^2 \mu_\mathrm{h}^2 B^2]},
\label{Equation_1b}
\end{equation}
\end{subequations}
where $B$ is the magnetic field, $e$ is the electron charge, and $n_\mathrm{e}$, $n_\mathrm{h}$, $\mu_\mathrm{e}$ and $\mu_\mathrm{h}$ are total electron concentration, total hole concentration, electron mobility and hole mobility, respectively~\cite{Wang2015,Fu2018,Wang2016,Fatemi2017,Luo2015}. 
%\section{Fit of the magnetoresistance (MR) to a two band model}
%\renewcommand{\thefigure}{S\arabic{figure}}
%\begin{figure}[h!]
%\centering
%\includegraphics[width=\linewidth]{Figure_Supp1.eps}
%\caption{Fit of the MR (fraction) at 5~K to a two band model for (a) WTe$_2$, (b) W$_{0.95}$Mo$_{0.05}$Te$_2$, (c) W$_{0.9}$Mo$_{0.1}$Te$_2$ and (d) W$_{0.8}$Mo$_{0.2}$Te$_2$, as described in the text of the main paper.}
%\label{Figure_Supp1}
%\end{figure}

%The MR (fraction) of all the compositions at 5~K are fitted to a two band model as described in the main text. The results of the fit are shown in figure~\ref{Figure_Supp1}. 

%\bibliography{References}

%